\documentclass[journal,comsoc]{IEEEtran}
\usepackage[T1]{fontenc}
\usepackage{textcomp}

\usepackage{amsthm}
\usepackage{bm}
\usepackage{xcolor}
\usepackage{graphicx}
\usepackage{graphics}
\usepackage{subfigure}
\usepackage{algorithm}
\usepackage{algorithmic}
\usepackage{setspace}
\usepackage[algo2e,linesnumbered,ruled]{algorithm2e}
\usepackage{float}
\usepackage{array}
\usepackage{multirow}
\usepackage{textcomp}

\newtheorem{definition}{Definition}

\newtheorem{example}{Example}

\SetKwProg{Fn}{Function}{}{end}

\usepackage{amsmath}
\usepackage{amssymb}

\usepackage[cmintegrals]{newtxmath}

\newcolumntype{M}[1]{>{\centering\arraybackslash}m{#1}}

\DeclareSymbolFont{symbolsC}{U}{txsyc}{m}{n}
\DeclareMathSymbol{\notniFromTxfonts}{\mathrel}{symbolsC}{61}

\ifCLASSOPTIONcompsoc
  \usepackage[nocompress]{cite}
\else
  \usepackage{cite}
\fi
\begin{document}
%

\title{Edge-assisted Parallel Uncertain Skyline Processing for Low-latency IoE Analysis}

\author{Chuan-Chi~Lai,~\IEEEmembership{Member,~IEEE,} Yan-Lin~Chen, Bo-Xin~Liu, and~Chuan-Ming~Liu,~\IEEEmembership{Member,~IEEE}
\IEEEcompsocitemizethanks{
\IEEEcompsocthanksitem The preliminary work was presented in 2020 IEEE GLOBECOM~\cite{9348055}.
\IEEEcompsocthanksitem This study was supported by the National Science and Technology Council, Taiwan, R.O.C. under Grant Nos. NSTC 112-2221-E-194-048-MY2, NSTC 113-2221-E-027-051-, and NSTC 114-2221-E-194-062-. This study was also partially supported by the Advanced Institute of Manufacturing with High-tech Innovations (AIM-HI) from The Featured Areas Research Center Program within the framework of the Higher Education Sprout Project by the Ministry of Education (MOE) in Taiwan. \emph{(Corresponding author: Chuan-Ming Liu.)}
\IEEEcompsocthanksitem C.-C. Lai is with the Department of Communications Engineering, National Chung Cheng University, Minxiong Township, Chiayi County 621301, Taiwan, and also with the Advanced Institute of Manufacturing with High-tech Innovations (AIM-HI), National Chung Cheng University, Minxiong Township, Chiayi County 621301, Taiwan (e-mail: chuanclai@ccu.edu.tw).  
\IEEEcompsocthanksitem Y.-L. Chen is with the R\&D Dept., Trend Micro Inc, Taipei 10669, Taiwan (e-mail: yanlin\_chen@trendmicro.com).
\IEEEcompsocthanksitem B.-X. Liu is with the Advanced Packaging Manufacturing System Section, Taiwan Semiconductor Manufacturing Company, Ltd (TSMC), Zhunan Township, Miaoli County 350012, Taiwan (e-mail: bxliub@tsmc.com).
\IEEEcompsocthanksitem C.-M. Liu is with the Department of Computer Science and Information Engineering, National Taipei University of Technology, Taipei 10618, Taiwan (e-mail: cmliu@ntut.edu.tw).
\IEEEcompsocthanksitem Copyright \copyright~2025 IEEE. Personal use of this material is permitted. However, permission to use this material for any other purposes must be obtained from the IEEE by sending a request to pubs-permissions@ieee.org.}
}

%
%

\markboth{Preprint submitted to IEEE Internet of Things Journal}%
{Lai \MakeLowercase{\textit{et al.}}: Bare Demo of IEEEtran.cls for Computer Society Journals}
%



\IEEEtitleabstractindextext{%
\begin{abstract}
Due to the Internet of Everything (IoE), data generated in our life become larger. As a result, we need more effort to analyze the data and extract valuable information. In the cloud computing environment, all data analysis is done in the cloud, and the client only needs less computing power to handle some simple tasks. However, with the rapid increase in data volume, sending all data to the cloud via the Internet has become more expensive. The required cloud computing resources have also become larger. To solve this problem, edge computing is proposed. Edge is granted with more computation power to process data before sending it to the cloud. Therefore, the data transmitted over the Internet and the computing resources required by the cloud can be effectively reduced. In this work, we proposed an Edge-assisted Parallel Uncertain Skyline (EPUS) algorithm for emerging low-latency IoE analytic applications. We use the concept of skyline candidate set to prune data that are less likely to become the skyline data on the parallel edge computing nodes. With the candidate skyline set, each edge computing node only sends the information required to the server for updating the global skyline, which reduces the amount of data that transfer over the internet. According to the simulation results, the proposed method is better than two comparative methods, which reduces the latency of processing two-dimensional data by more than 50\%. For high-dimensional data, the proposed EPUS method also outperforms the other existing methods.
\end{abstract}
\begin{IEEEkeywords}
Skyline Query, Internet of Everything, Uncertain Data, Edge Computing, Latency
\end{IEEEkeywords}}

\maketitle

\IEEEdisplaynontitleabstractindextext

%
\IEEEpeerreviewmaketitle

\section{Introduction}\label{sec:introduction}
\IEEEPARstart{T}{he}
\emph{Internet of Everything} (IoE) devices in life have grown rapidly and have generated larger and faster data. How to effectively process these massive IoE data and provide real-time analysis is an important challenge. \emph{Edge Computing} therefore has become a promising computing model that can process big data streams in a distributed or parallel manner to provide rapid response to meet the low-latency requirement of emerging IoE applications~\cite{9348055,7488250,8089336,8630994,9123504,9162084,9330566}. The edge computing model allocates more computing resources to edge servers to deal with big data problems, rather than cloud computing models that use large computing server clusters. Services implemented through edge computing can effectively reduce the response time for processing big data streams, and can quickly answer user queries. Therefore, this motivates us to propose a query processing method for stream computing services based on edge computing environments~\cite{8731646}~\cite{9206577}.

For IoE data analytic applications, the uncertainty of collected big data is a challenge~\cite{8767322,8288619,9044717}. For example, the collected data may include a lot of errors or incomplete information. Also, the data entering into the analytic system may be very dynamic in IoE environments. These characteristics of data can be modeled as uncertain data objects with a probabilistic data model\cite{Wang2013}. The probabilistic data model can make the system more effective in query processing and statistical analysis. 
We hence consider how to process and monitor the skyline query over uncertain data streams to support low-latency IoE data analytic applications in the forthcoming IoE era. In fact, uncertain data is more complex and require more computation to process in comparison to certain data. 
Also, data streams are time-sensitive, which means the process time are not allowed to be long, or the result might not be usable for low-latency IoE applications~\cite{8476168,9032170,9499122,9502154}.

We consider a kind of spatial query, \emph{Skyline}, in this work. Skyline query is a common data processing technique for searching the candidate result of multiple criteria decision making (also known as multi-objective optimization or Pareto optimization) problems~\cite{8920092}~\cite{8935189}. Skyline is also called the \emph{Pareto frontier} in Pareto optimization.
Skyline query has been successfully applied to many well-known applications, such as location-based services~\cite{6341729}~\cite{AL_Jawarneh2020}, transportation~\cite{9123533}, crowd-sourcing~\cite{8865657}, mobile crowd-sensing~\cite{9354847}, and cloud computing~\cite{9252150}.

Most skyline query processing methods~\cite{10.1145/1061318.1061320}~\cite{10.1145/1559845.1559898} are designed based on centralized computing environments.
Recently, some research~\cite{7219441,KOH2017114,9186333} discussed skyline query using MapReduce framework based on parallel Hadoop system in cloud computing environments. However, there is no or little research about skyline query based on edge computing environment. As the result, this give us the motivation of proposing a skyline query method based on an edge computing environment.

Hence, we proposed a workable solution, \emph{Edge-assisted Parallel Uncertain Skyline} (EPUS) algorithm, for efficiently processing skyline queries over uncertain data streams based on a parallel edge computing environment. 
Using EPUS, \emph{Edge Computing Nodes} (ECN) can collaboratively prune input data that cannot be the skyline. In this way, the average latency (or average computation time) is reduced a lot, which is obviously better than the brute-force method. In addition, the proposed approach is not only suitable for uncertain data but also can process some data with slight modification.


The main contributions of our work are summarized as follows.
\begin{itemize}
	\item Currently, this work is one of the pioneers in discussing the design of real-time skyline query processing algorithms that combine uncertain data streams and edge computing environments.
	\item We propose an Edge-assisted Parallel Uncertain Skyline (EPUS) algorithm to effectively filter out irrelevant information, thereby improving the efficiency of skyline query processing for uncertain data streams.
	\item The proposed EPUS can be developed as a low-latency service/tool to help feature/parameter selection of big uncertain data streams, thereby reducing the expensive data preprocessing overhead of artificial intelligence (AI) model training in large-scale IoE data analysis applications in the future.
	\item Building upon our preliminary results in~\cite{9348055}, this work offers a more in-depth explanation of the proposed solution, thorough analysis, and extensive comparative simulation results for the edge-assisted IoE environment, specifically considering the enhanced Machine Type Communication (eMTC) model.
	\item The simulation results indicate that the proposed EPUS significantly improves the system performance of probabilistic skyline query processing in terms of average transmission cost and average system latency.
\end{itemize}

The rest of paper is organized as follows. Some literature on skyline query processing are presented in Section~\ref{sec:related_works}.
Section~\ref{sec:preliminary_and_problem} introduces the preliminary and problem statement of this work.
The proposed approach with algorithms and examples are explained in Section~\ref{sec:proposed_approach}. 
Section~\ref{sec:analysis} introduces the analysis and discussion of the time complexity and transmission cost of the proposed method. In Section~\ref{sec:simulation}, we conducts some simulations in various situations to validate the performance of the proposed EPUS algorithm. Finally, we make the conclusion remarks in Section~\ref{sec:conclusion}.

\section{Related Work}
\label{sec:related_works}
With the help of virtualization technology, edge computing has become a popular computing model to most emerging IoE applications~\cite{10.1145/3461841} in recent years. The service provider can deploy a series of micro services based on virtualization technology to edge computing environments and then provide services to users. Compared with traditional centralized computing and cloud computing, the edge computing framework~\cite{8731646}~\cite{9206577} is more suitable for processing big data brought by massive IoE devices due to its low-latency advantage. 

Although the edge computing framework can provide an efficient computing environment for IoE data analytic applications, there are still some challenges that need to be resolved. For example, if we want to deploy a multi-criteria decision making service based on the edge environment for IoE data analytic applications, how to design the data processing procedure? However, most of the existing works are not designed based on the edge computing environment.

Skyline is one of the most popular queries and is widely used in decision-making applications~\cite{10.1145/1559845.1559898}~\cite{DBLP:journals/corr/KalyvasT17}. Christos~\textit{et al.}~\cite{DBLP:journals/corr/KalyvasT17} investigated skyline query processing, which also contains many variants of skyline query. 
For snapshot skyline queries on certain data, Papadias~\textit{et al.}~\cite{10.1145/1061318.1061320} proposed the \emph{Branch-and-Bound Skyline} (BBS) method. BBS was designed based on the best-first nearest neighbor search~\cite{Hjaltason:1999:DBS:320248.320255} to optimize the handling of skyline's I/O overhead. However, snapshot skylines are not very useful in streaming environments because they keep changing over time.
Zhang~\textit{et al.}~\cite{10.1145/1559845.1559898} focused on frequent skyline queries and proposed a two-step framework, \emph{Frequent Skyline Query over a sliding Window} (FSQW), including filtering and sampling steps. FSQW was proposed to minimize the transmission cost of processing frequent skyline queries in a client-server computing architecture.

In addition, several distributed solutions for skyline query processing have been explored~\cite{KOH2017114,Sun2010}. 
Sun~\textit{et al.}~\cite{Sun2010} proposed a tree-based grid partition indexing approach, \emph{GridSky}, for master-slave computing clusters to process skyline queries over distributed certain data streams. In this framework, the master node incrementally updates the final skyline after receiving all local skylines from the slave nodes. The combination of GridSky indexing and the master-slave model enables both slave and master nodes to prune irrelevant data, thereby reducing transmission costs between nodes. 
Koh~\textit{et al.}~\cite{KOH2017114} introduced a parallel skyline processing method based on the \emph{MapReduce} framework, called MR-Sketch. This approach applies middle-split partitioning in both the mapper and reducer steps, and evaluates the performance of skyline query processing under different partitioning strategies. However, these methods do not address uncertain data, which is generally more complex and requires higher computational costs than certain data.

There are some other research focusing on uncertain data streams. Due to the uncertainty of data attributes, there are many combinations of skyline results, and the possible worlds of these combinations are called probabilistic skylines.
For handling continuous probabilistic skyline, Zhang~\textit{et al.}~\cite{ZHANG20131212} proposed a threshold-based method, Probabilistic Skyline Operator (PSO). PSO first retrieves ``top-k" skyline data objects using multiple given probability thresholds. Then, PSO uses the maximum probability of the top-k skyline as the threshold to filter out irrelevant data, thereby effectively maintaining the probability skyline.
Pan~\textit{et al.}~\cite{doi:10.1155/2014/365064} used a \emph{Candidate List} (CL) to store data that are possible to be skyline and this design significant reduced a lot of required computing resource. However, this work did not consider distributed edge computing environments.
A MapReduce-based parallel skyline processing with bucket partition method was proposed by Gavagsaz~\cite{Gavagsaz2020}, but this work did not support continuous skyline query in streaming based real-time data analytic applications.

Although many studies are mentioned above, none of them considers continuous skyline queries based on uncertain data streams in edge computing environments. Therefore, we propose a parallel processing method based on an edge computing environment to effectively handle continuous skyline queries. In order to highlight our novelty and contribution, the comparative summary of the related work is presented in Table~\ref{table:comparison_of_methods}. We hope that our proposed method can attract more research on related topics and contribute to this field in the future. 

\begin{table*}[!t]
	\renewcommand{\arraystretch}{1.1}
	\centering
	\small
	\caption{Comparative Summary of The Related Work}
	\label{table:comparison_of_methods} 
	\begin{tabular}{|M{2.4em}|M{12.3em}|M{9em}|M{4.7em}|M{4.5em}|M{4.85em}|M{4.85em}|M{3.55em}|}
		\hline
		\textbf{Refer. No.} & \textbf{Method} & \textbf{Performance Metrics} & \textbf{Query Type} & \textbf{Data Type} & \textbf{Computing Model} & \textbf{Edge Computing} & \textbf{Data Streams} \\ \hline\hline  
		\cite{10.1145/1061318.1061320} & \raggedright Branch-and-Bound Skyline (BBS) & \raggedright Minimize I/O cost & Snapshot & Certain & Centralized & \texttimes & \texttimes \\ \hline 
		\cite{10.1145/1559845.1559898} & \raggedright Frequent Skyline Query over a sliding Window (FSQW) & \raggedright Minimize transmission cost & Frequent & Certain & Centralized & \texttimes & \texttimes \\ \hline 	
		\cite{KOH2017114} & \raggedright MapReduce framework with middle-split partitioning (MR-Sketch) & \raggedright Minimize computation time & Snapshot & Certain & Distributed & \texttimes & $\checkmark$ \\ \hline 	
		\cite{Sun2010} & \raggedright Tree-based grid partition indexing (GridSky) & \raggedright Minimize transmission cost and computation time & Continuous & Certain & Distributed & \texttimes & $\checkmark$ \\ \hline
		\cite{ZHANG20131212} & \raggedright Probabilistic Skyline Operator (PSO) with multiple given thresholds for data pruning & \raggedright Minimize computational delay & Continuous & Uncertain & Centralized & \texttimes & $\checkmark$ \\ \hline
		\cite{doi:10.1155/2014/365064} & \raggedright Candidate List (CL) for maintaining the possible skyline set & \raggedright Minimize computation time & Continuous & Uncertain & Centralized & \texttimes & $\checkmark$ \\ \hline
		\cite{Gavagsaz2020} & \raggedright Parallel computation
		of probabilistic skyline query (PCPS) based on MapReduce framework with bucket partition & \raggedright Minimize computation time & Snapshot & Uncertain & Distributed & \texttimes & \texttimes \\ \hline
		\textbf{This work} & \raggedright \textbf{Edge-assisted parallel uncertain skyline (EPUS) with edge candidate sets} & \raggedright \textbf{Minimize transmission cost and computation time} & \textbf{Continuous} & \textbf{Uncertain}& \textbf{Distributed} & $\bm\checkmark$ & $\bm\checkmark$ \\ 
		\hline 
	\end{tabular}
\end{table*}

\section{Preliminaries and Problem Statement}
\label{sec:preliminary_and_problem}

\subsection{Preliminary}
Uncertain data refers to information whose value is not precisely known, and it can be modeled in three main ways: the fuzzy model~\cite{Fuzzy_Databases_2006}, the evidence-oriented model~\cite{10.5555/645918.672335,542025}, and the probabilistic model~\cite{1617375}. The probabilistic model~\cite{Prabhakar2009} is further classified into continuous and discrete probabilistic data models. In the continuous probabilistic data model, each uncertain data object $u_i$ is characterized by a \emph{Probability Density Function} (PDF), denoted as $\text{pdf}(u_i)$, where $\text{pdf}(u_i)=\int_{x\in u_i}\text{pdf}(x)dx=1$.

In this work, we consider uncertain data with the discrete probabilistic data model and it can be defined as follows:
\begin{definition}[Discrete Probabilistic Data Model]
	\label{def:DPDM}
	Given an uncertain data object $u_i=\{u_{i,1},u_{i,2},\dots,u_{i,j}\}$, which includes $j$ instances. Each instance is with its own probability of occurrence $\text{Pr}(u_{i,j})$. Hence, the occurrence probability of uncertain data object $u_i$ is the sum of all instances' occurrence probabilities and it can be denoted as
	\begin{equation*}
		\text{Pr}(u_{i})=\sum_{u_{i,j}\in u_{i},\forall j}u_{i,j} \leq 1.
	\end{equation*}
\end{definition} 

\begin{figure}[!t]
	\centering
	\includegraphics[width=.4\columnwidth]{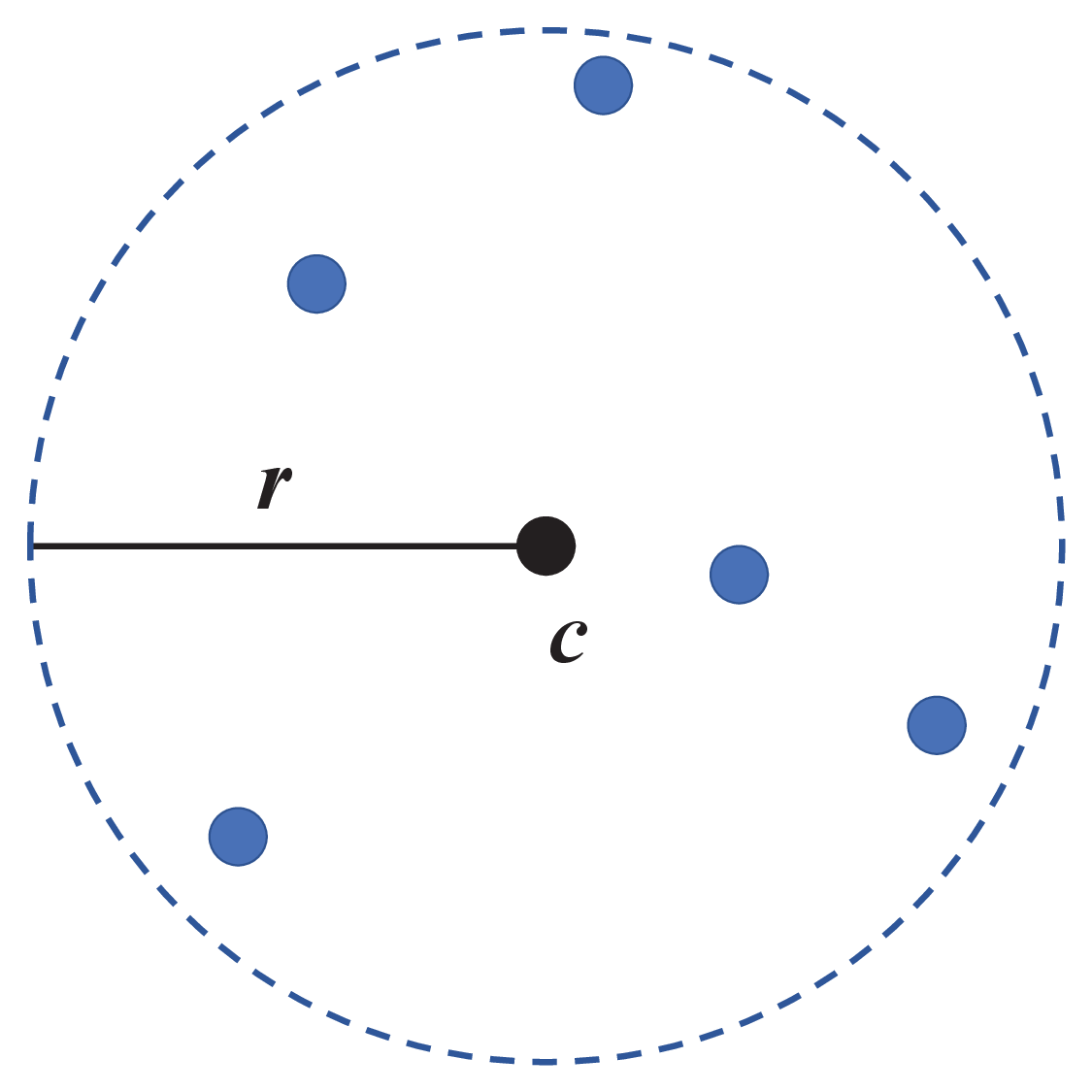}
	\caption{A simple example of a two-dimensional uncertain data object.}
	\label{fig:ex:2d_uncertain_data_object}
\end{figure}

An incoming data object can be represented as a $d$-sphere according to the number of data dimension $d$. Given a center point $c$ of uncertain data object $u_i$, and the corresponding radius $r$, all instance of $u_i$ will locate inside the $d$-sphere. In another words, the Euclidean distance between $c$ and any instance of $u_i$ will not exceed $r$. 
An example of a two-dimensional uncertain data object is shown as in Fig.~\ref{fig:ex:2d_uncertain_data_object}. Each blue point indicates one data instance. 
An example in Table~\ref{table:2d_uncertain_data_set} shows a data set including three two-dimensional uncertain data objects. Each data object has three instances and each instance contains two attributes with its own existing (or occurrence) probability.


In a data stream environment, new data objects continuously arrive, each typically associated with a timestamp and a limited lifespan. Once data becomes obsolete or expires, it may no longer provide useful information and can even lead to inaccurate analysis results. Therefore, it is essential to filter out obsolete or irrelevant data to ensure the reliability and value of analytical insights. To support continuous monitoring and processing of data streams, the \emph{sliding window} technique is commonly employed. Sliding windows can be categorized as either \emph{time-based} or \emph{count-based}. In this work, we adopt the count-based sliding window approach to implement our proposed solution. The count-based sliding window is formally defined as follows:
\begin{definition}[Count-Based Sliding Window]
	\label{def:count_sw}
	A sliding window is denote as $SW$. The sliding window has a maximum size $n$, denote as $|SW|=n$. The size of sliding window at time $t$ is denote as $|SW(t)|$. In any time, $|SW(t)|$ will not exceed the maximum size $n$. That is $|SW(t)|\leq n, \forall t$. The sliding window processes the incoming data objects in a \emph{First-In-First-Out} (FIFO) manner.
\end{definition} 

\begin{example}
Assume data objects $u_1$ comes at $t=1$, $u_2$ comes at $t=2$, and so on. The maximum size of sliding window $SW$ is 4. That is $|SW|=4$. Table~\ref{table:sliding_window} gives a example to show the changes of sliding window from time $t=1$ to $t=6$. 
\end{example}

\begin{table}[!t]
	\centering
	\small
	\caption{A 2D Uncertain Data Set Example}
	\label{table:2d_uncertain_data_set} 
	\begin{tabular}{|c|c|c|c|}
		\hline
		\textbf{Object}            & \textbf{Instance} & \textbf{Probability} & \textbf{Attributes} \\ \hline
		\multirow{3}{*}{$u_1$} &   $u_{1,1}$    &    0.4    &    [28,37]   \\ \cline{2-4} 
		&   $u_{1,2}$    &    0.1    &   [27,35]    \\ \cline{2-4} 
		&   $u_{1,3}$   &     0.5     &   [25,38]   \\ \hline
		\multirow{3}{*}{$u_2$} &   $u_{2,1}$    &    0.1      &    [9,35]    \\ \cline{2-4} 
		&   $u_{2,2}$   &     0.2     &   [9,38]   \\ \cline{2-4} 
		&   $u_{2,3}$   &     0.7    &   [10,37]   \\ \hline
		\multirow{3}{*}{$u_3$} &   $u_{3,1}$    &     0.5     &     [24,92]    \\ \cline{2-4} 
		&   $u_{3,3}$   &     0.3     &   [22,91]     \\ \cline{2-4} 
		&   $u_{3,3}$   &     0.2     &   [22,88]   \\ \hline
	\end{tabular}
\end{table}

\begin{table}[!t]
	\centering
	\small
	\caption{A Sliding Window Example}
	\label{table:sliding_window} 
	\begin{tabular}{|c|c|c|}
		\hline
		\textbf{Time} & \textbf{Sliding Window} & \textbf{Size} \\ \hline
		1	&  $SW(1)=\{u_1\}$  &  $|SW(1)|=1$    \\ \hline
		2	&  $SW(2)=\{u_1,u_2\}$  &  $|SW(2)|=2$    \\ \hline
		3	&  $SW(3)=\{u_1,u_2,u_3\}$  &  $|SW(3)|=3$    \\ \hline
		4	&  $SW(4)=\{u_1,u_2,u_3,u_4\}$  &  $|SW(4)|=4$    \\ \hline
		5	&  $SW(5)=\{u_2,u_3,u_4,u_5\}$  &  $|SW(5)|=4$    \\ \hline
		6	&  $SW(6)=\{u_3,u_4,u_5,u_6\}$  &  $|SW(6)|=4$    \\ \hline
	\end{tabular}
\end{table}

To search the probabilistic skyline, the system needs to calculate the dominant c between different uncertain objects and instances. According to the uncertain data model considered in Definition~\ref{def:DPDM}, the dominant relationship will be modeled as a probability and it can be divided into two levels: instance-level dominance probability and object-level dominance probability. The instance-level dominance probability can be defined as follows:
\begin{definition}[Instance-Level Dominance Probability]
	\label{def:instance_dominant}
	Given two instances, $o_x$ and $o_y$, of two different data objects $u_x$ and $u_y$, $x\neq y$. Instance $o_x$ dominates $o_y$, denote as $o_x\prec o_y$, if and only if all the attributes of $o_x$ are less or equal to $o_y$'s corresponding attributes, and exists at least one attribute that $o_x$ is less than $o_y$. That is, the instance-level dominance probability for $o_x$ with respect to $o_y$ is derived by	
	\begin{align*}\label{eq:instance_dominant_probability}
		\text{Pr}(o_x \prec& o_y)=\\
		&\begin{cases}
			\text{Pr}(o_x)\cdot \text{Pr}(o_y), & \text{if $(o_x.attr(i)\leq o_y.attr(i), \forall i)$}\\
			& \wedge (o_x.attr(j)<o_y.attr(j), \exists j);\\
			0, & \text{otherwise}.
		\end{cases}
	\end{align*}
\end{definition} 

\begin{example}
Given two 3D uncertain data instances, $o_1=[10,4,7]$ and $o_2=[15,4,9]$. We can say that $o_1$ dominates $o_2$ which is denoted as $o_1\prec o_2$. Please note that we assume that an instance with smaller attribute values is a better one.
\end{example}

Since each uncertain data object may contain multiple instances, some instances of one object may dominate instances of another object, while others may not. Each instance also has its own probability of existence. Therefore, the object-level dominance relationship is represented as a dominance probability, which is the sum of the instance-level dominance probabilities between all pairs of instances from the two objects. The object-level dominance probability can be formally defined as follows:
\begin{definition}[Object-Level Dominance Probability]
	\label{def:object_dominant}
	Given two uncertain data objects $u_1$ and $u_2$, the dominance probability of $u_1\prec u_2$ can be derived by
	\begin{equation*}\label{eq:obj_dominant_probability}
		\text{Pr}(u_1\prec u_2)=\sum_{o_{1,i}\in u_{1},o_{2,j}\in u_{2},\forall i,j}\text{Pr}(o_{1,i}\prec o_{2,j}).
	\end{equation*}
\end{definition}

\begin{example}
Consider uncertain data objects $u_1$ and $u_2$ in Table~\ref{table:2d_uncertain_data_set}. Both of them have 3 instances and each instance is with two attributes and its existing probability. According to the above assumptions and definitions, if we would like to calculate the probability that $u_2\prec u_1$. We have to sum up the probability as
follow:
\begin{align*}
	\text{Pr}(u_2 \prec u_1)&=\sum_{u_{2,i}\in u_{2},u_{1,j}\in u_{1},\forall i,j}\text{Pr}(u_{2,i}\prec u_{1,j})\\
	&=\text{Pr}(u_{2,1}\prec u_{1,1})+\text{Pr}(u_{2,1}\prec u_{1,2})\\
	&+\text{Pr}(u_{2,1}\prec u_{1,3})+\text{Pr}(u_{2,2}\prec u_{1,3})\\
	&+\text{Pr}(u_{2,3}\prec u_{1,1})+\text{Pr}(u_{2,3}\prec u_{1,3})\\
	&=\text{Pr}(u_{2,1})\times\left(\text{Pr}(u_{1,1})+\text{Pr}(u_{1,2})+\text{Pr}(u_{1,3})\right)\\
	&+\text{Pr}(u_{2,2})\times\text{Pr}(u_{1,3})\\
	&+\text{Pr}(u_{2,3})\times\left(\text{Pr}(u_{1,1})+\text{Pr}(u_{1,3})\right)\\
	&=0.1\times 1+0.2\times 0.5+0.7\times 0.9\\
	&=0.83.
\end{align*}
\vspace{-12pt}
\end{example}

With the above definitions, the probabilistic skyline is defined as follows:
\begin{definition}[Probabilistic Skyline]
	\label{def:p_skyline}
	With the notations defined above, given a sliding window $SW$ is full of uncertain data objects, the probabilistic skyline of $SW$ is
	\begin{equation*}\label{eq:p_skyline}
		Skyline(SW)=\{u|\forall u,u'\in SW, u\neq u', \nexists u',\text{Pr}(u' \prec u)=1\}.
	\end{equation*}
\end{definition} 

\subsection{Problem Statement}
\begin{figure}[!t]
	\centering
	\includegraphics[width=\columnwidth]{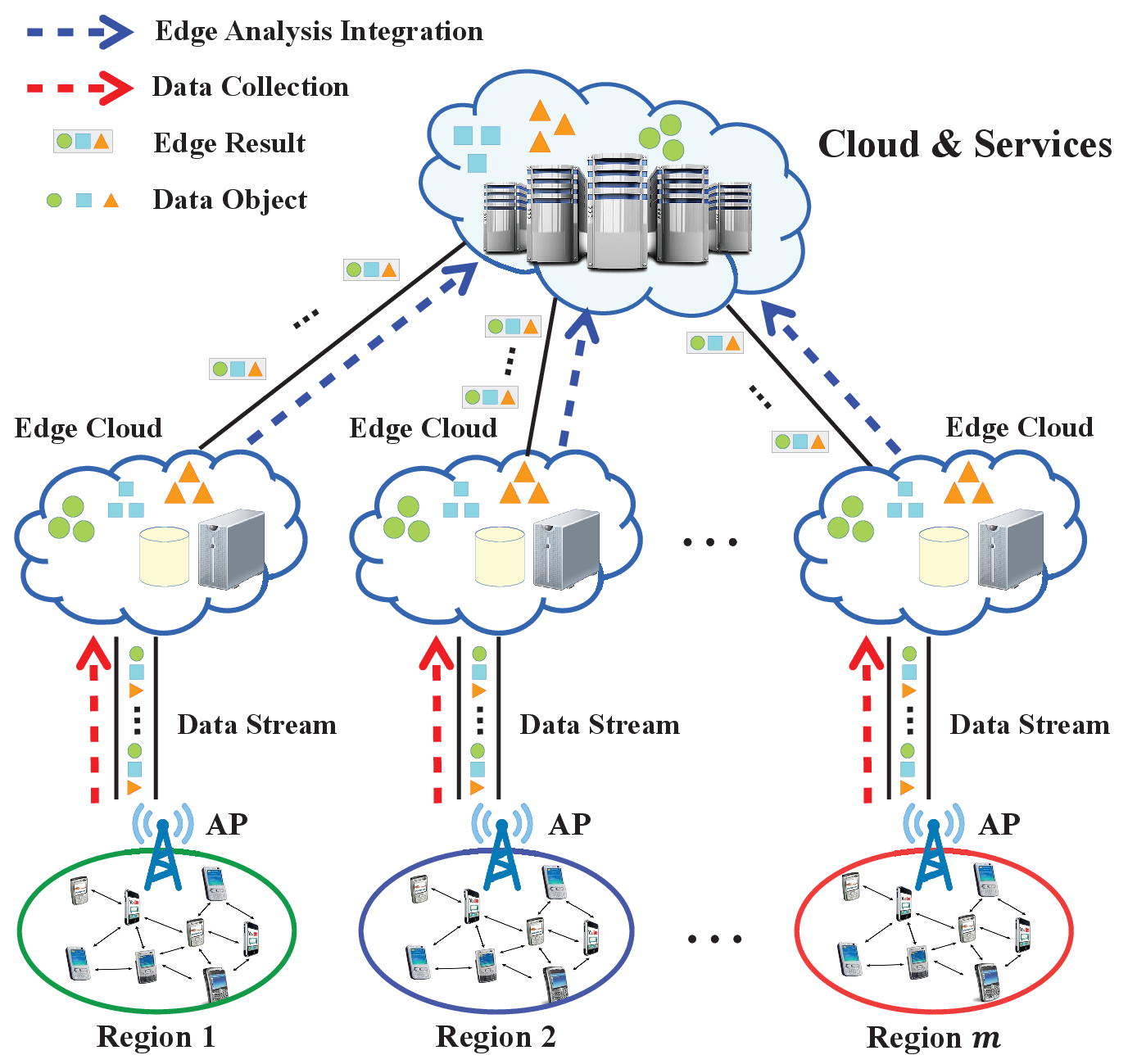}
	\caption{The considered edge computing environment.}
	\label{fig:system_model}
\end{figure}

As shown in Fig.~\ref{fig:system_model}, we consider an edge computing environment consisting of $m$ edge computing nodes (ECNs), $E_1, E_2,\dots, E_m$, with adequate computing resources and a main server node, $S$. All the data comes into ECNs are treated as uncertain data streams. Each ECN, $E_k$, examines the dominance probabilities of all objects in the edge sliding window, $SW_k$, and then reports the edge skyline set to the server node $S$, where $k=1,2,\dots,m$. The server node $S$ uses the edge skyline sets received to calculate the global skyline. The above procedure will be repeated until there is no incoming data.

Since this research focuses on the edge computing environment, it is necessary to reduce the data transmission cost between ECNs and the main server node as much as possible. The time to calculate the probabilistic skyline is also an important factor because the data streams are time-sensitive and the average latency must be minimized. In short, our goal is to propose a new parallel algorithm based on the aforementioned edge computing environment to maintain the global probabilistic skyline with low average latency and low transmission cost.

\section{The Proposed Edge-assisted Parallel Uncertain Skyline (EPUS)}
\label{sec:proposed_approach}
In this section, we will introduce the design of the proposed \emph{Edge-assisted Parallel Uncertain Skyline} (EPUS) algorithm in detail. The frequently used notations in the proposed algorithm are depicted in Table~\ref{table:notations}.

\begin{table}[!t]
	\centering
	\small
	\caption{Frequently Used Notations}
	\label{table:notations} 
	\begin{tabular}{|c|l|}
		\hline
		\textbf{Notation} & \textbf{Meaning} \\ \hline
		$u_i$ & Uncertain data object $i$ \\ \hline
		$u_{i,j}$ & Instance $j$ of uncertain object $u_i$  \\ \hline
		$S$ & The main server \\ \hline
		$E_k$ & The $k$th edge computing node \\ \hline	
		$m$ & The number of edge computing nodes \\ \hline
		$SW_S$ & The sliding window on $S$ \\ \hline
		$SW_k$ & The sliding window on $E_k$ \\ \hline	
		$ESK_{k,1}$ & The edge skyline set on $E_k$  \\ \hline
		$ESK_{k,2}$ & The edge candidate skyline set on $E_k$  \\ \hline
		$SK_1$ & The global skyline set on $S$ \\ \hline
		$SK_2$ & The global candidate skyline set on $S$ \\ \hline
		$D_{\rm obsolete}$ & A temporary data set to record obsolete data objects \\ \hline
		$D_{\rm new}$ & A temporary data set to record new data objects \\ \hline
	\end{tabular}
\end{table}

\subsection{Data Indexing}

\begin{figure}[!t]
	\centering
	\includegraphics[width=.85\columnwidth]{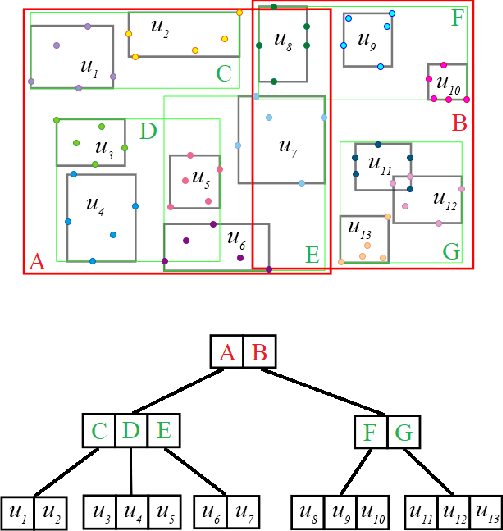}
	\caption{An example of R-tree including 13 two-dimensional data objects with 5 instances.}
	\label{fig:ex:r_tree}
\end{figure}

To accelerate the computation of dominance probabilities, we employ R-tree~\cite{10.1145/602259.602266} as the data indexing structure in the proposed EPUS algorithm. Each uncertain data object, characterized by multiple instances, is represented by its minimum bounding rectangle (MBR), which captures the maximum and minimum values across all dimensions. The MBRs are stored as index entries (leaf nodes) in the R-tree, allowing uncertain data to be treated as certain data during the pruning stage. This approach enables efficient exclusion of irrelevant data and reduces the average latency for dominance probability calculations. Fig.~\ref{fig:ex:r_tree} illustrates an example of R-tree indexing for 13 two-dimensional data objects, each with 5 instances; rectangles denote the MBRs stored in the R-tree. Since overlapping MBRs can degrade search performance, we utilize bulk-loading techniques~\cite{Arge2002} to minimize overlap at each tree level and optimize query efficiency.

\subsection{Candidate Skyline Set}
In the EPUS algorithm, we introduce the concept of the \emph{Candidate Skyline Set} (CSS) to efficiently reduce both the computational overhead for probabilistic skyline calculation and the volume of data transmitted across the network. Each edge computing node (ECN) is responsible for maintaining two sets: the \emph{edge skyline set} and the \emph{edge candidate skyline set}. When an update is triggered at an ECN, the node transmits only the necessary update information to the main server. The formal definitions of the edge skyline set and the edge candidate skyline set are as follows:
\begin{definition}[Edge Skyline Set]
	\label{def:edge_skyline}
	Given a sliding window $SW_k$ on ECN $E_k$ and the corresponding edge skyline set $ESK_{k,1}=Skyline(SW_k)$, where $k=1,2,\dots,m$. 
\end{definition} 
\begin{definition}[Edge Candidate Skyline Set]
	\label{def:edge_candidate_skyline}
	Suppose that the notations are defined as above, the edge candidate skyline set will be $ESK_{k,2}=Skyline(SW_k-ESK_{k,1})$, where $k=1,2,\dots,m$. 
\end{definition} 

The main server is responsible for maintaining both the global skyline set and the global candidate skyline set, updating them as necessary based on information received from the ECNs. Leveraging the R-tree indexing structure described above, the server can efficiently retrieve relevant data without accessing unnecessary data points. Once the global skyline set is determined from the uncertain data objects in the server's sliding window, the global candidate skyline set is defined as the skyline of all uncertain data objects in the sliding window that are not part of the global skyline set. The formal definitions of the global skyline set and the global candidate skyline set are as follows:
\begin{definition}[Global Skyline Set]
	\label{def:global_skyline}
	Given a sliding window $SW_S$ on the main server $S$ and $SW_S=\bigcup_{k=1}^{m}ESK_{k,1}$, the corresponding global skyline set $SK_1=Skyline(SW_S)$.
\end{definition} 
\begin{definition}[Global Candidate Skyline Set]
	\label{def:global_candidate_skyline}
	With the notations defined as above, the global candidate skyline set will be $SK_2=Skyline(SW_S-SK_1)$.
\end{definition} 

We utilize the Candidate Skyline Set (CSS) as a distributed pruning mechanism on both the edge computing nodes and the main server. By design, CSS significantly reduces the amount of data that must be examined during updates, since any object not present in either the skyline set or the CSS cannot become a skyline object. Consequently, irrelevant data are excluded from consideration during update operations, improving overall efficiency.

\subsection{The Tasks of an Edge Computing Node}
Each edge computing node (ECN) $E_k$ is responsible for maintaining its local skyline information. After computing the edge skyline set $ESK_{k,1}$ and the edge candidate skyline set $ESK_{k,2}$, $E_k$ compares the previous and current skyline sets to detect any changes. If an update is required, $E_k$ sends an update message to the main server $S$. This message includes: (1) new data in $ESK_{k,1}$, (2) new data in $ESK_{k,2}$, and (3) obsolete data from $E_k$. The server then updates the global skyline set based on the received information from all ECNs. In summary, each ECN performs two main tasks: \emph{Receive} and \emph{Update}, which are described in detail below.


We denote the EPUS procedure on each ECN as EEPUS. The main operations of EEPUS are described in Algorithm~\ref{alg:EEPUS}. Initially, each $E_k$ stores the current state of $ESK_{k,1}$ and $ESK_{k,2}$ before processing incoming data. After updating the edge skyline, $E_k$ compares the previous and current states and sends any update information to the server. When a new data stream $s_k$ arrives at $E_k$, the node checks whether the sliding window $SW_k$ is full. If $SW_k$ has reached its maximum size, $E_k$ removes obsolete data from $SW_k$ and then adds the new data objects from $s_k$ into the window. These operations are performed by the function $\mathsf{ReceiveData}(s_k,SW_k)$, as shown in Algorithm~\ref{alg:EEPUS:receive}. ECN $E_k$ invokes this function at line~\ref{alg:EEPUS:3} of Algorithm~\ref{alg:EEPUS}.

\begin{algorithm2e}[!t]
	\small
	\SetAlgoLined
	\KwIn{Uncertain data stream $s_k$, Sliding window $SW_k$}
	\While{true}{	\label{alg:EEPUS:1}
		\If{$|s_k|>0$}{		\label{alg:EEPUS:2}
			$oldESK_{k,1}\leftarrow \mathsf{getSkyline}(SW_k)$\;
			$oldESK_{k,2}\leftarrow \mathsf{getCandidateSkyline}(SW_k)$\;
			\tcc{invoke Algorithm~\ref{alg:EEPUS:receive}}
			$D_{\rm obsolete}\leftarrow\mathsf{ReceiveData}(s_k,SW_k)$ \label{alg:EEPUS:3}\;
			\tcc{invoke Algorithm~\ref{alg:EEPUS:update}}
			$ESK_{k,1},ESK_{k,2}\leftarrow\mathsf{UpdateSkyline}(oldESK_{k,1},oldESK_{k,2},s_k,$ $D_{\rm obsolete})$\label{alg:EEPUS:4}\;	
			$newESK_{k,1}\leftarrow ESK_{k,1}\setminus oldESK_{k,1}$\label{alg:EEPUS:5}\;
			$newESK_{k,2}\leftarrow ESK_{k,2}\setminus oldESK_{k,2}$\;
			$\mathsf{SendResult}(D_{\rm obsolete},newESK_{k,1},newESK_{k,2})$\label{alg:EEPUS:7}\;
		}
	}
	\caption{The procedure of EEPUS on ECN $E_k$}
	\label{alg:EEPUS}	
\end{algorithm2e}
%

\begin{algorithm2e}[!t]
	\small
	\SetAlgoLined
	\KwIn{Uncertain data stream $s_k$, Sliding window $SW_k$}
	\KwOut{Obsolete data set $D_{\rm obsolete}$}
	\If{$|SW_k|\geq n$}{
		$D_{\rm obsolete}\leftarrow SW_k.\mathsf{collectObsoleteData}()$\;
		\ForEach{data object $o$ in $D_{\rm obsolete}$}{
			Remove data object $o$ from $SW_k$\;
		}					
	}
	$SW_k.\mathsf{addData}(s_k)$\;
	\Return $D_{\rm obsolete}$\;
	\caption{$\mathsf{ReceiveData}(s_k,SW_k)$}
	\label{alg:EEPUS:receive}
\end{algorithm2e}
%


After ECN $E_k$ obtains the obsolete data, it proceeds to update both the edge skyline set $ESK_{k,1}$ and the edge candidate skyline set $ESK_{k,2}$ at line~\ref{alg:EEPUS:4} of Algorithm~\ref{alg:EEPUS}. The detailed update operations are implemented in the function $\mathsf{UpdateSkyline}(ESK_{k,1},ESK_{k,2},s_k,D_{\rm obsolete})$, as shown in Algorithm~\ref{alg:EEPUS:update}.
Since removing obsolete data from $ESK_{k,2}$ does not affect the skyline result, this step can be performed directly without additional checks (see lines~\ref{alg:EEPUS:update:1} to~\ref{alg:EEPUS:update:3} in Algorithm~\ref{alg:EEPUS:update}). For $ESK_{k,1}$, obsolete data must also be removed. When doing so, it is necessary to examine whether any objects in $ESK_{k,2}$—previously dominated by the obsolete data—should be promoted to $ESK_{k,1}$, as they may now qualify as skyline objects. This process is handled in lines~\ref{alg:EEPUS:update:4} to~\ref{alg:EEPUS:update:12} of Algorithm~\ref{alg:EEPUS:update}.
After all necessary updates and promotions from $ESK_{k,2}$ to $ESK_{k,1}$, new incoming data are added to $ESK_{k,1}$ for further evaluation and updates, as described in line~\ref{alg:EEPUS:update:13} of Algorithm~\ref{alg:EEPUS:update}.

\begin{algorithm2e}[!t]
	\small
	\SetAlgoLined
	\KwIn{Edge skyline set $ESK_{k,1}$ and Edge candidate skyline set $ESK_{k,2}$, Uncertain data stream $s_k$, Obsolete Data Set $D_{\rm obsolete}$}
	\KwOut{Updated $ESK_{k,1}$ and $ESK_{k,2}$}	
	\ForEach{data object $o$ in $D_{\rm obsolete}$}{\label{alg:EEPUS:update:1}
		Remove data object $o$ from $ESK_{k,2}$\;
	}\label{alg:EEPUS:update:3}
	\ForEach{data object $o$ in $ESK_{k,1}$}{\label{alg:EEPUS:update:4}
		\If{$o.\mathsf{isObsolete}()$}{
			\ForEach{data object $o'$ in $ESK_{k,2}$}{
				\If{$o\prec o'$}{
					Move data object $o'$ into $ESK_{k,1}$\;
				}
			}
		}				
	}\label{alg:EEPUS:update:12}
	$ESK_{k,1}.\mathsf{append}(s_k)$\label{alg:EEPUS:update:13}\;
	\ForEach{data object $o$ in $ESK_{k,1}$}{\label{alg:EEPUS:update:14}
		\If{$o\prec o', \forall o'\neq o, o'\in ESK_{k,1}$}{
			Move data object $o'$ into $ESK_{k,2}$\;
		}				
	}
	\ForEach{data object $o$ in $ESK_{k,2}$}{\label{alg:EEPUS:update:19}
		\If{$o\prec o', \forall o'\neq o, o'\in ESK_{k,2}$}{
			Remove data object $o'$ from $ESK_{k,2}$\;
		}				
	}
	\tcc{remove all the obsolete data in this ECN}	
	$\mathsf{RemoveAllObsoleteData}()$\;
	\Return $ESK_{k,1},ESK_{k,2}$
	\caption{\protect \\ $\mathsf{UpdateSkyline}(ESK_{k,1},ESK_{k,2},s_k,D_{\rm obsolete})$}
	\label{alg:EEPUS:update}
\end{algorithm2e}
%


After updating $ESK_{k,1}$ and $ESK_{k,2}$, ECN $E_k$ must verify the membership of objects in these sets. Some objects moved from $ESK_{k,2}$ to $ESK_{k,1}$, as well as newly added objects, may not actually satisfy the skyline property. Therefore, $E_k$ checks for dominance relationships within $ESK_{k,1}$: if any object in $ESK_{k,1}$ is dominated by another, it is moved to $ESK_{k,2}$, as it no longer qualifies as a skyline object. This process is implemented by the for-loop at line~\ref{alg:EEPUS:update:14} in Algorithm~\ref{alg:EEPUS:update}.
A similar procedure applies to $ESK_{k,2}$: if any object in $ESK_{k,2}$ is dominated by another object in $ESK_{k,2}$, it is removed from the set. This is handled by the for-loop at line~\ref{alg:EEPUS:update:19} in Algorithm~\ref{alg:EEPUS:update}.
Finally, ECN $E_k$ removes all obsolete data previously collected, resulting in the updated $ESK_{k,1}$ and $ESK_{k,2}$ as produced by the last two operations in Algorithm~\ref{alg:EEPUS:update}.

With the obtained latest $ESK_{k,1}$ and $ESK_{k,2}$, ECN $E_k$ will compute two data sets, $newESK_{k,1}$ and $newESK_{k,2}$, including new skyline objects that are not in the original $oldESK_{k,1}$ and $oldESK_{k,2}$. Finally, $E_k$ send the update messages including the information of $D_{\rm obsolete}$, $newESK_{k,1}$ and $newESK_{k,2}$ to the server node $S$. The above three operations are done from line~\ref{alg:EEPUS:5} to line~\ref{alg:EEPUS:7} of Algorithm~\ref{alg:EEPUS}. 
With the while loop at line~\ref{alg:EEPUS:1}, each $E_k$ will repeat the procedure of EEPUS and send the update information to the server node if $|s_k|>0$ at line~\ref{alg:EEPUS:2}.

\subsection{The Tasks of the Main Sever}

The main server is responsible for maintaining the global skyline set. Upon receiving update information from any ECN, the server initiates the update procedure. The tasks performed by the main server node $S$ are described in the following pseudo-code algorithms. We refer to the EPUS procedure on the server side as SEPUS. Algorithm~\ref{alg:SEPUS} outlines the operations of SEPUS. When the server $S$ receives update information from any ECN, the data stream $s$ becomes non-empty and the update procedure begins. After the update is completed, the server waits for the next update information.

The update procedure includes two steps. The first step is to call $\mathsf{ReceiveEdgeUpdate}(s,SW_S,SK_1,SK_2)$ for updating the global skyline set $SK_1$ and the global candidate skyline set $SK_2$ with the received update information from any ECNs. The detailed operations of $\mathsf{ReceiveEdgeUpdate}(s,SW_S,SK_1,SK_2)$ are described in Algorithm~\ref{alg:SEPUS:reseive}. 
When the server node $S$ receives the message form ECN $E_k$, it will remove obsolete data from the sliding window $SW_S$ first. Such operations will be done by the for-loop procedure at line~\ref{alg:SEPUS:reseive:2}.

\begin{algorithm2e}[!t]
	\small
	\SetAlgoLined
	\KwIn{Uncertain data stream $s$, Sliding window $SW_S$, Global skyline set $SK_1$, Global skyline candidate set $SK_2$}
	\While{true}{
		\If{$|s|>0$}{
			\tcc{invoke Algorithm~\ref{alg:SEPUS:reseive}}
			$D_{\rm obsolete},D_{\rm new}\leftarrow\mathsf{ReceiveEdgeUpdate}(s,SW_S,SK_1,SK_2)$\;
			\tcc{invoke Algorithm~\ref{alg:SEPUS:update}}
			$\mathsf{UpdateGlobalSkyline}(s,SW_S,SK_1,SK_2,D_{\rm obsolete},$
			$D_{\rm new})$\;
		}				
	}
	\caption{The procedure of SEPUS on server $S$}
	\label{alg:SEPUS}
\end{algorithm2e}

%
\begin{algorithm2e}[!t]
	\small
	\SetAlgoLined
	\KwIn{Uncertain data stream $s$, Sliding window $SW_S$, Global skyline set $SK_1$, Global skyline candidate set $SK_2$}
	\KwOut{Obsolete data set $D_{\rm obsolete}$, New data set $D_{\rm new}$}
	Parse the receive data stream $s$ and then get the edge obsolete data set $D_{\rm obsolete}$, the set of new edge skyline objects $ESK_1$, and the set of new edge candidate skyline objects $ESK_2$\;	
	\ForEach{data object $o$ in $D_{\rm obsolete}$}{
		\label{alg:SEPUS:reseive:2}
		Remove data object $o$ from $SW_S$\;
	}
	\ForEach{data object $o$ in $ESK_1$}{\label{alg:SEPUS:reseive:5}
		\uIf{data object $o$ is not in $SW_S$}{
			$SW_S.\mathsf{add}(o)$\;	
			$D_{\rm new}.\mathsf{add}(o)$\;
		}
		\ElseIf{data object $o$ is in $SK_2$}{
			Move data object $o$ from $SK_2$ to $D_{\rm new}$\;
		}
	}
	\ForEach{data object $o$ in $ESK_2$}{\label{alg:SEPUS:reseive:15}
		\uIf{data object $o$ is not in $SW_S$}{
			$SW_S.\mathsf{add}(o)$\;	
			$SK_2.\mathsf{add}(o)$\;		
		}
		\ElseIf{data object $o$ is in $SK_1$}{	
			Move data object $o$ from $SK_1$ to $SK_2$\;
		}
	}
	\Return $D_{\rm obsolete}, D_{\rm new}$\;
	\caption{$\mathsf{ReceiveEdgeUpdate}(s,SW_S,SK_1,SK_2)$}
	\label{alg:SEPUS:reseive}
\end{algorithm2e}
%

\begin{algorithm2e}[!t]
	\small
	\SetAlgoLined
	\KwIn{Sliding window $SW_S$, Global skyline set $SK_1$, Global skyline candidate set $SK_2$}
	\ForEach{data object $o$ in $D_{\rm obsolete}$}{
		Remove data object $o$ from $SK_1$ and $SK_2$\;
	}	
	\ForEach{data object $o$ in $D_{\rm new}$}{
		\uIf{$o\prec o', \forall o'\neq o, o'\in SK_1$}{
			Move data object $o$ from $D_{\rm new}$ to $SK_1$\;
			Move data object $o'$ from $SK_1$ to $SK_2$\;
		}
		\ElseIf{$o\prec o', \forall o'\neq o, o'\in D_{\rm new}$}{
			Move data object $o'$ from $D_{\rm new}$ to $SK_2$\;				
		}				
	}
	\ForEach{data object $o$ in $SK_1$}{
		\If{$o\prec o', \forall o'\neq o, o'\in D_{\rm new}$}{
			Remove data object $o'$ from $D_{\rm new}$\;
		}	
	}
	Add the rest of objects in $D_{\rm new}$ to $SK_1$\;	
	\ForEach{data object $o$ in $SK_2$}{
		\If{$o\prec o', \forall o'\neq o, o'\in SK_2$}{
			Remove data object $o'$ from $SK_2$\;
		}				
	}
	\caption{\protect\\
		$\mathsf{UpdateGlobalSkyline}(SW_S,SK_1,SK_2,D_{\rm obsolete},$ $D_{\rm new})$}
	\label{alg:SEPUS:update}
\end{algorithm2e}
%


For new data in the set of new edge skyline objects $ESK_1$ and the set of new edge candidate skyline objects $ESK_2$, the server checks whether each data object is already present in its sliding window $SW_S$. If a received object is already in $SW_S$, this indicates the object has been moved, and the server updates its membership in $ESK_1$ or $ESK_2$ according to the dominance relationships with existing objects. If the received objects from $ESK_1$ or $ESK_2$ are not present in $SW_S$, the server adds them to the sliding window. These operations are performed by the for-loop procedures at line~\ref{alg:SEPUS:reseive:5} and line~\ref{alg:SEPUS:reseive:15}, respectively.

After that, the SEPUS will call $\mathsf{UpdateGlobalSkyline}(SW_S,SK_1,SK_2, D_{\rm obsolete},D_{\rm new})$ for updating the global skyline set $SK_1$ and the global candidate skyline set $SK_2$ according to the received information of obsolete data and new data.
The procedure of $\mathsf{UpdateGlobalSkyline}(SW_S,SK_1,SK_2, D_{\rm obsolete},D_{\rm new})$ is presented in Algorithm~\ref{alg:SEPUS:update}.
In fact, the procedure of updating global skyline is almost identical to the procedure of updating edge global skyline on each ECN.

\subsection{Running Examples}
Assume there is an ECN $E_1$ with $|SW_1|=10$ and one new data object comes into $E_1$ at each time step.  
To illustrate how an ECN processes incoming data in detail, we described 6 different minimal running examples with the corresponding visualized figures as follows:
\begin{figure*}[!t]
	\centering
	\subfigure[$SW_1(t=11)=\{u_2,u_3,\dots,u_{11}\}$]{
		\label{fig:running_ex_1:t_11:a} 
		\includegraphics[width=0.325 \textwidth]{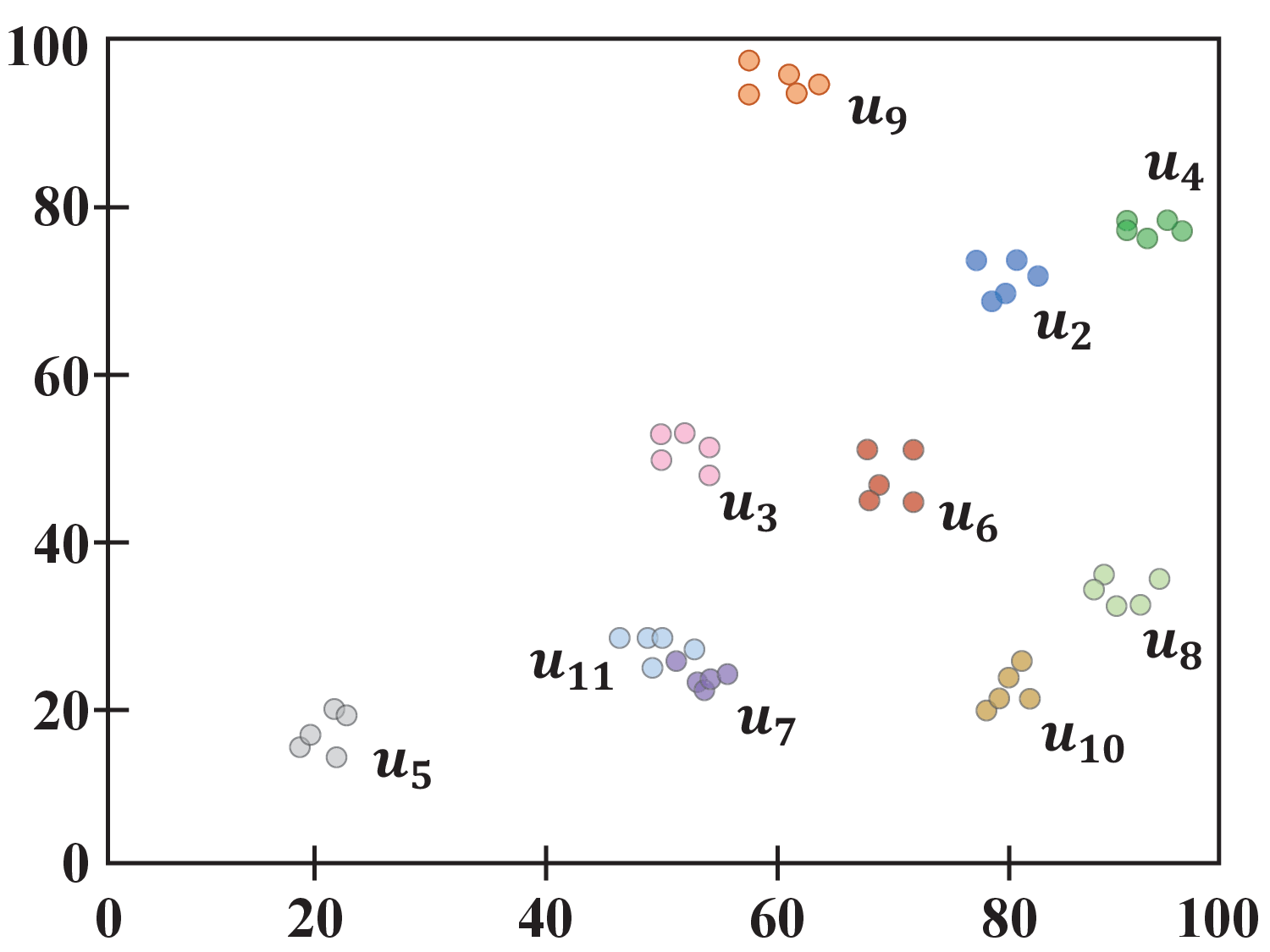}}%
	\subfigure[$ESK_{1,1}(t=11)=\{u_5\}$]{
		\label{fig:running_ex_1:t_11:b} 
		\includegraphics[width=0.325 \textwidth]{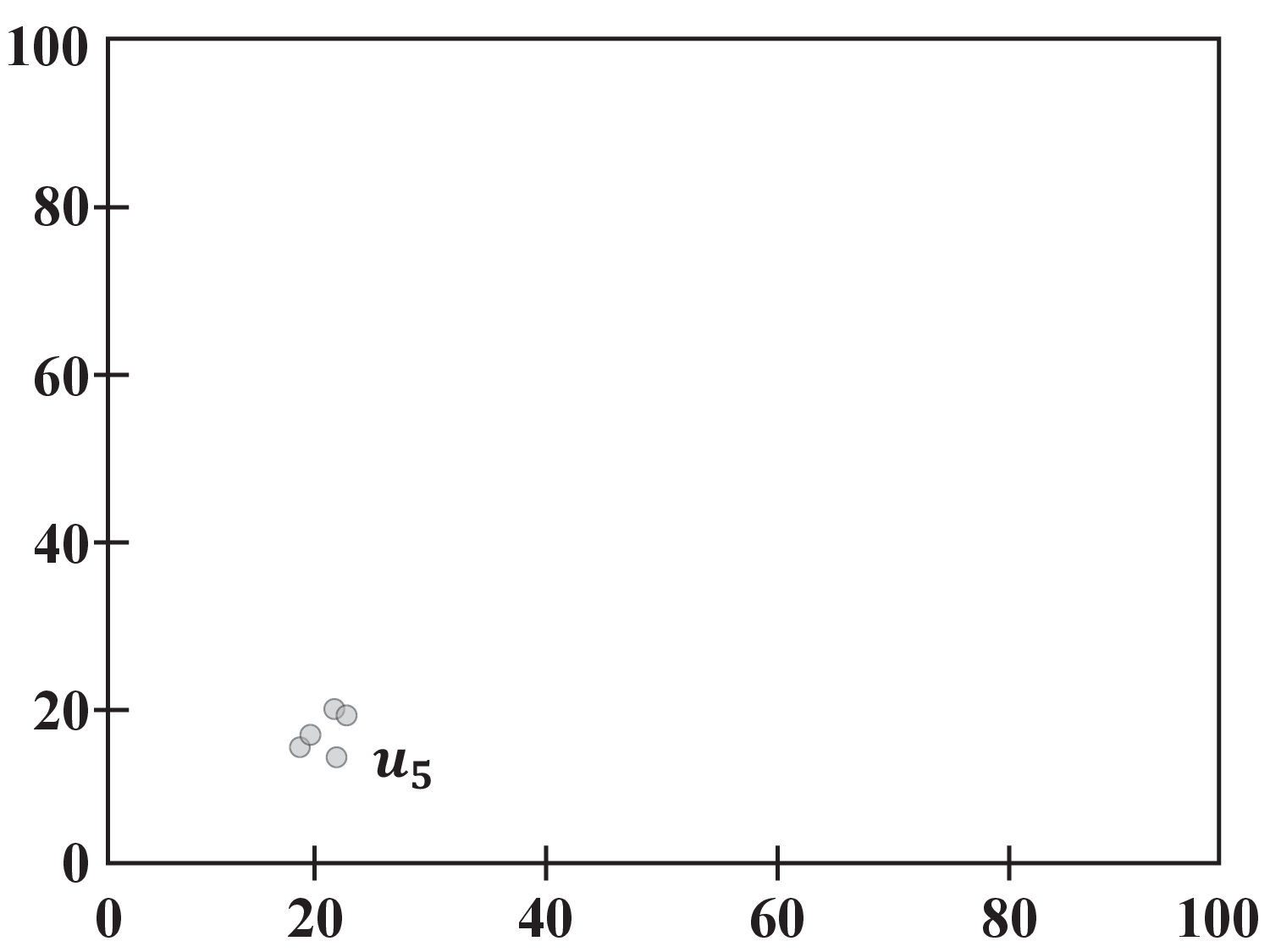}}%
	\subfigure[$ESK_{1,2}(t=11)=\{u_7,u_{10},u_{11}\}$]{
		\label{fig:running_ex_1:t_11:c} 
		\includegraphics[width=0.325 \textwidth]{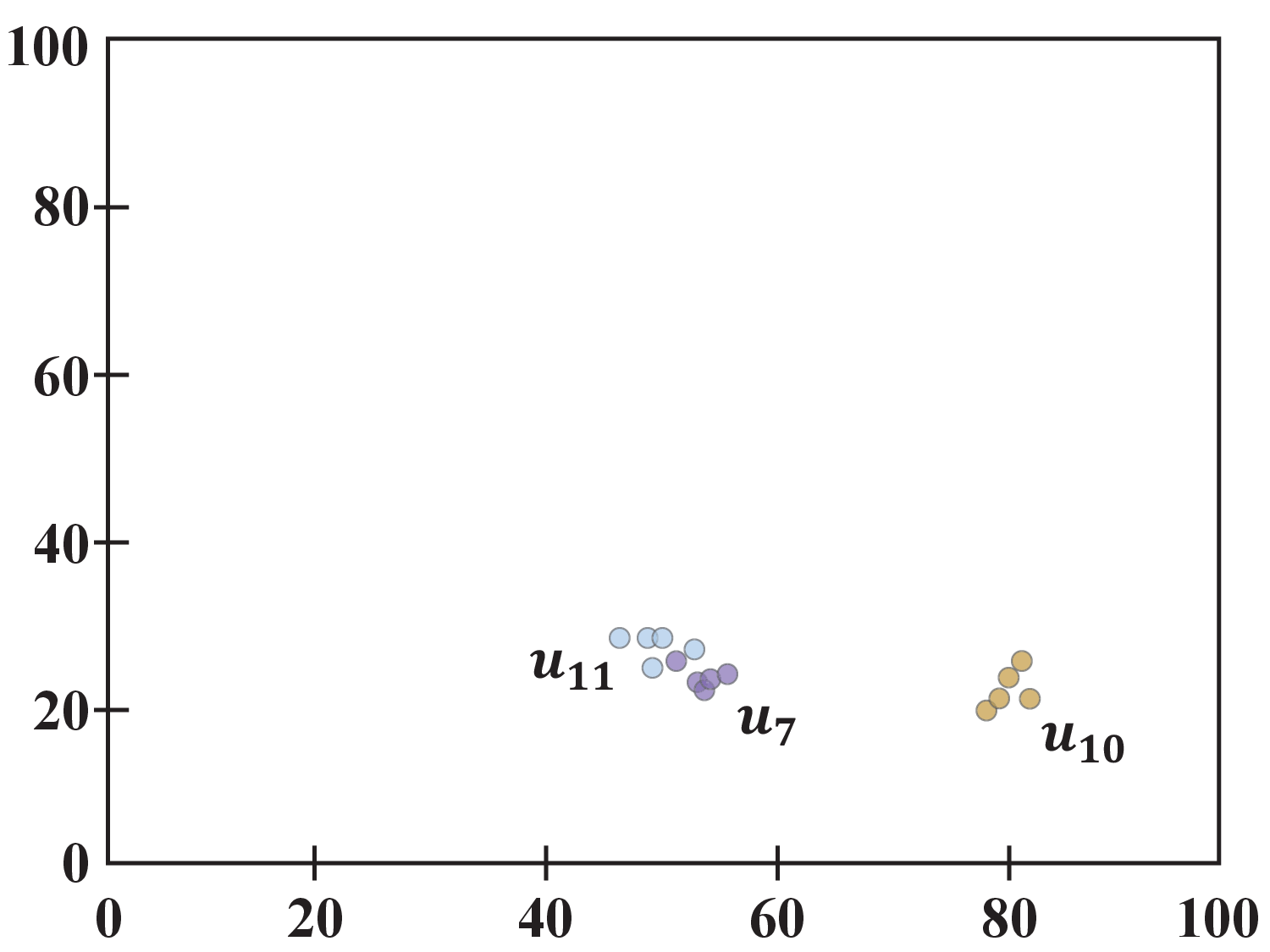}}%
	\caption{The content snapshot of~\subref{fig:running_ex_1:t_11:a} $SW_1(t=11)=\{u_2,u_3,\dots,u_{11}\}$,~\subref{fig:running_ex_1:t_11:b} $ESK_{1,1}(t=11)=\{u_5\}$, and~\subref{fig:running_ex_1:t_11:c} $ESK_{1,2}(t=11)=\{u_7,u_{10},u_{11}\}$ in $E_1$.
	}
	\label{fig:running_ex_1:t_11} 
\end{figure*}

\begin{figure*}[!t]
	\centering
	\subfigure[$SW_1(t=12)=\{u_3,u_4,\dots,u_{12}\}$]{
		\label{fig:running_ex:t_12:a} 
		\includegraphics[width=0.325 \textwidth]{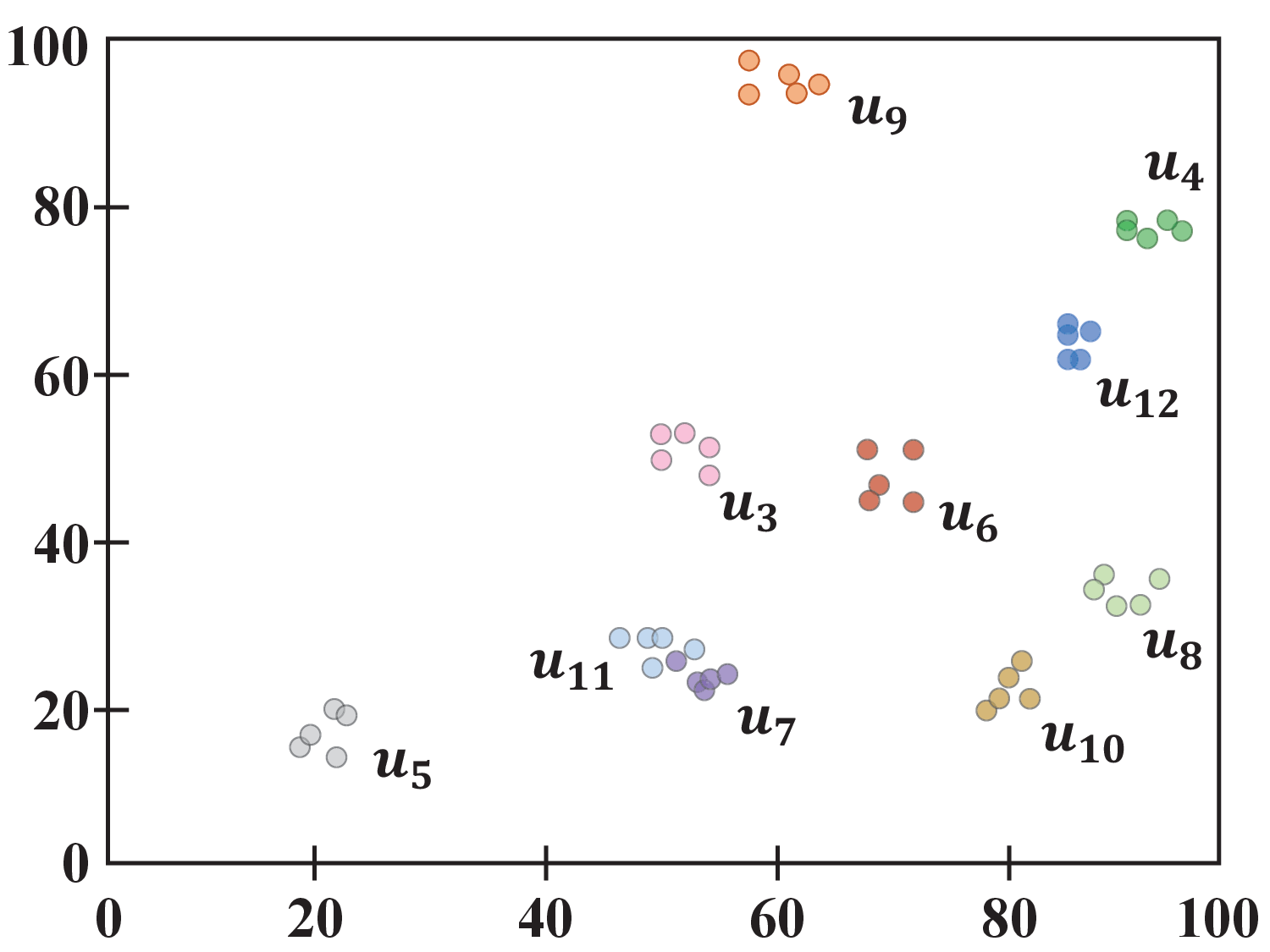}}%
	\subfigure[$ESK_{1,1}(t=12)=\{u_5\}$]{
		\label{fig:running_ex:t_12:b} 
		\includegraphics[width=0.325 \textwidth]{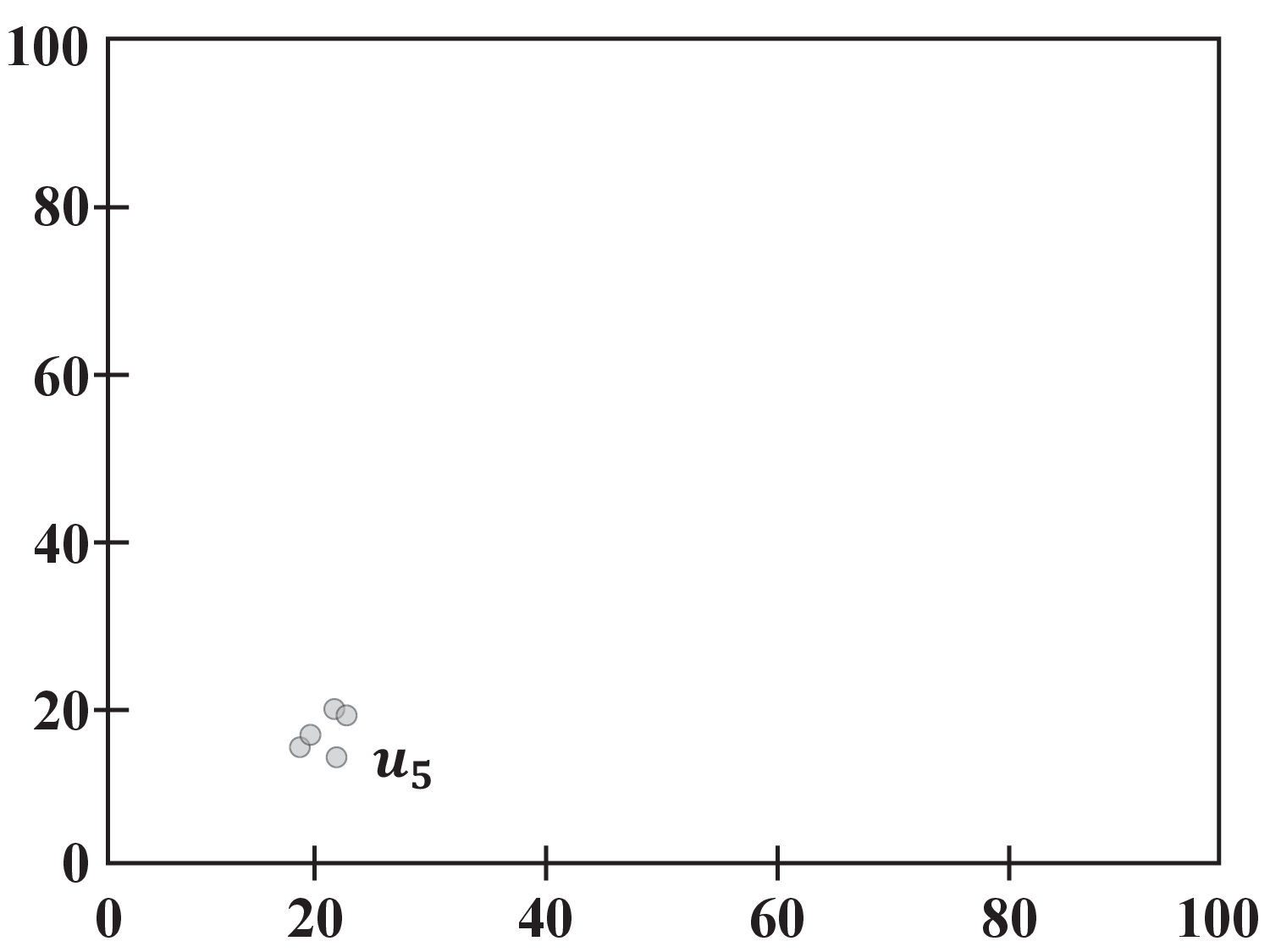}}%
	\subfigure[$ESK_{1,2}(t=12)=\{u_7,u_{10},u_{11}\}$]{
		\label{fig:running_ex:t_12:c} 
		\includegraphics[width=0.325 \textwidth]{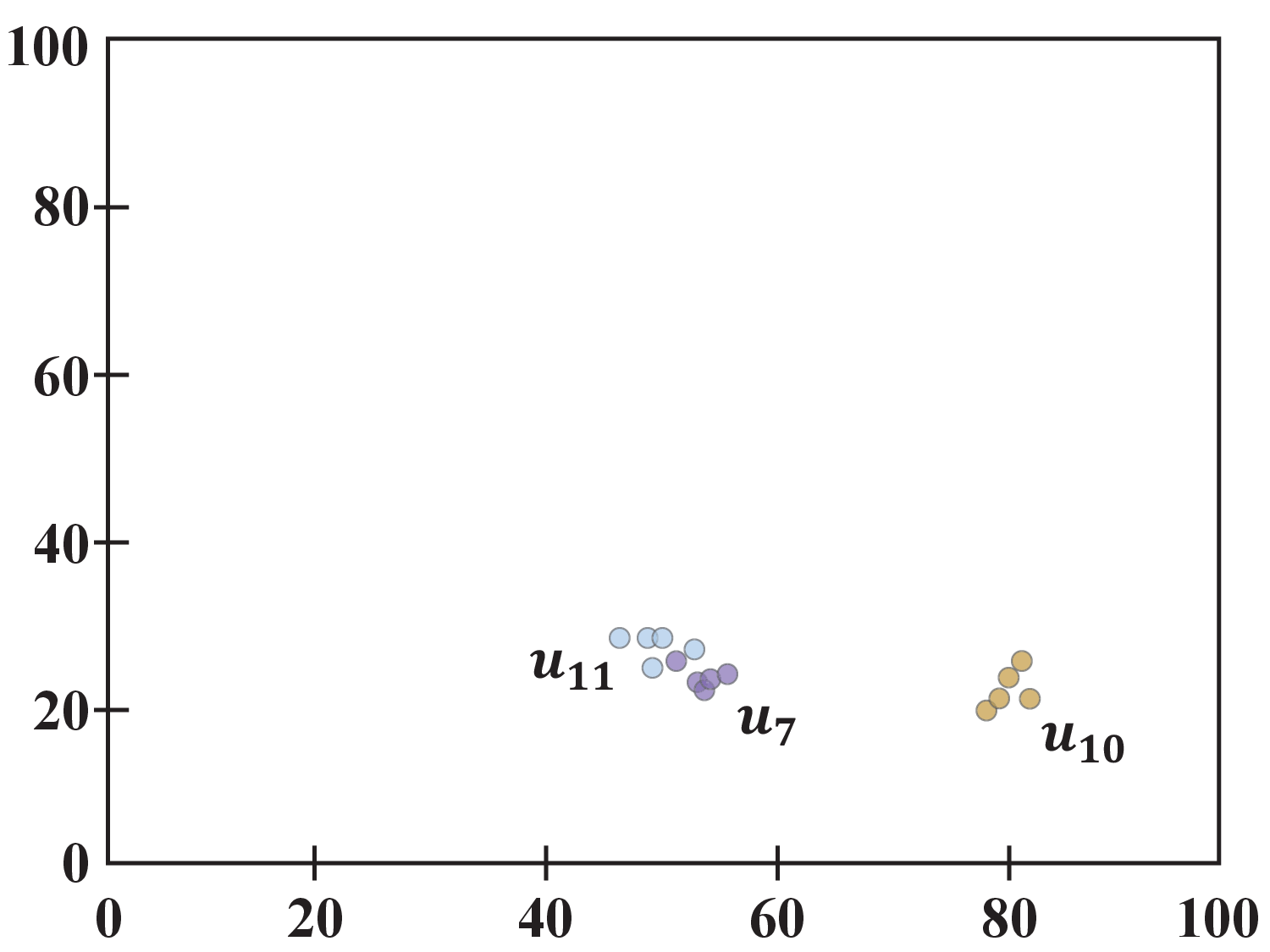}}%
	\caption{The content snapshot of~\subref{fig:running_ex:t_12:a} $SW_1(t=12)=\{u_3,u_4,\dots,u_{12}\}$,~\subref{fig:running_ex:t_12:b} $ESK_{1,1}(t=12)=\{u_5\}$, and~\subref{fig:running_ex:t_12:c} $ESK_{1,2}(t=12)=\{u_7,u_{10},u_{11}\}$ in $E_1$.
	}
	\label{fig:running_ex:t_12} 
\end{figure*}

\begin{figure*}[!t]
	\centering
	\subfigure[$SW_1(t=13)=\{u_4,u_5,\dots,u_{13}\}$]{
		\label{fig:running_ex:t_13:a} 
		\includegraphics[width=0.325 \textwidth]{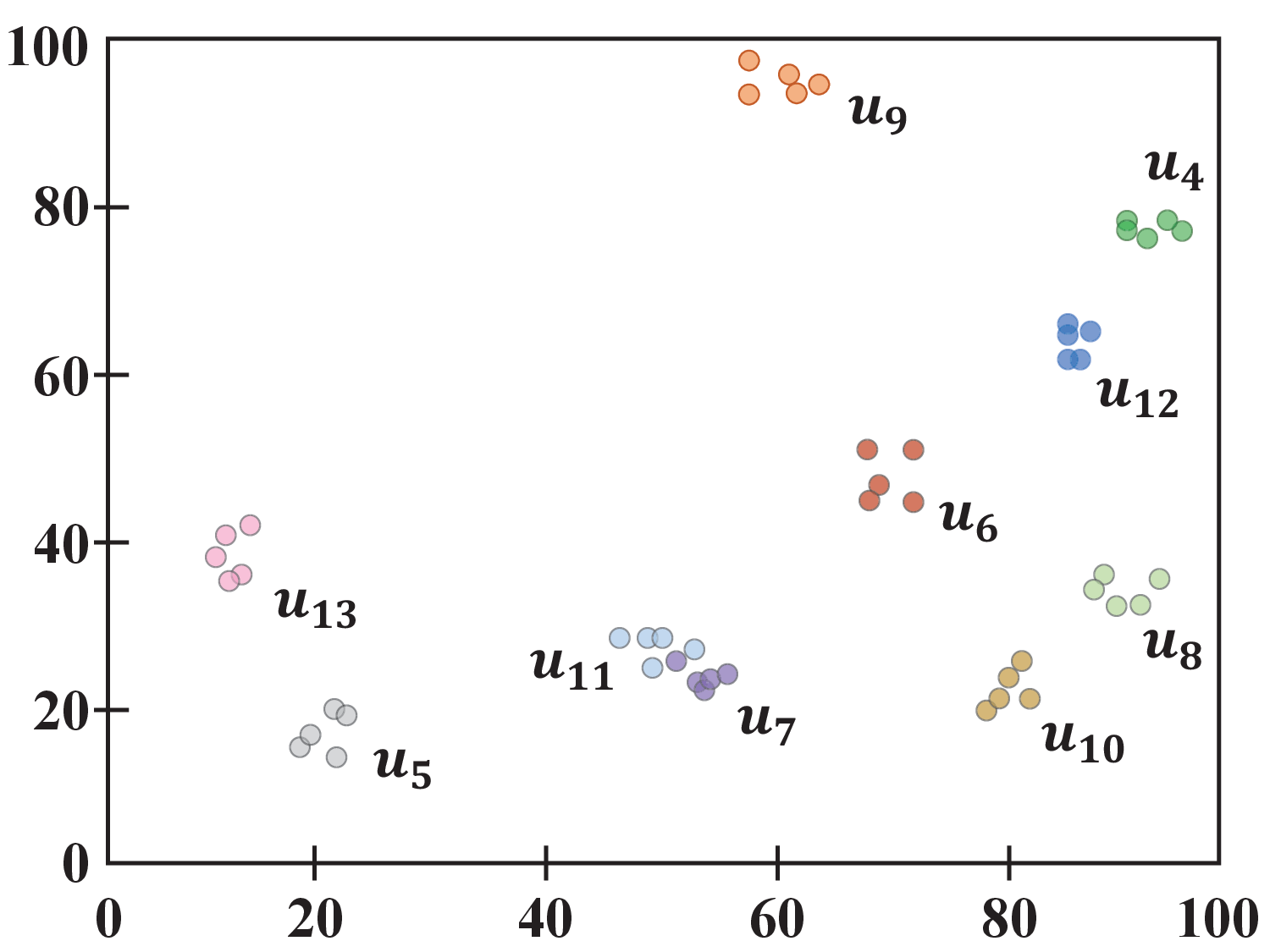}}%
	\subfigure[$ESK_{1,1}(t=13)=\{u_5,u_{13}\}$]{
		\label{fig:running_ex:t_13:b} 
		\includegraphics[width=0.325 \textwidth]{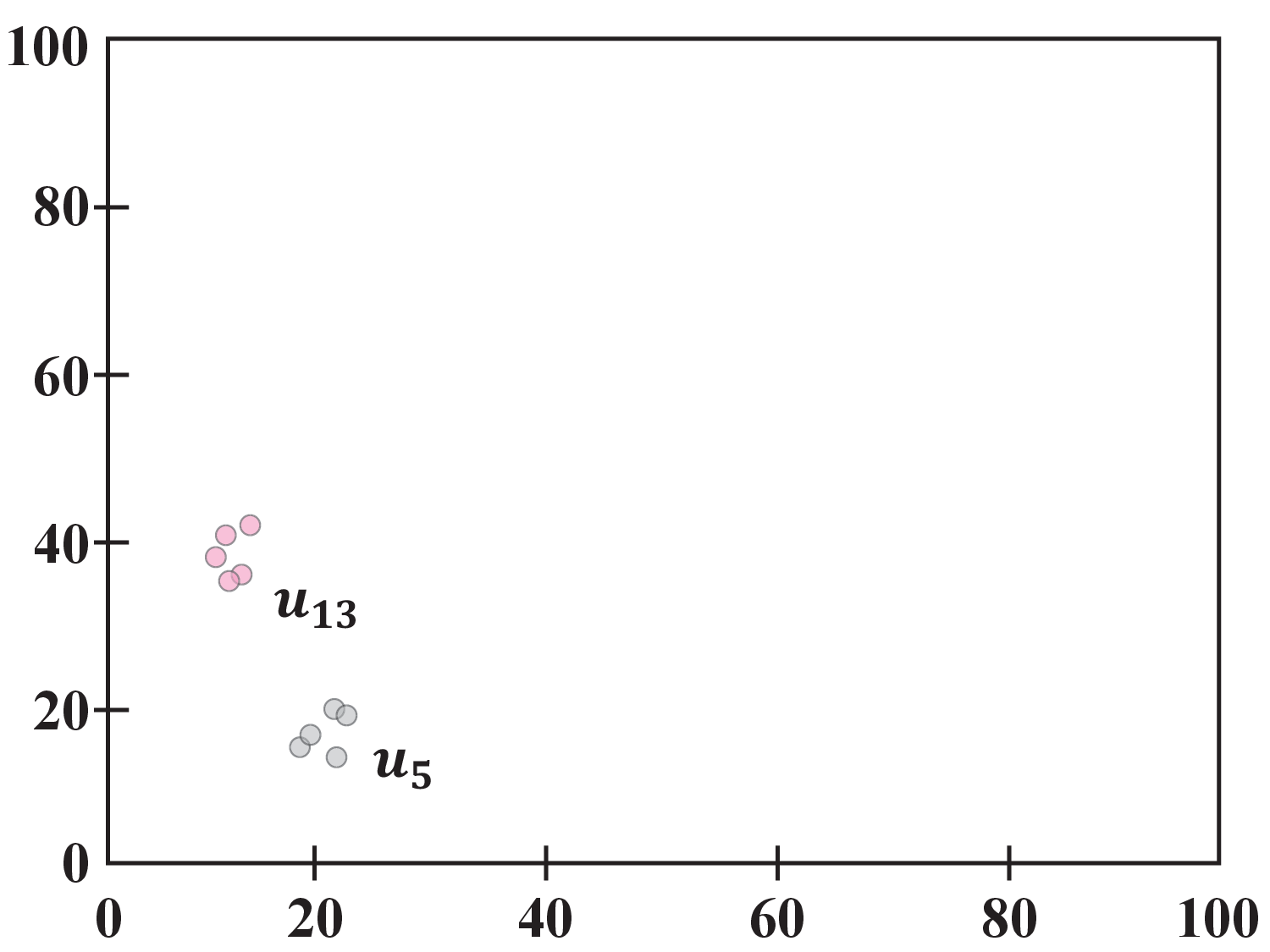}}%
	\subfigure[$ESK_{1,2}(t=13)=\{u_7,u_{10},u_{11}\}$]{
		\label{fig:running_ex:t_13:c} 
		\includegraphics[width=0.325 \textwidth]{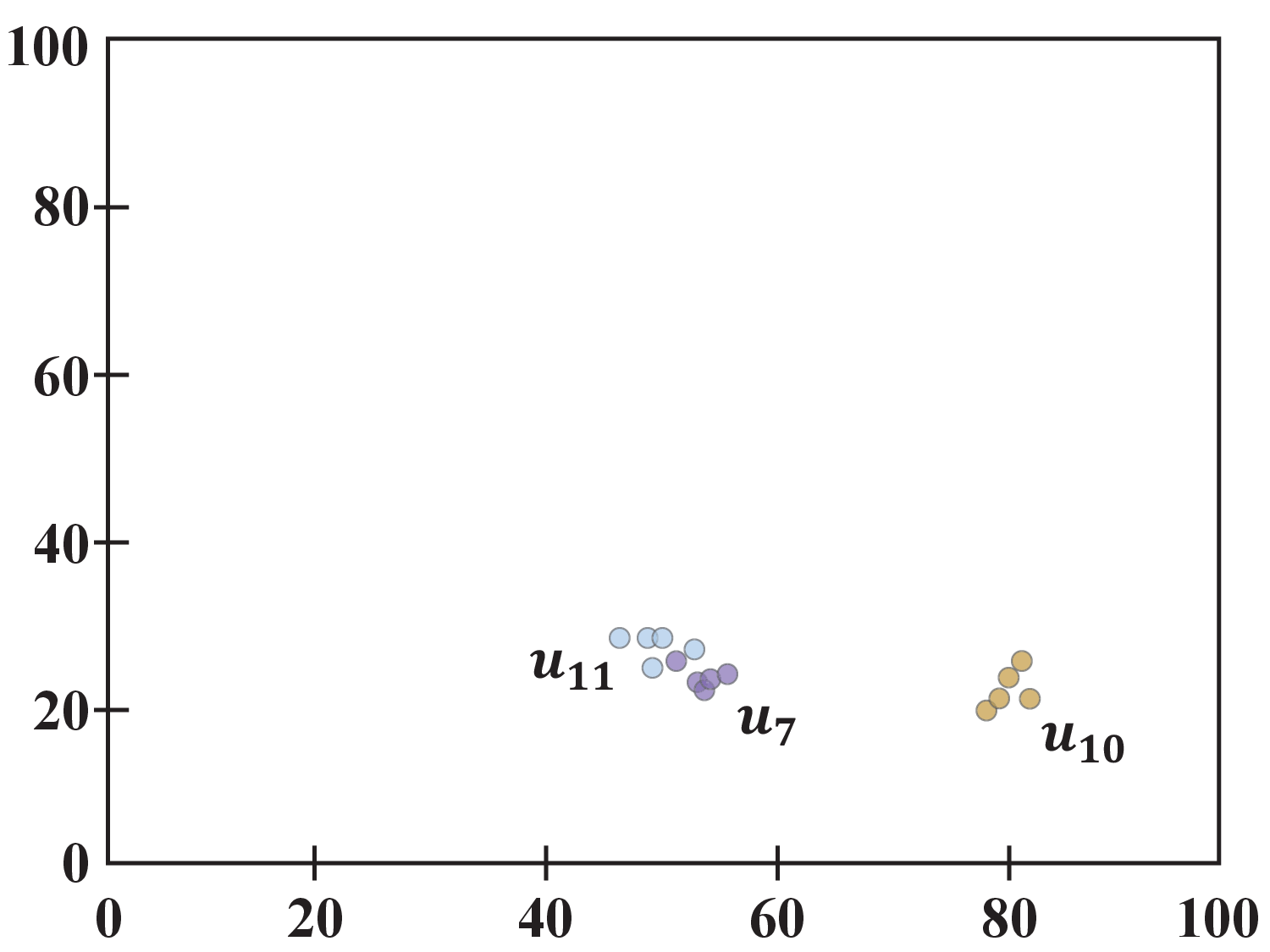}}%
	\caption{The content snapshot of~\subref{fig:running_ex:t_13:a} $SW_1(t=13)=\{u_4,u_5,\dots,u_{13}\}$,~\subref{fig:running_ex:t_13:b} $ESK_{1,1}(t=13)=\{u_5,u_{13}\}$, and~\subref{fig:running_ex:t_13:c} $ESK_{1,2}(t=13)=\{u_7,u_{10},u_{11}\}$ in $E_1$.
	}
	\label{fig:running_ex:t_13} 
\end{figure*}

\begin{figure*}[!ht]
	\centering
	\subfigure[$SW_1(t=14)=\{u_5,u_6,\dots,u_{14}\}$]{
		\label{fig:running_ex:t_14:a} 
		\includegraphics[width=0.325 \textwidth]{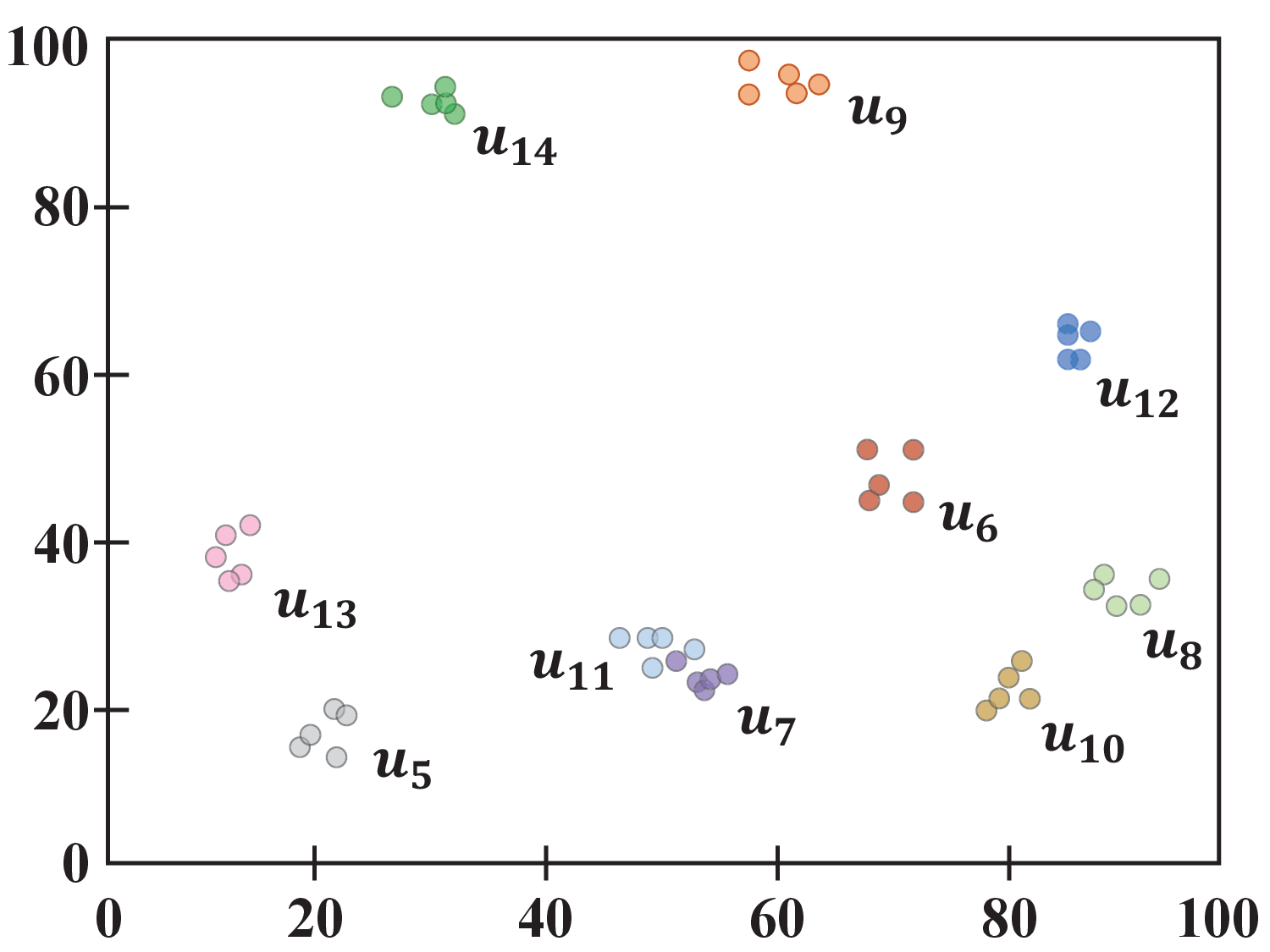}}%
	\subfigure[$ESK_{1,1}(t=14)=\{u_5,u_{13}\}$]{
		\label{fig:running_ex:t_14:b} 
		\includegraphics[width=0.325 \textwidth]{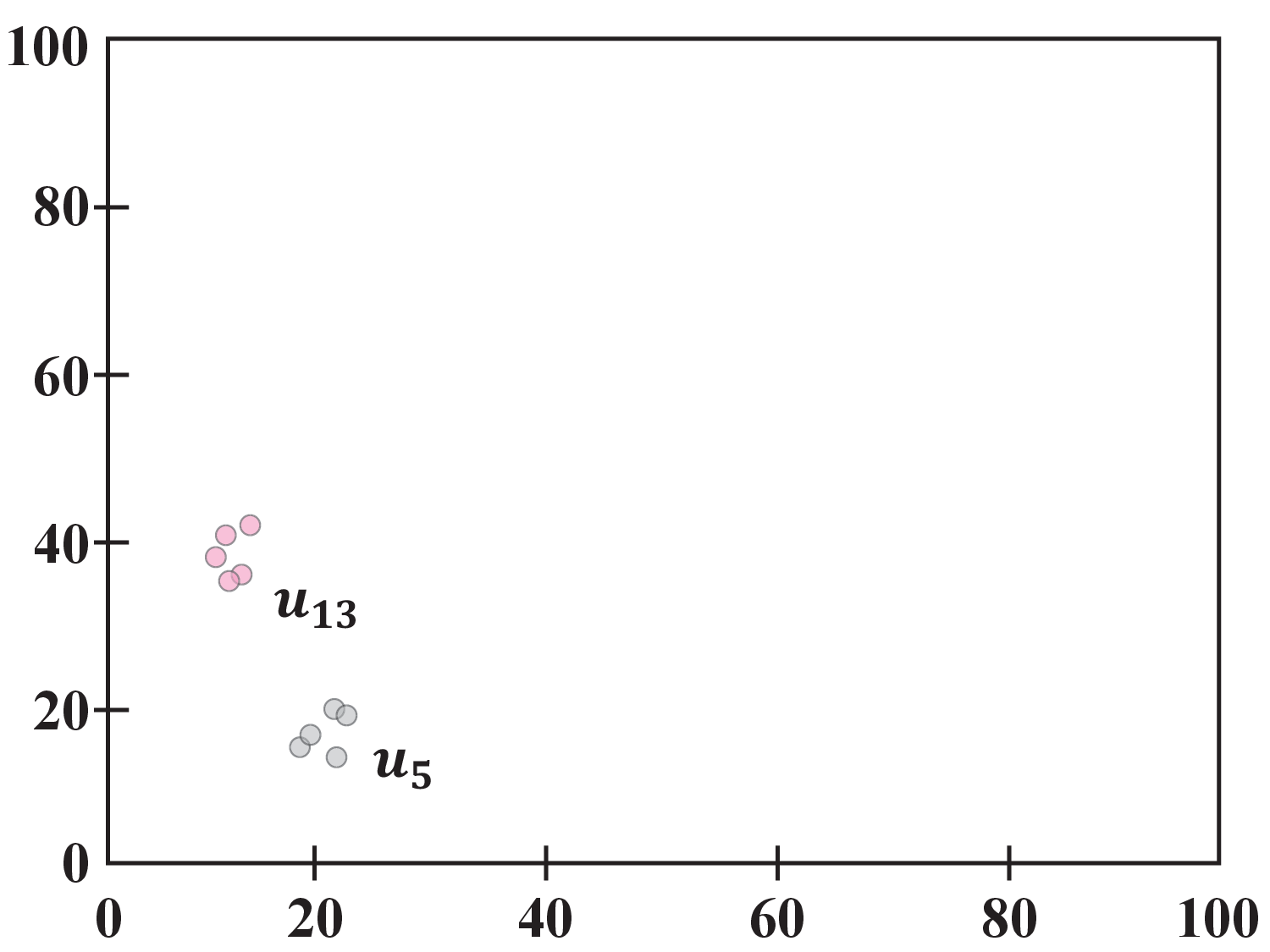}}%
	\subfigure[$ESK_{1,2}(t=14)=\{u_7,u_{10},u_{11},u_{14}\}$]{
		\label{fig:running_ex:t_14:c} 
		\includegraphics[width=0.325 \textwidth]{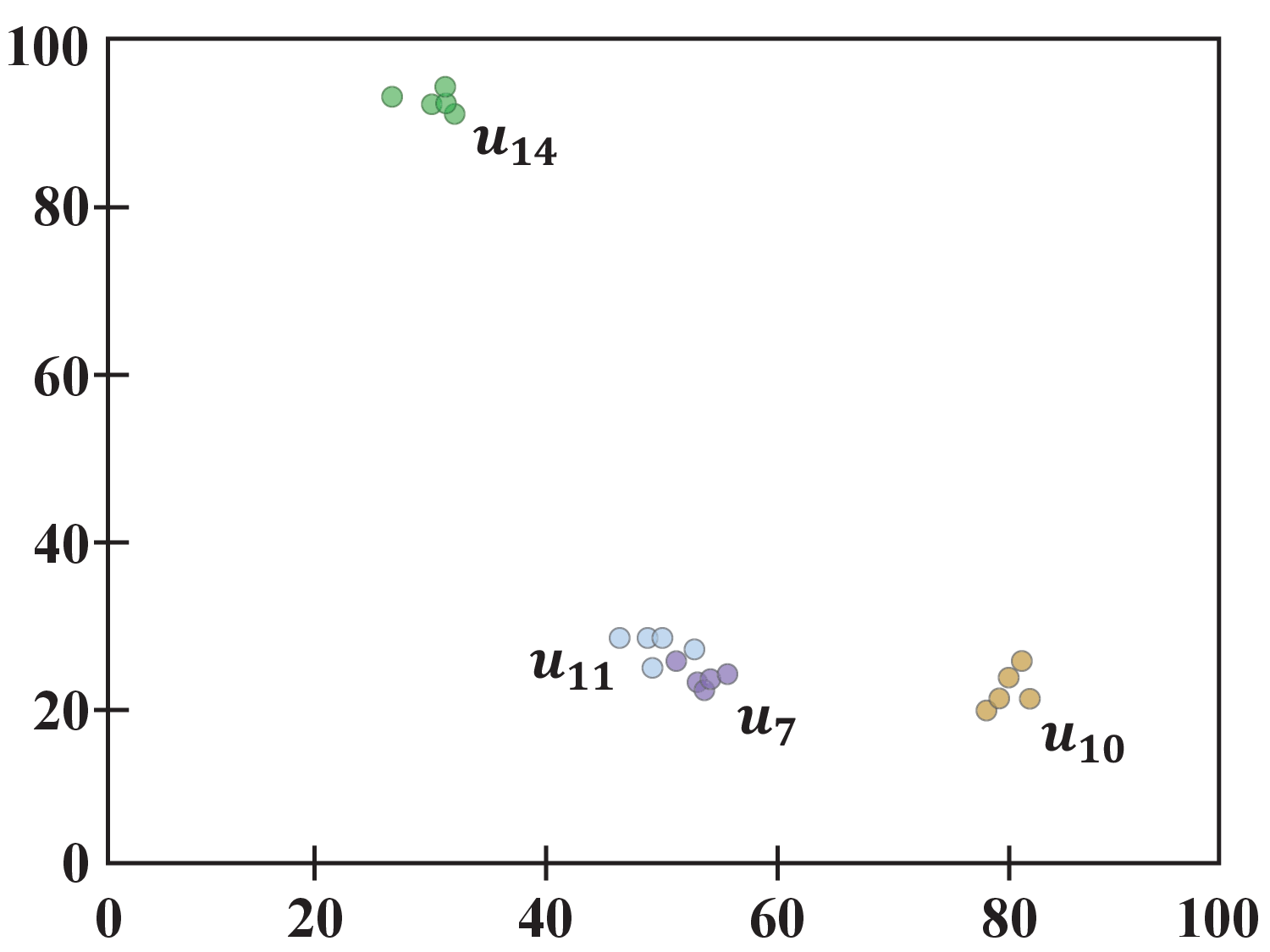}}%
	\caption{The content snapshot of~\subref{fig:running_ex:t_14:a} $SW_1(t=14)=\{u_5,u_6,\dots,u_{14}\}$,~\subref{fig:running_ex:t_14:b} $ESK_{1,1}(t=14)=\{u_5,u_{13}\}$, and~\subref{fig:running_ex:t_14:c} $ESK_{1,2}(t=14)=\{u_7,u_{10},u_{11},u_{14}\}$ in $E_1$.
	}
	\label{fig:running_ex:t_14} 
\end{figure*}

\begin{figure*}[!ht]
	\centering
	\subfigure[$SW_1(t=15)=\{u_6,u_7,\dots,u_{15}\}$]{
		\label{fig:running_ex:t_15:a} 
		\includegraphics[width=0.325 \textwidth]{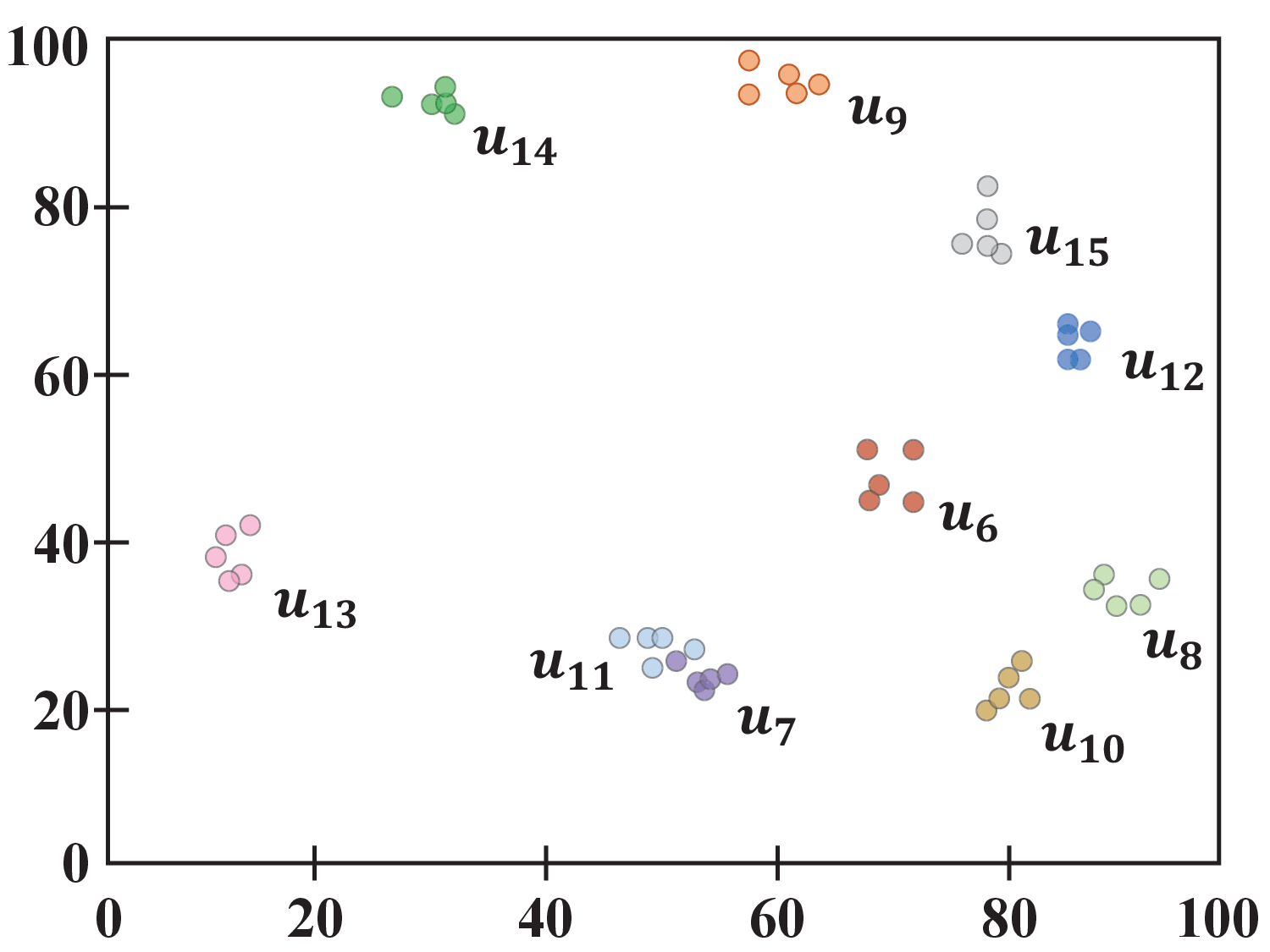}}%
	\subfigure[$ESK_{1,1}(t=15)=\{u_7,u_{10},u_{11},u_{13}\}$]{
		\label{fig:running_ex:t_15:b} 
		\includegraphics[width=0.325 \textwidth]{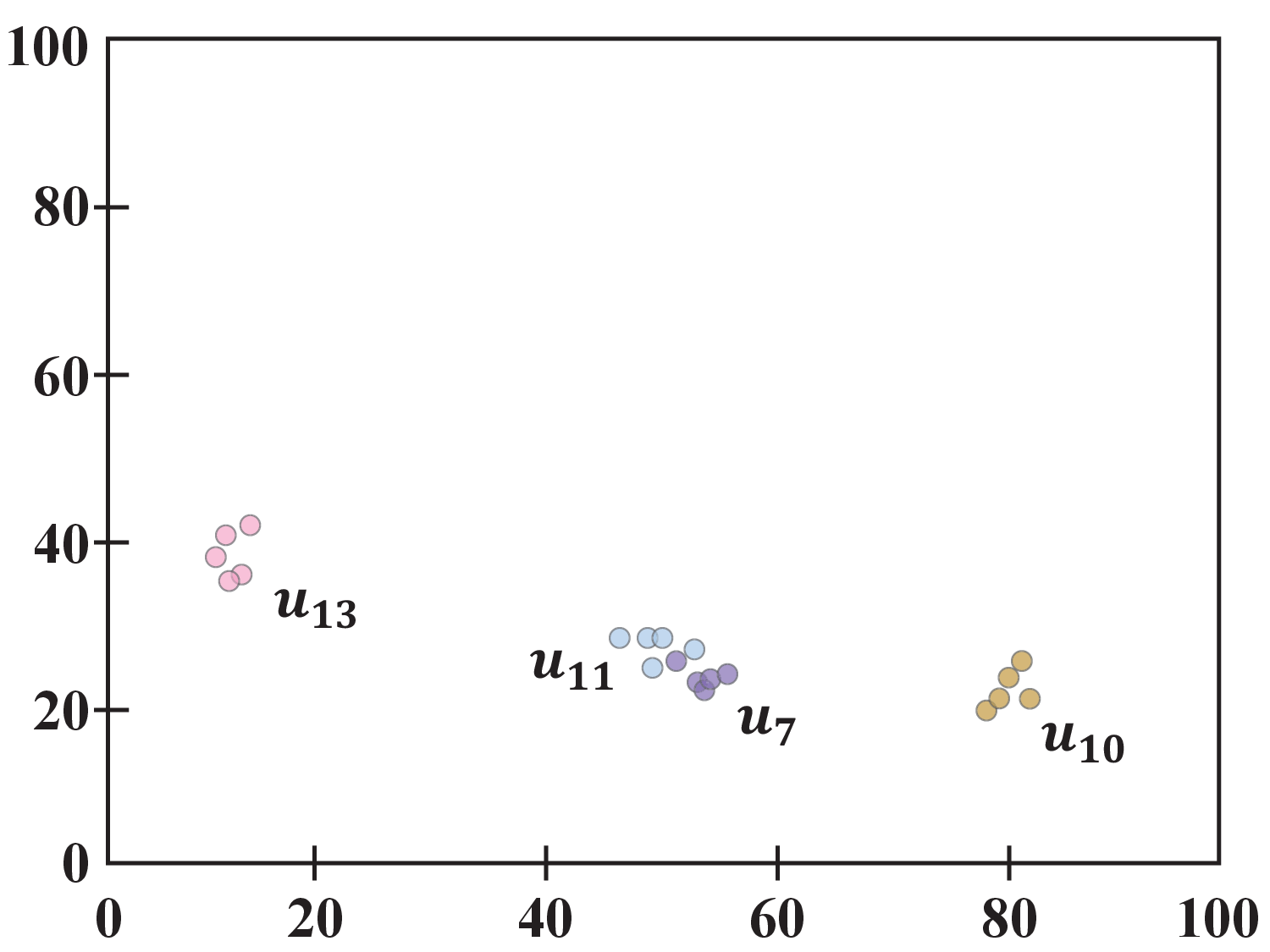}}%
	\subfigure[$ESK_{1,2}(t=15)=\{u_6,u_8,u_{14}\}$]{
		\label{fig:running_ex:t_15:c} 
		\includegraphics[width=0.325 \textwidth]{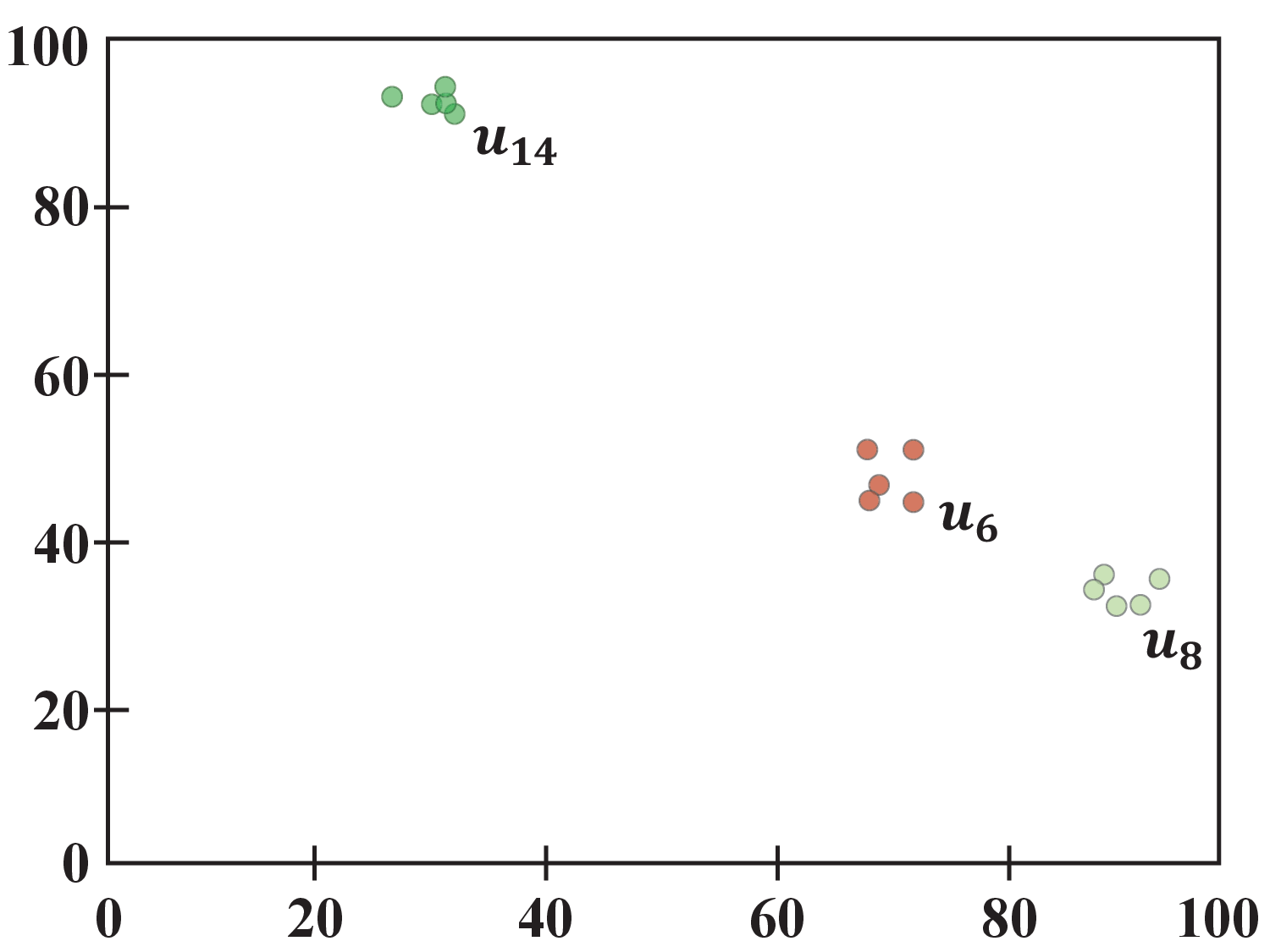}}%
	\caption{The content snapshot of~\subref{fig:running_ex:t_15:a} $SW_1(t=15)=\{u_6,u_7,\dots,u_{15}\}$,~\subref{fig:running_ex:t_15:b} $ESK_{1,1}(t=15)=\{u_7,u_{10},u_{11},u_{13}\}$, and~\subref{fig:running_ex:t_15:c} $ESK_{1,2}(t=15)=\{u_6,u_8,u_{14}\}$ in $E_1$.
	}
	\label{fig:running_ex:t_15} 
\end{figure*}

\begin{figure*}[!t]
	\centering
	\subfigure[$SW_1(t=16)=\{u_7,u_8,\dots,u_{16}\}$]{
		\label{fig:running_ex:t_16:a} 
		\includegraphics[width=0.325 \textwidth]{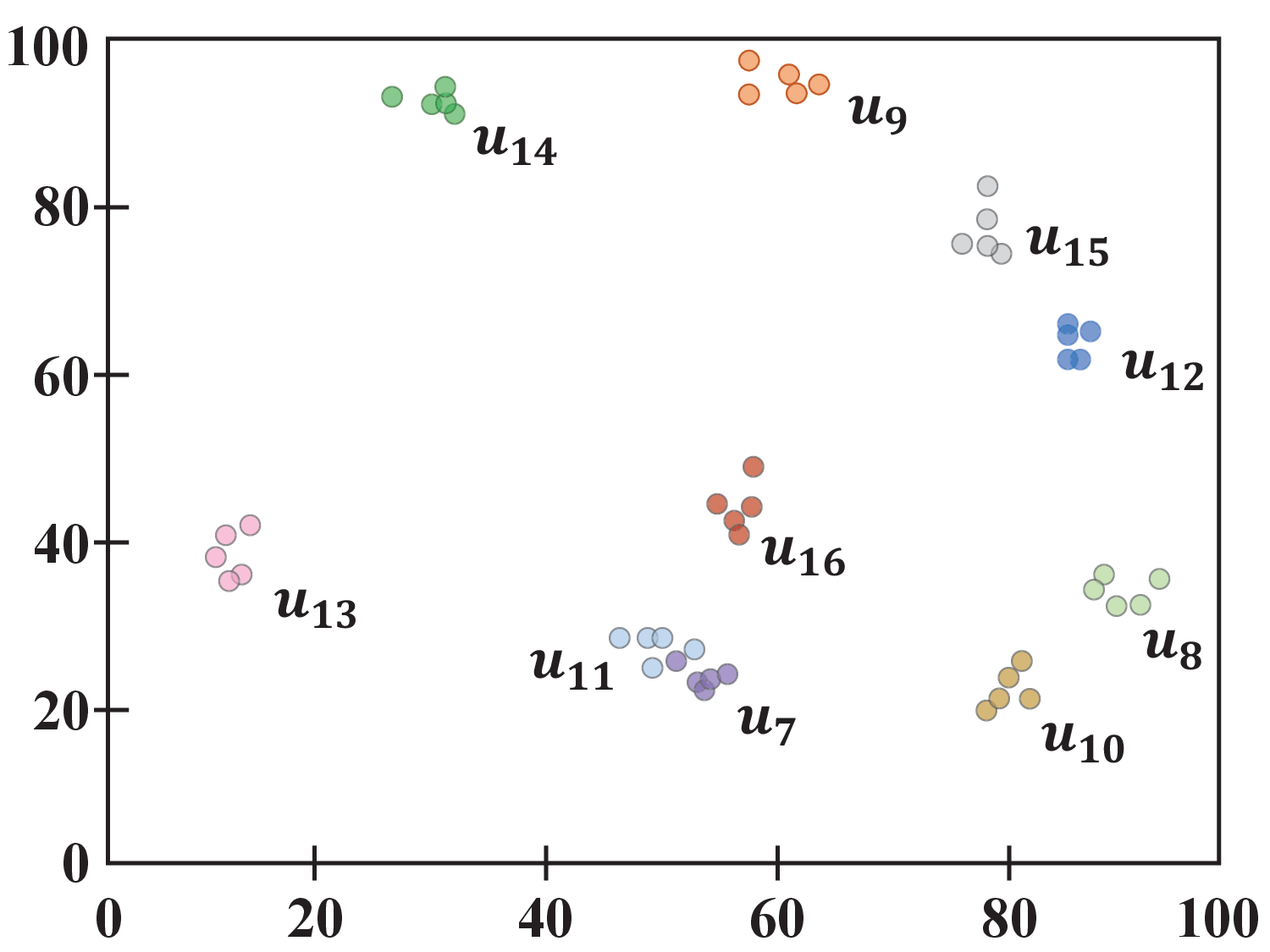}}%
	\subfigure[$ESK_{1,1}(t=16)=\{u_7,u_{10},u_{11},u_{13}\}$]{
		\label{fig:running_ex:t_16:b} 
		\includegraphics[width=0.325 \textwidth]{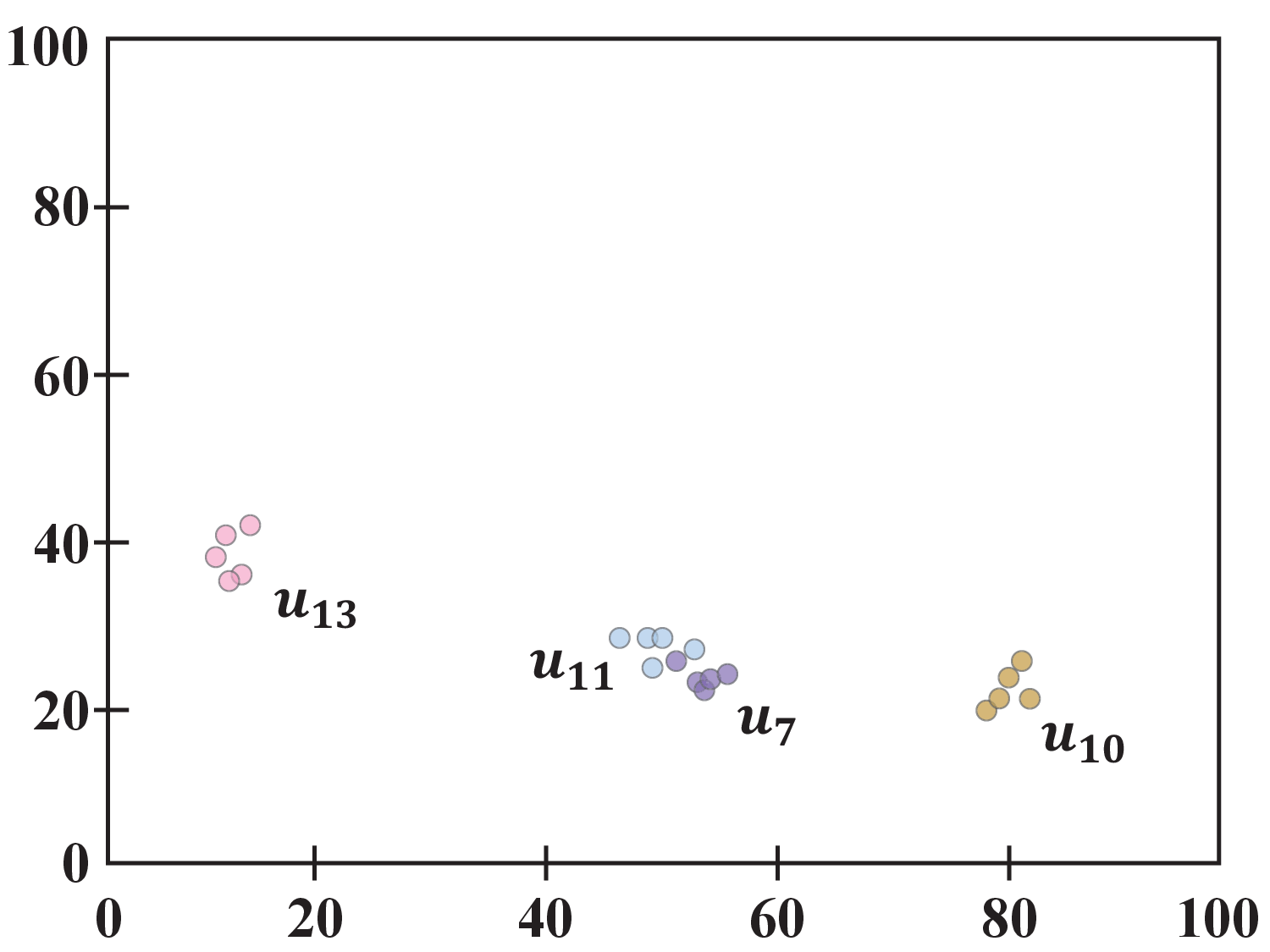}}%
	\subfigure[$ESK_{1,2}(t=16)=\{u_8,u_{14},u_{16}\}$]{
		\label{fig:running_ex:t_16:c} 
		\includegraphics[width=0.325 \textwidth]{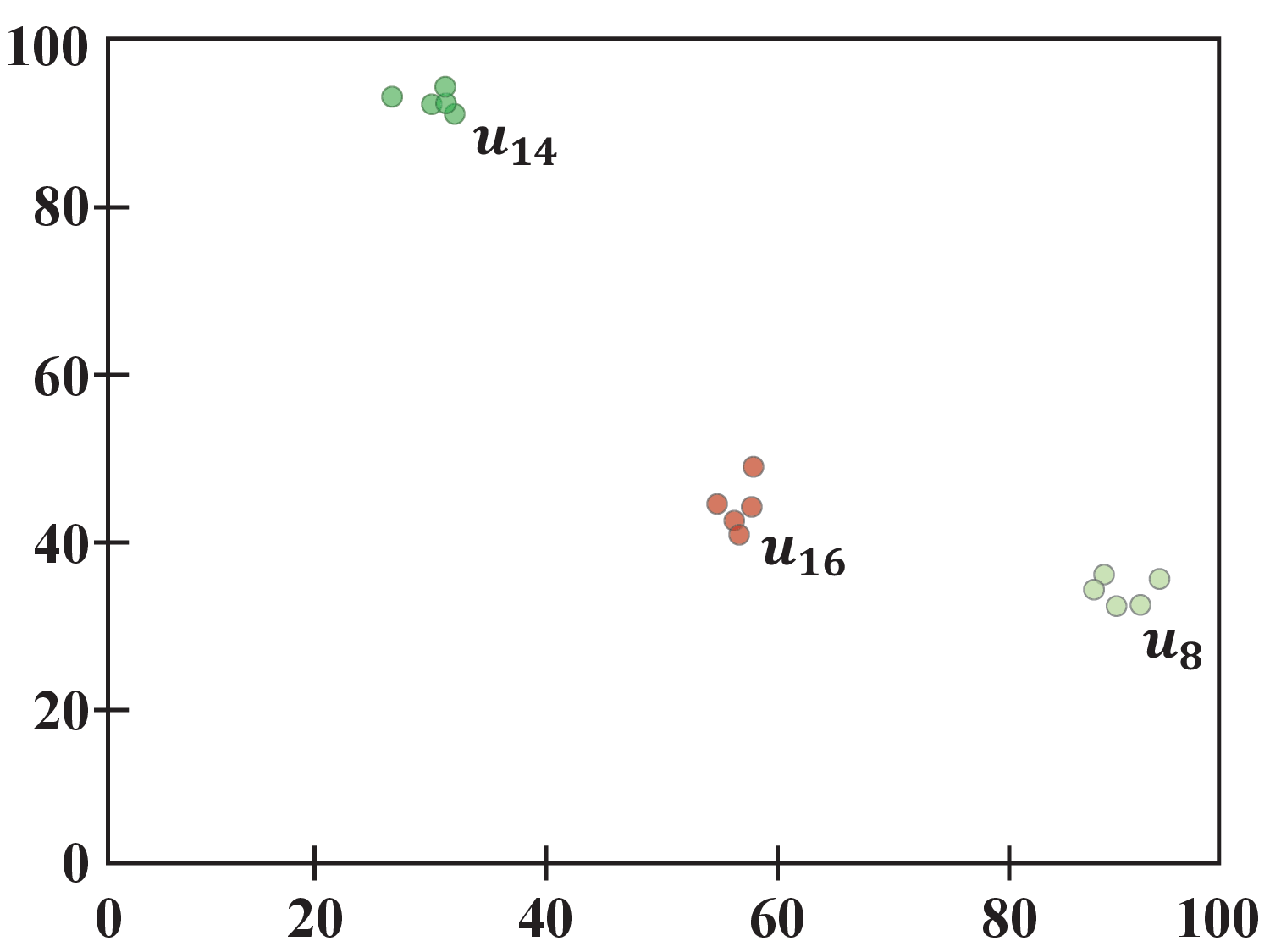}}%
	\caption{The content snapshot of~\subref{fig:running_ex:t_16:a} $SW_1(t=16)=\{u_7,u_8,\dots,u_{16}\}$,~\subref{fig:running_ex:t_16:b} $ESK_{1,1}(t=16)=\{u_7,u_{10},u_{11},u_{13}\}$, and~\subref{fig:running_ex:t_16:c} $ESK_{1,2}(t=16)=\{u_8,u_{14},u_{16}\}$ in $E_1$.
	}
	\label{fig:running_ex:t_16} 
\end{figure*}

\begin{enumerate}
	\item \textbf{Obsolete data is not the edge skyline (in $\bm{ESK_{1,1}}$) and the edge candidate skyline (in $\bm{ESK_{1,2}}$).} When $t=11$, the sliding window $SW_1$ contains data $\{u_2,u_3,\dots,u_{11}\}$. The edge skyline set $ESK_{1,1}$ is $\{u_5\}$ and the edge candidate skyline set $ESK_{1,2}$ is $\{u_7,u_{10},u_{11}\}$. The status of $E_1$ at $t=11$ is depicted in Fig.~\ref{fig:running_ex_1:t_11}.
	When $t=12$, $u_2$ is obsolete and a new data $u_{12}$ comes into the system. Since $SW_1$ is full, the obsolete data $u_2$ will be removed from sliding window $SW_1$. Data object $u_{12}$ will be added to the $SW_1$ after $u_2$ is removed. After data receive procedure (Algorithm~\ref{alg:EEPUS:receive}) is finished, $E_1$ will start the update procedure (Algorithm~\ref{alg:EEPUS:update}). First, it is obvious that both $ESK_{1,1}$ and $ESK_{1,2}$ do not change after removing $u_2$ because $u_2$ is dominated by the other objects. Thus, $u_2$ will be directly removed from $E_1$. 
	\item \textbf{New data is not the edge skyline (in $\bm{ESK_{1,1}}$) and the edge candidate skyline (in $\bm{ESK_{1,2}}$).} Continue the above case of $t=12$, the system will try to add $u_{12}$ to $ESK_{1,1}$ since $u_2$ is not in both $ESK_{1,1}$ and $ESK_{1,2}$. Then $E_1$ will start to update $ESK_{1,1}$. By comparing $u_{12}$ with $u_5$ in $ESK_{1,1}$, $u_{12}$ is dominated by $u_5$ in $ESK_{1,1}$, so $E_1$ will try to add $u_{12}$ to $ESK_{1,2}$ and check whether $u_{12}$ is edge candidate skyline or not. Again, $u_{12}$ is dominated by $u_7$ and $u_{11}$ in $ESK_{1,2}$. As a result, the incoming data $u_{12}$ will not be the edge skyline or the edge candidate skyline on $E_1$. Finally, $E_1$ will send the update message, $\{D_{\rm obsolete}:[u_2],newESK_{1,1}:[],newESK_{1,2}:[]\}$, to the main server node $S$. The content snapshot of $E_1$ at $t=12$ is presented in Fig.~\ref{fig:running_ex:t_12}.
	\item \textbf{New data is the edge skyline (in $\bm{ESK_{1,1}}$).} Continue the above example, when $t=13$, a new data object $u_{13}$ arrives. Object $u_3$ is obsolete and will be removed from $SW_1$. The new data $u_{13}$ will be added to the $SW_1$. $E_1$ will check whether $u_3$ is already in $ESK_{1,1}$ or $ESK_{1,2}$ or not. Since $u_3$ is not the member of the edge skyline $ESK_{1,1}$ and the edge candidate skyline $ESK_{1,2}$, $E_1$ will directly remove $u_3$ from $SW_1$ and then insert $u_{13}$ into $SW_1$. Then, $E_1$ will added $u_{13}$ to $ESK_{1,1}$ because none of objects in $SW_1$ can dominate $u_{13}$. It means that $u_13$ is a member of the edge skyline in $E_1$. Until this step, no further changes will occur afterwards. In this case, $E_1$ finally will send the update massage, $\{D_{\rm obsolete}:[u_3],newESK_{1,1}:[u_{13}],newESK_{1,2}:[]\}$, to the server node $S$ and the content snapshot of $E_1$ at $t_13$ is shown in Fig.~\ref{fig:running_ex:t_13}. 
	\item \textbf{New data is the edge candidate skyline (in $\bm{ESK_{1,2}}$).} If data object $u_{14}$ arrives at $t=14$, object $u_4$ is not in both $ESK_{1,1}$ and $ESK_{1,2}$, so $E_1$ will directly remove the obsolete data $u_4$ from $SW_1$. As usual, $E_1$ will try to add $u_{14}$ to $ESK_{1,1}$ for further update. We can find that $u_5$ and $u_{13}$ in $ESK_{1,1}$ dominate $u_{14}$, so $u_{14}$ will not be the edge skyline. Next, $E_1$ will try to add $u_{14}$ to $ESK_{1,2}$ and check the dominance relations. None of objects in $ESK_{1,2}$ can dominate $u_{14}$, so $u_{14}$ will be the edge candidate skyline and kept in $ESK_{1,2}$. The final content snapshot of $E_1$ at $t_{14}$ is shown in Fig.~\ref{fig:running_ex:t_14}. ECN $E_1$ will set the update message, $\{D_{\rm obsolete}:[u_4],newESK_{1,1}:[],newESK_{1,2}:[u_{14}]\}$, to the main server $S$.
	\item \textbf{Obsolete data is the edge skyline (in $\bm{ESK_{1,1}}$).} As shown in Fig.~\ref{fig:running_ex:t_14:b}, $u_5$ is the member of the edge skyline set $ESK_{1,1}$ at $t=14$. Since $u_{5}$ becomes obsolete at $t=15$, $E_1$ will remove $u_5$ from $SW_1$ and $ESK_{1,1}$. However, some data objects in edge candidate skyline set $ESK_{1,2}$ dominated by $u_5$ may have probability to be the edge skyline. According to Fig.~\ref{fig:running_ex:t_14:c}, objects $u_7$, $u_{10}$ and $u_{11}$ are not dominated by any other data objects in $ESK_{1,1}$, so these objects will become the edge skyline. $E_1$ then moves $u_7$, $u_{10}$ and $u_{11}$ to $ESK_{1,1}$ and the content of $ESK_{1,1}$ is depicted in Fig.~\ref{fig:running_ex:t_15:b}. Next, $E_1$ will examine the remaining objects in the set $SW_1\setminus ESK_{1,1}$ to find the objects that are the edge candidate skyline. As a result, $u_6$ and $u_8$ becomes the edge candidate skyline. $E_1$ then adds $u_6$ and $u_8$ to $ESK_{1,2}$ and the content of $ESK_{1,2}$ is depicted in Fig.~\ref{fig:running_ex:t_15:c}. After that, $E_1$ will examine the new data object $u_{15}$. As shown in Fig.~\ref{fig:running_ex:t_15:a}, $u_{15}$ is dominated by $u_7$, $u_{11}$, and $u_{13}$ in $ESK_{1,1}$, and $u_6$ in $ESK_{1,2}$, respectively. Hence, there is no further change on $ESK_{1,1}$ and $ESK_{1,2}$. Finally, $E_1$ will send the message, $\{D_{\rm obsolete}:[u_5],newESK_{1,1}:[u_7,u_{10},u_{11}],newESK_{1,2}:[u_6,u_8]\}$, to the sever node $S$ for updating the global skyline.
	\item \textbf{Obsolete data is the edge candidate skyline (in $\bm{ESK_{1,2}}$).} At $t=16$, data object $u_{16}$ enters $E_1$ and $u_6$ becomes obsolete. Since $u_6$ is in $ESK_{1,2}$ at $t=15$, as shown in Fig.~\ref{fig:running_ex:t_15:c}, $E_1$ will remove $u_6$ from $SW_1$ and $ESK_{1,2}$, and then try to add $u_{12}$ and $u_{15}$ to $ESK_{1,2}$. However, the new data object $u_{16}$ is dominated by $u_{11}$ in $ESK_{1,1}$ but not dominated by any objects in $ESK_{1,2}$. Object $u_{16}$ thus becomes the edge candidate skyline.
	Fig.~\ref{fig:running_ex:t_16} shows the final states of $SW_1$, $ESK_{1,1}$, and $ESK_{1,2}$. As shown in Fig.~\ref{fig:running_ex:t_16:c}, $E_1$ will add $u_{16}$ to $ESK_{1,2}$ instead of $u_{12}$ and $u_{15}$. ECN $E_1$ will send the message, $\{D_{\rm obsolete}:[u_6],newESK_{1,1}:[],newESK_{1,2}:[u_{16}]\}$, to the sever node $S$.
\end{enumerate}

The above examples have described the scenarios on an ECN. In fact, the main server node $S$ also handles the incoming data streams received from each ECN in the same way except for sending update messages.


\section{Analysis and Discussion}
\label{sec:analysis}

In this section, we are going to analyze and discuss its time complexity and transmission cost of the proposed EPUS in the average case.

\subsection{Time Complexity of EEPUS on Each ECN}
With the above assumptions and notations of the considered system model and our proposed EPUS algorithm, consider the case that a new query, $q(\varDelta_t)$, is issued for monitoring the skyline for a time period $\varDelta_t=[0,\varDelta_t-1]$. At the initial step, $t=0$, each ECN $E_k$ will firstly construct a temporary R-tree, $R_k$, for indexing the data objects in the sliding window, $SW_k$, and then derive the edge skyline, $ESK_{k,1}$, and the edge candidate skyline, $ESK_{k,2}$. So the time complexity of the initial step for $E_k$ will be
\begin{equation}\label{eq:1}
	T_k^{\text{initial}}=T^{\text{construction}}(R_k)+T(ESK_{k,1})+T(ESK_{k,2}).
\end{equation}
where $T^{\text{construction}}(R_k)$ is the time for  the construction of $R_k$, $T(ESK_{k,1})$ is the time for deriving $ESK_{k,1}$, and $T(ESK_{k,2})$ is the time for deriving $ESK_{k,2}$.

According to~\cite{ALBORZI20076}~\cite{Arge:2008:PRP:1328911.1328920}, the time for constructing a $d$-dimensional R-tree for a dataset $U$ is $\mathcal{O}(\frac{|U|}{b}\log_{r_k/b}\frac{|U|}{b})$, where $b$ is the I/O block size of the data on the memory and $r_k$ is the degree fan-out of $R_k$. In our work, we handle uncertain data objects in a object-oriented model ($b=1$), so the time for $E_k$ to construct $R_k$ is 
\begin{equation}\label{eq:2}
	T_k^{\text{construction}}(R_k) = |SW_k|\log_{r_k}|SW_k|.
\end{equation}

\begin{figure}[!t]	
	\centering
	\includegraphics[width=.45\columnwidth]{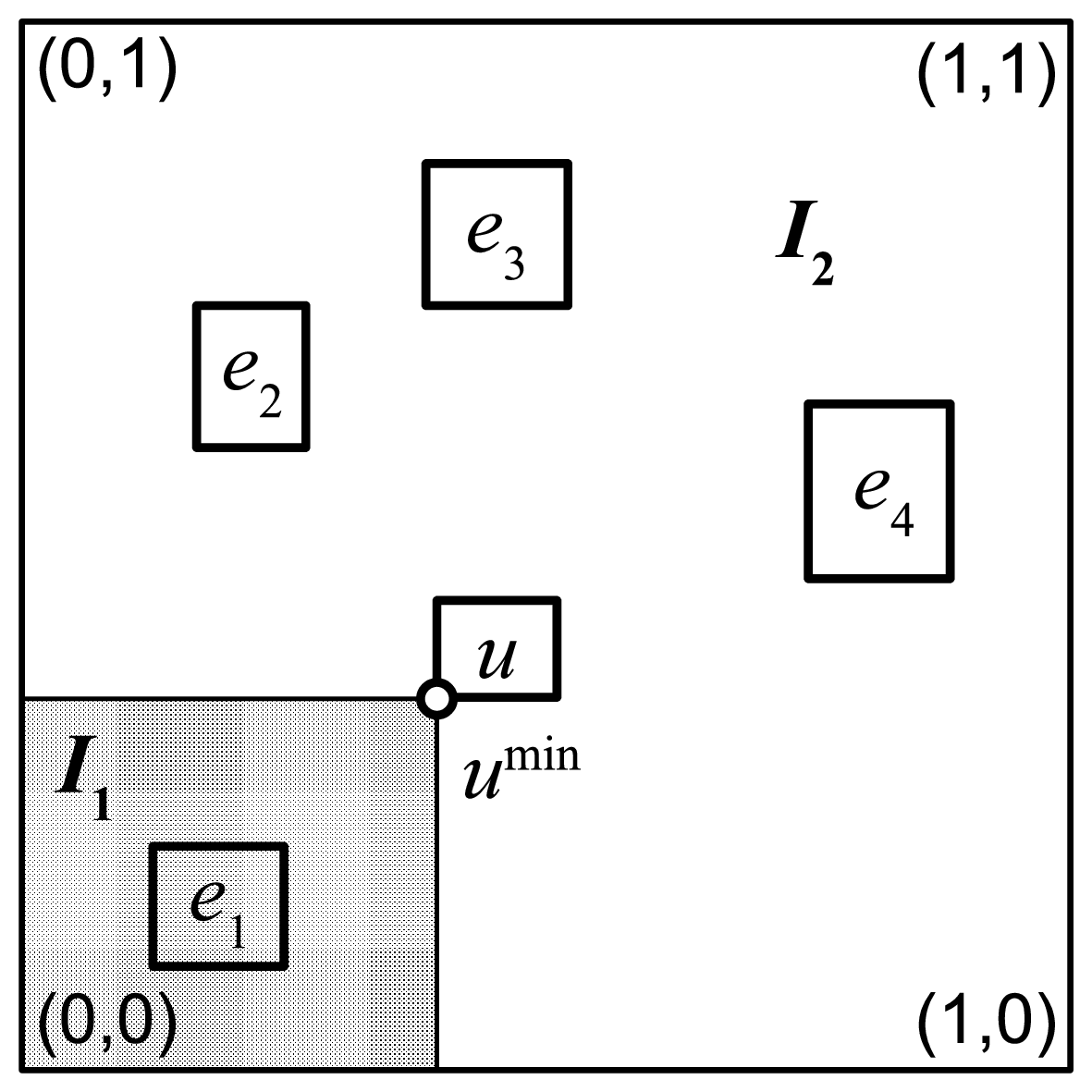}
	\caption{The approximate upper bound of the computational cost on computing $ESK_{k,1}$ and $ESK_{k,2}$ at the level $l$ of $R_k$.}
	\label{fig:complexity_upper_bound}
\end{figure}%

In practice, $T(ESK_{k,1})$ and $T(ESK_{k,2})$ are very difficult to model because they are dependent to the distribution of input data streams. In this work, we assume that all input data objects are uniformly and independently distributed in the domain space $[0, 1000]^d$ where $d$ is the data dimension. For ease of analysis, we can normalize the domain space into $[0, 1]^d$. Let the leaf level be $l_k=0$, the height of $R_k$ will approximate $h_k=1+\lceil\log_{r_k}\left(|SW_k|/r_k\right)\rceil$ and the number of nodes at level $l_k$ of $R_k$ will be $N_{l_k}=|SW_k|/\left(r_k\right)^{l_k+1}$. Besides, the extent $\theta_l$ (i.e., length of any 1-D projection) of a node at the $l$-th level can be estimated by $l=\left(1/N_{l_k}\right)^{1/d}$.
For deriving $ESK_{k,1}$ at the initial step, $E_k$ will examine the dominance relations between the data objects in $|SW_k|$. With the help of the R-tree, Fig.~\ref{fig:complexity_upper_bound} shows that the gray region $I_1$ corresponds to the maximal region, entirely covering nodes (at level $l$) that can dominate the uncertain data object $u$. Note that $u_{\min}$ in Fig.~\ref{fig:complexity_upper_bound} represents the best instance (or minimum boundary) of $u$. In this case, $u$ is dominated by $e_1$, so $u$ will not be the member of $ESK_{k,1}$. Then, the average complexity of required node accesses in $R_k$ for checking whether each object $u\in SW_k$ is the member of $ESK_{k,1}$ will be
\begin{equation}\label{eq:3}
	T_k^{\text{skyline}}(u)=\sum_{l_k=0}^{h_k-1}N_{l_k}\times n^2\times v_{u_{\min}}^d,
\end{equation}
where $v_{u_{\min}}$ is the value of $u_{\min}$ after 1-D projection and $n$ is the number of instances in an uncertain object. Hence, the time complexity of deriving the edge skyline $ESK_{k,1}$ on $E_k$ will be
\begin{equation}\label{eq:4}
	T_k^{\text{skyline}}(SW_k)=|SW_k|\times T_k^{\text{skyline}}(u),
\end{equation}
where $u\in SW_k$.
After obtaining $ESK_{k,1}$, ECN $E_k$ then use $R_k$ find the edge candidate skyline $ESK_{k,2}$ in the same way. To avoid wasting the memory space, we can append a flag value to each node on $R_k$ to indicate whether a node is the edge skyline. That is, the approximate time complexity of deriving $ESK_{k,2}$ will be 
\begin{equation}\label{eq:eq:5}
	T_k^{\text{skyline}}(\overline{SW}_k)=|\overline{SW}_k|\times T_k^{\text{skyline}}(u),
\end{equation}
where $\overline{SW}_k=SW_k\setminus ESK_{k,1}$ and $u\in\overline{SW}_k$.

However, during the time period of the skyline query, $\varDelta_t$, ECN $E_k$ needs to continuously update both $ESK_{k,1}$ and $ESK_{k,2}$, so we can use HashMap to store these two sets to reduce the complexity of further searches and comparisons. With~\eqref{eq:4} and~\eqref{eq:eq:5}, the average time complexities of constructing $ESK_{k,1}$ and $ESK_{k,2}$ at the initial step can be respectively formulated as
\begin{align}
	T(ESK_{k,1})&=T_k^{\text{skyline}}(SW_k)+T^{\text{construction}}(ESK_{k,1})\nonumber\\
	&=|SW_k|\times T_k^{\text{skyline}}(u)+|ESK_{k,1}|\label{eq:eq:6}
\end{align}
and
\begin{align}
	T(ESK_{k,2})&=T_k^{\text{skyline}}(\overline{SW}_k)+T^{\text{construction}}(ESK_{k,2})\nonumber\\
	&=|\overline{SW}_k|\times T_k^{\text{skyline}}(u)+|ESK_{k,2}|,\label{eq:eq:7}
\end{align}
where $T^{\text{construction}}(ESK_{k,1})=|ESK_{k,1}|$ and $T^{\text{construction}}(ESK_{k,2})=|ESK_{k,2}|$.

According to~\eqref{eq:2} to~\eqref{eq:eq:5} and $|\overline{SW}_k|\leq|SW_k|$, the time complexity of the initial step for $E_k$, $T_k^{\text{initial}}$, in~\eqref{eq:1} can be rewritten as
\begin{align}
	T_k^{\text{initial}}&\leq|SW_k|\times\left(\log_{r_k}|SW_k|+2\times T_k^{\text{skyline}}(u)\right) \nonumber\\
	&+|ESK_{k,1}|+|ESK_{k,2}|.\label{eq:8}
\end{align}

If $t>0$, $E_k$ will continuously update $ESK_{k,1}$ and $ESK_{k,2}$. Suppose that $F_k^{\text{in}}$ is the buffer $E_k$ for receiving data at each time slot, so the buffer size $|F_k^{\text{in}}|$ is also the data arrival rate. Since the size of $SW_k$ is fixed and $|F_k^{\text{in}}|$ oldest data objects $SW_k$ will leave at each time slot, some data objects in $ESK_{k,1}$ and $ESK_{k,2}$ need to be removed in advance if they becomes obsolete data. 
We assume that $F_k^{\text{out}}$ is the set of obsolete data and $|F_k^{\text{out}}|=|F_k^{\text{in}}|$. The time complexity for handling the departure of these data from $ESK_{k,1}$ will be
\begin{equation}\label{eq:9}
	T_{\text{update}}^{\text{departure}}(ESK_{k,1})=|F_k^{\text{out}}|+T_k^{\text{skyline}}(\overline{ESK}_{k,2}),
\end{equation}
where the complexity of checking and removing obsolete data objects from $ESK_{k,1}$ is $\mathcal{O}(1)\times|F_k^{\text{out}}|$ since $ESK_{k,1}$ is stored in HashMaps, $T_k^{\text{skyline}}(\overline{ESK}_{k,2})$ is the time for examining whether the remaining valid edge candidate skyline objects can become edge skyline or not. Note that $\overline{ESK}_{k,2}=ESK_{k,2}\setminus F_k^{\text{out}}$ is the set of remaining valid edge candidate skyline objects. For handling the departure of these data from $ESK_{k,2}$, the time complexity will be
\begin{equation}\label{eq:10}
	T_{\text{update}}^{\text{departure}}(ESK_{k,2})=|F_k^{\text{out}}|+T_k^{\text{skyline}}(\overline{SW}_k),
\end{equation}
where $T_k^{\text{skyline}}(\overline{SW}_k)$ is the time for selecting objects that may become edge candidate skyline from $\overline{SW}_k=SW_k\setminus (F_k^{\text{out}}\cup ESK_{k,1}\cup ESK_{k,2})$.

Hence, with~\eqref{eq:9} and~\eqref{eq:10}, the bound of time complexity for maintaining $ESK_{k,1}$ and $ESK_{k,2}$ at time $t$ will be 
\begin{align}
	T_k^{\text{departure}}(F_k^{\text{out}},t)&=T_{\text{update}}^{\text{departure}}(ESK_{k,1})+T_{\text{update}}^{\text{departure}}(ESK_{k,2}) \nonumber\\
	&= 2|F_k^{\text{out}}|+\left(|\overline{ESK}_{k,2}|+|\overline{SW}_k|\right)\times T_k^{\text{skyline}}(u).\label{eq:11}
\end{align}

After handling the removal of obsolete data, according to~\eqref{eq:3} to~\eqref{eq:eq:5}, the time complexity for handling input new data and updating $ESK_{k,1}$ and $ESK_{k,2}$ at time $t$ will be
\begin{align}
	T_k^{\text{arrival}}(F_k^{\text{in}},t)&=T_{\text{update}}^{\text{arrival}}(ESK_{k,1})+T_{\text{update}}^{\text{arrival}}(ESK_{k,2})\nonumber\\
	&=\left(|F_k^{\text{in}}|+T_k^{\text{skyline}}(F_k^{\text{in}})\right)+\left(|F_k^{\text{in}}|+T_k^{\text{skyline}}(\overline{F}_{\text{in}})\right)\nonumber\\
	&=2|F_k^{\text{in}}|+\left(|F_k^{\text{in}}|+|\overline{F}_{\text{in}}|\right)\times T_k^{\text{skyline}}(u)\label{eq:12}
\end{align}
where $\overline{F}_k^{\text{in}}=F_k^{\text{in}}\setminus ESK_{k,1}$. 

With~\eqref{eq:11} and~\eqref{eq:12}, the time complexity of the update step when $t>0$ will be
\begin{align*}
	T_k^{\text{update}}&=T_k^{\text{departure}}(F_k^{\text{out}},t)+T_k^{\text{arrival}}(F_k^{\text{in}},t)\\
	&=2\left(|F_k^{\text{in}}|+|F_k^{\text{out}}|\right)\\
	&+\left(|\overline{ESK}_{k,2}|+|\overline{SW}_k|+|F_k^{\text{in}}|+|\overline{F}_k^{\text{in}}|\right)\times T_k^{\text{skyline}}(u).
\end{align*}
Since $|F_k^{\text{in}}|=|F_k^{\text{out}}|$, $|\overline{ESK}_{k,2}|\leq|ESK_{k,2}|$, $|\overline{SW}_k|\leq|SW_k|$, and $|\overline{F}_k^{\text{in}}|\leq|F_k^{\text{in}}|$, the time complexity bound of the update step when $t>0$ can be simplified as
\begin{align}
	T_k^{\text{update}}\leq 4|F_k^{\text{in}}|+\left(|ESK_{k,2}|+|SW_k|+2|F_k^{\text{in}}|\right)\times T_k^{\text{skyline}}(u).\label{eq:13}
\end{align}

In summary, the average time complexity of executing the EEPUS procedure on $E_k$ for a skyline query $q(\varDelta_t)$ can be summarized as
\begin{equation}\label{eq:14}
	T_k\left(q(\varDelta_t)\right)=\frac{1}{\varDelta_t}\left(T_k^{\text{initial}}+\sum_{t=1}^{\varDelta_t-1}T_k^{\text{update}}\right).
\end{equation}
Since the time complexity of SEPUS on the server node $S$ depends on the amount of data received from each ECN in each time slot, we will discuss this in detail after analyzing the transmission cost.

\subsection{Transmission Cost}
In the proposed EPUS approach, the server node, $S$, uses almost the same way to compute the global skyline, $SK_1$, and the global candidate skyline, $SK_2$ with consideration of the data objects in its sliding window, $SW_S$. Note that we focus on the computation latency in this work, so we ignore the issue of transmission time between each ECN, $E_k$, and the server node, $S$.
Since $S$ receive edge skyline sets from every ECNs for compute $SK_1$ and $SK_2$ at the initial step, the input buffer of $S$ at $t=0$ will be
\begin{equation}\label{eq:15}
	F_S^{\text{in}}(t=0)=\bigsqcup_{k=1}^{m}\left(ESK_{k,1}^{t=0}\cup ESK_{k,2}^{t=0}\right),
\end{equation}
and $|F_S^{\text{in}}(t)|$ is the data arrival rate for $S$, where $ESK_{k,1}^{t=0}$ and $ESK_{k,2}^{t=0}$ are the sets of $ESK_{k,1}$ and $ESK_{k,2}$ at $t=0$, respectively. For simplicity, we only consider the case of $|F_S^{\text{in}}(t)|\leq SW_S$ in our work. It means that the server node $S$ always has sufficient space resource for $SW_S$ to handle the receive information from each $E_k$. According to the design of our proposed EPUS, when $t>0$, $E_k$ does not always transmit the whole sets $ESK_{k,1}^t$ and $ESK_{k,2}^t$ for updating the global result. Each $E_k$ transmit the new objects in $ESK_{k,1}^t$ and $ESK_{k,2}^t$ only. The set of new objects in $ESK_{k,1}^t$ and $ESK_{k,2}^t$ are denoted as $newESK_{k,1}^t=ESK_{k,1}^t\setminus ESK_{k,1}^{t-1}$ and $newESK_{k,2}^t=ESK_{k,2}^t\setminus ESK_{k,2}^{t-1}$, respectively. In addition, each $E_k$ needs to notify $S$ with the information of the obsolete data set, $F_k^{\text{out}}$. 
Hence, for each update step $t>0$, the input buffer of $S$ will be
\begin{equation}\label{eq:16}
	F_S^{\text{in}}(t>0)=\bigsqcup_{k=1}^{m}\left(F_k^{\text{out}}\cup newESK_{k,1}^t\cup newESK_{k,2}^t\right).
\end{equation}

In summary, the average transmission cost for processing a skyline query $q(\varDelta_t)$ during a time period $\varDelta_t$ can be estimated by
\begin{equation}\label{eq:17}
	C_{\text{average}}= \dfrac{1}{\varDelta_t}\left(\left|F_S^{\text{in}}(t=0)\right|+\sum_{t=1}^{\varDelta_t-1}\left|F_S^{\text{in}}(t>0)\right|\right).
\end{equation}

\subsection{Time Complexity of SEPUS on The Server Node}
The server node also constructs an R-tree, $R_S$, for maintaining the objects in $SW_S$ received from each $E_k$, where $k=1,2,\dots,m$. Then, the average complexity of required node accesses in $R_S$ for checking whether each object $u\in SW_S$ is the member of $SK_k$ will be
\begin{equation}\label{eq:18}
	T_S^{\text{skyline}}(u)=\sum_{l_S=0}^{h_s-1}N_{l_S}\times n^2\times v_{u_{\min}}^d,
\end{equation}
where $h_s=1+\lceil\log_{r_S}\left(|SW_S|/r_S\right)\rceil$ is the height of $R_S$, $r_S$ is the degree fan-out of $R_S$, and $v_{u_{\min}}$ is the value of $u_{\min}$ after 1-D projection and $n$ is the number of instances in an uncertain object. With~\eqref{eq:18}, the time complexity of the initial step for the server node, $S$, is similar to~\eqref{eq:8} and it can be written as
\begin{align}
	T_S^{\text{initial}}&\leq|SW_S|\times\left(\log_{r_S}|SW_S|+2\times T_S^{\text{skyline}}(u)\right) \nonumber\\
	&+|SK_1|+|SK_2|.\label{eq:19}
\end{align}

According to~\eqref{eq:17}, both the average arrival rate and departure rates of data objects for $S$ are $C_{\text{average}}$. Similar to the equations from~\eqref{eq:9} to~\eqref{eq:13}, the time complexity of the update step for $S$ will be
\begin{align}
	T_S^{\text{update}}&\leq 4 C_{\text{average}}+\left(|SK_2|+|SW_S|+2C_{\text{average}}\right)\times T_S^{\text{skyline}}(u).\label{eq:20}
\end{align}	

With~\eqref{eq:19} and~\eqref{eq:20}, the average time complexity of executing SEPUS procedure on $S$ for monitoring a skyline query $q(\varDelta_t)$ will approximate
\begin{equation}\label{eq:21}
 	T_S\left(q(\varDelta_t)\right)=\frac{1}{\varDelta_t}\left(T_S^{\text{initial}}+\sum_{t=1}^{\varDelta_t-1}T_S^{\text{update}}\right).
\end{equation}

\subsection{System Latency}
According the above assumptions and~\eqref{eq:14}, suppose that the computing power of each ESN is $P_k$ (objects/sec), the average computation latency on parallel edge nodes will be
\begin{align}\label{eq:22}
	L^{\rm comp}_{\rm edge} = \frac{1}{m}\sum_{k=1}^{m}\dfrac{T_k\left(q(\varDelta_t)\right)}{P_k}.
\end{align}	
According to~\eqref{eq:21}, suppose that the computing power of the server node is $P_S$, the average computation latency of the server node is estimated by
\begin{align}\label{eq:23}
	L^{\rm comp}_S = \dfrac{T_S\left(q(\varDelta_t)\right)}{P_S}.
\end{align}	
In the system model considered, network bottlenecks generally occur on the server side. According to~\eqref{eq:17}, suppose that the receiving data rate of the server node $S$ is $B_S$ (bps) and the unit size of a data object is $|u|$ (bits), the average transmission latency will be
\begin{align}\label{eq:24}
	L^{\rm comm}_{S} = \dfrac{C_{\text{average}}\times |u|}{B_S}.
\end{align}	

Finally, according to~\eqref{eq:22},~\eqref{eq:23}, and~\eqref{eq:24}, the average system latency will be expressed as 
\begin{align}\label{eq:25}
	L^{\rm system} = L^{\rm comp}_{\rm edge} + L^{\rm comm}_{S} + L^{\rm comp}_S.
\end{align}

\section{Simulation Results}
\label{sec:simulation}
To evaluate the performance of our proposed EPUS, we conducted simulations on a computer equipped with an Intel Xeon W-1250 CPU and 48GB RAM running Ubuntu 20.04.4 LTS. All simulations were implemented in Python 3.7. The default transmission rate (uplink/downlink) was set to 1 Mbps, consistent with the peak uplink rate of the enhanced Machine Type Communication communication (eMTC) architecture. Additionally, the upload information for each data object was encapsulated in an MQTT packet with a size of 3 KB (0.003 MB). The detailed simulation parameter settings are shown in Table~\ref{table:simulation_settings}.

In this section, we conduct several simulations to verify the performance of the proposed EPUS. We perform the following three different methods to compare in the edge computing environment:
\begin{itemize}
	\item \textbf{Parallel Brute-Force (PBF).} For this baseline method, each ECN $E_k$ calculates $ESK_{k,1}$ in a straightforward manner without any index-based or tree-based pruning design, where $k=1,2,\dots,m$. When $E_k$ receives new data, $E_k$ will always use all the data in sliding window $SW_k$ to re-calculate the edge skyline set, $ESK_{k,1}$. After that, $E_k$ sends the whole updated $ESK_{k,1}$ to the server node. The server node $E_S$ will continuously update the global skyline set, $SK_1$, as it receives updated $ESK_{k,1}$ from each $E_k$, where $k=1,2,\dots,m$.		
	\item \textbf{Parallel R-tree Pruning Only (PRPO).} This comparative approach uses an R-tree index structure~\cite{10.1145/602259.602266} to prune out irrelevant data. With PRPO, each ECN $E_k$ uses the Minimum Bounding Rectangles (MBRs) from the R-tree index to prune out the irrelevant data in $SW_k$. When the new data comes into the sliding window $SW_k$, $E_k$ with PRPO recalculates new edge skyline $ESK_{k,1}$ with the help of R-tree Pruning, where $k=1,2,\dots,m$. After that, $E_k$ sends the whole updated $ESK_{k,1}$ to the server node $E_S$. When $E_S$ receives $ESK_{k,1}$ from edge nodes, $E_S$ will first save the received candidate data objects into sliding window $SW_S$. Once $SW_S$ changes, $E_S$ will use all the data in sliding window $SW_S$ to recalculate the global skyline. $E_S$ also utilizes the MBR information in the R-tree to perform data pruning so as to accelerate global skyline processing. 
	\item \textbf{Edge-assisted Parallel Uncertain Skyline (EPUS).} For the proposed EPUS, when $E_k$ receives new data, $E_k$ will consider the dominance relations between new/obsolete data, the current edge skyline set, $ESK_{k,1}$, and current edge skyline candidate set, $ESK_{k,2}$, and then update $ESK_{k,1}$ and $ESK_{k,2}$ as needed. The difference between EPUS and PRPO is that PRPO uploads the complete information of $ESK_{k,1}$, while EPUS only uploads the updated information of $ESK_{k,1}$ and $ESK_{k,2}$. The main server node uses the received update information from each ECN to update the global skyline set, $SK_1$, and skyline candidate set, $SK_2$, in the same way. 
\end{itemize}

\begin{table}[!t]
	\renewcommand{\arraystretch}{1.1}
	\centering
	\small
	\caption{Parameter Settings}
	\label{table:simulation_settings} 
	\begin{tabular}{|p{4cm}|l|p{1.2cm}|}
		\hline
		\textbf{Parameter} & \textbf{Values} & \textbf{Default Value} \\ \hline 
		Number of ECNs, $m$ & 2,4,\dots,10 & 6 \\  \hline
		Number of data objects, $N$ & - & 10000 \\  \hline
		Data dimensionality, $d$ & 2,3,\dots,10 & 2 \\  \hline
		Number of data instances in each data object, $n$  & 3,4,\dots,10 & 5 \\  \hline
		Radius of each data object, $r$ & 4,6,\dots,20 & 5 \\  \hline
		Domain range of data attribute & [0,1000] & - \\  \hline
		Size of sliding window on each ECN, $|SW_k|$ & 100,300,500,700 & 300 \\ 	\hline
		Size of sliding window on the server, $|SW_S|$ & - & $m*|SW_k|$ \\ 	\hline
		Size of each data object, $|u|$ (KB) & - & 3 \\ 	\hline
		Receiving data rate of the server node, $B_S$ (Mbps) & - & 1 \\ 	\hline
	\end{tabular}
\end{table}




The simulation results will be discussed below from three perspectives: 1) system architecture perspective and 2) data engineering perspective.



%
\begin{figure*}[!t]
	\centering
	\subfigure[Average System Latency]{
		\label{fig:number_of_ECNs:a} 
		\includegraphics[width=0.33 \textwidth]{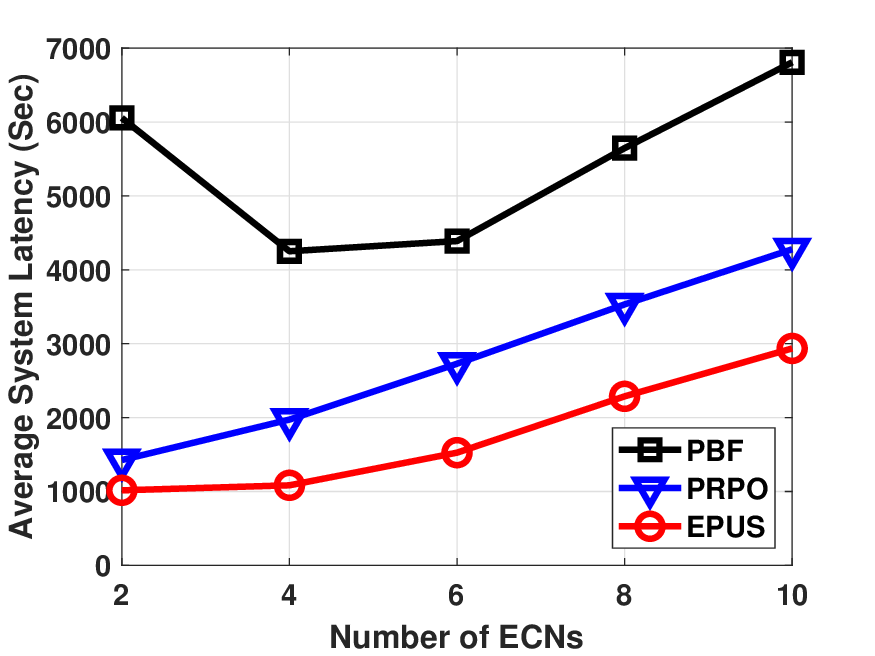}}%
	\subfigure[Average Transmission Latency]{
		\label{fig:number_of_ECNs:b} 
		\includegraphics[width=0.33 \textwidth]{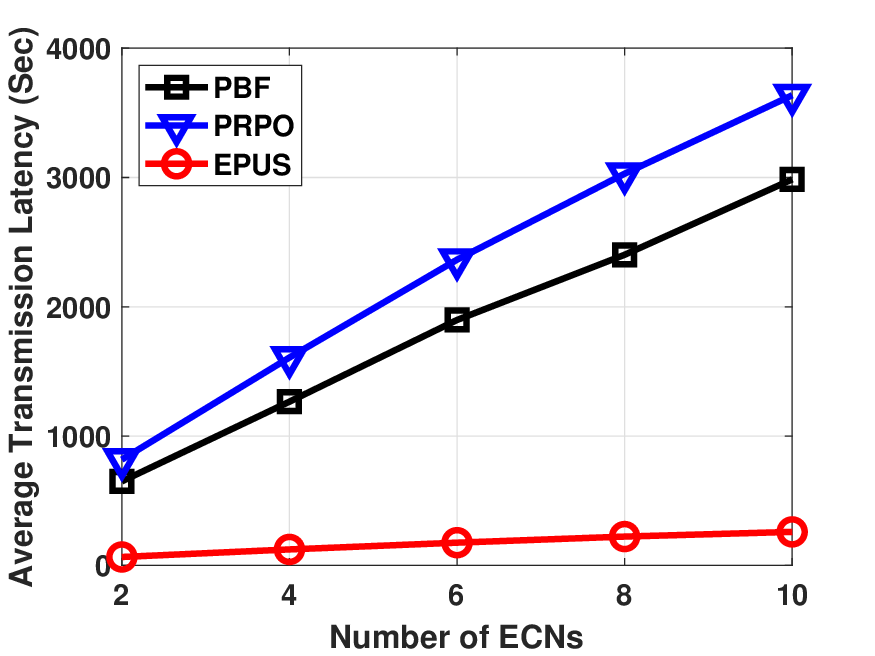}}%
	\subfigure[Average Computation Latency]{
		\label{fig:number_of_ECNs:c} 
		\includegraphics[width=0.33 \textwidth]{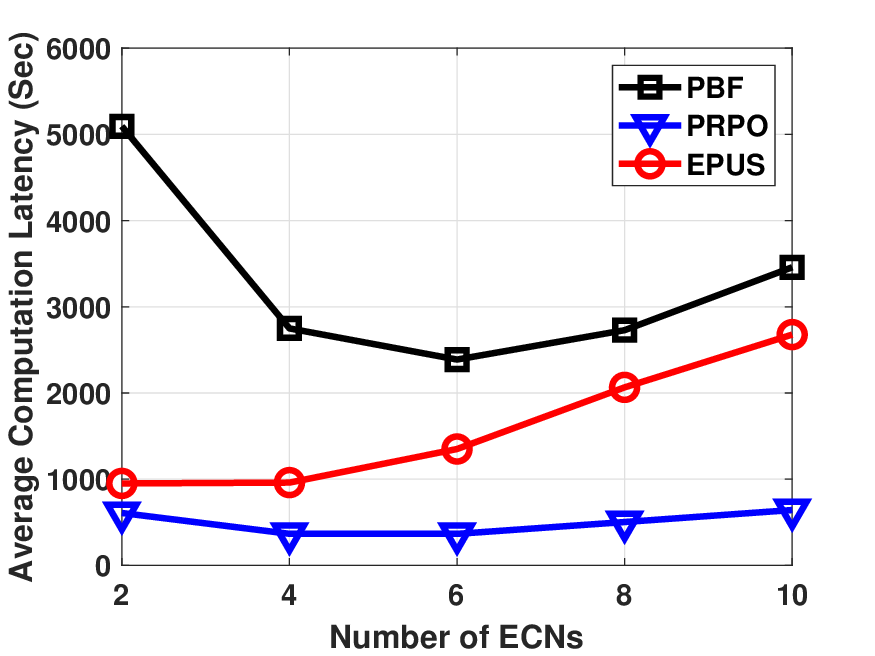}}%
	\caption{The effect of the number of ECNs on~\subref{fig:number_of_ECNs:a} the average system latency,~\subref{fig:number_of_ECNs:b} the average transmission latency, and~\subref{fig:number_of_ECNs:c} the average computation latency.}
	\label{fig:number_of_ECNs} 
\end{figure*}

\begin{figure*}[!t]
	\centering
	\subfigure[Average System Latency]{
		\label{fig:size_of_sw:a} 
		\includegraphics[width=0.33 \textwidth]{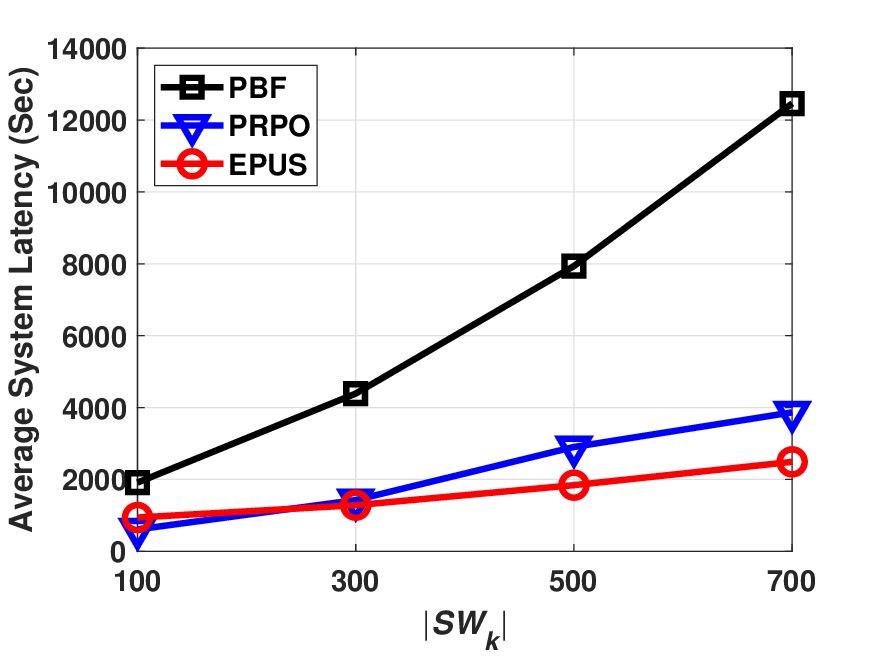}}%
	\subfigure[Average Transmission Latency]{
		\label{fig:size_of_sw:b} 
		\includegraphics[width=0.33 \textwidth]{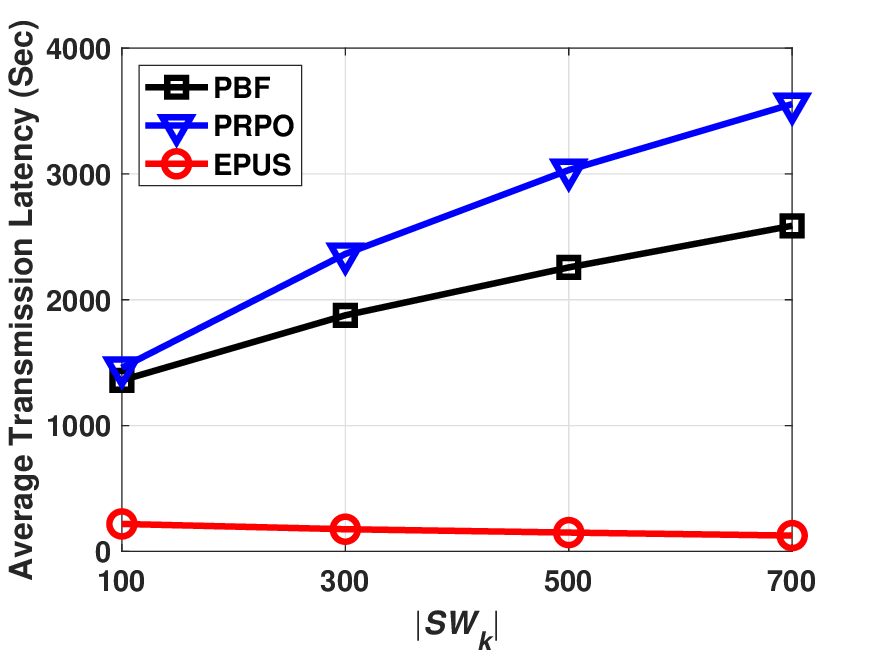}}%
	\subfigure[Average Computation Latency]{
		\label{fig:size_of_sw:c} 
		\includegraphics[width=0.33 \textwidth]{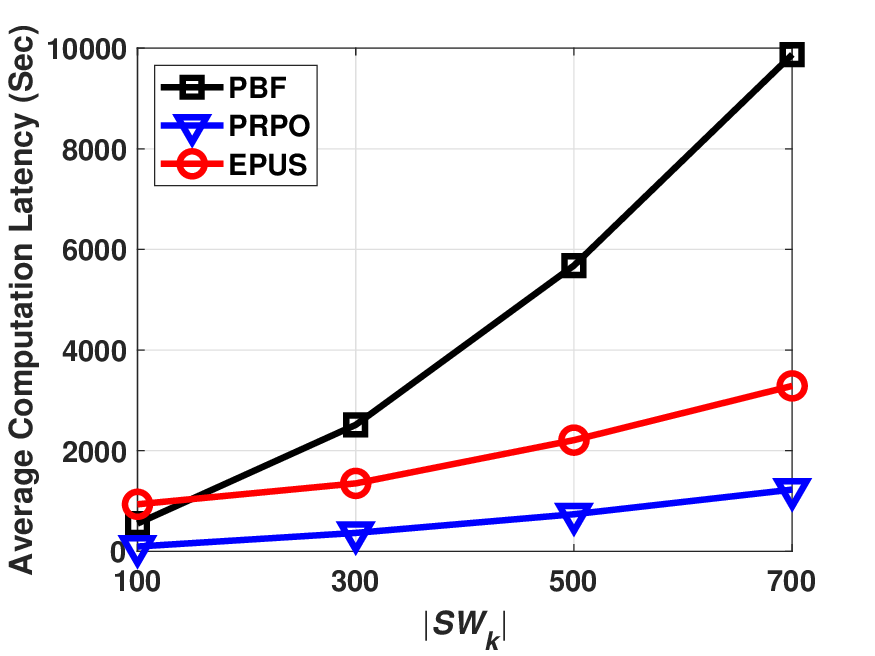}}%
	\caption{The effect of the size of $|SW_k|$ on~\subref{fig:size_of_sw:a} the average system latency and~\subref{fig:size_of_sw:b} the average transmission latency, and~\subref{fig:size_of_sw:c} the average computation latency, where $k=1,2,\dots,m$.}
	\label{fig:size_of_sw} 
\end{figure*}

\subsection{Results From A System Architecture Perspective}
From the perspective of a distributed edge computing system, we will show performance results on average system latency, average transmission latency, and average computation latency when varying the number of ECNs and the size of sliding window on each ECN.

\subsubsection{Number of Edge Computing Nodes}
As shown in Fig.~\ref{fig:number_of_ECNs}, we compare the performance of the proposed EPUS with the baseline methods, PBF and PRPO, as the number of ECNs, $m$, increases from 2 to 10. The average system latency results are presented in Fig.~\ref{fig:number_of_ECNs:a}. As the number of ECNs grows, EPUS exhibits a much slower increase in average system latency compared to PBF and PRPO, which both rise sharply.

Fig.~\ref{fig:number_of_ECNs:b} shows the average transmission latency, representing the time required to transmit data between ECNs and the server node. EPUS consistently achieves the lowest transmission latency among all methods, as it transmits only the updated portions of the edge skyline sets, $ESK_{k,1}$ and $ESK_{k,2}$, rather than the entire $ESK_{k,1}$ set as in PBF and PRPO. This selective transmission significantly reduces the amount of data sent, resulting in improved efficiency. Although PRPO employs R-tree pruning, it still transmits the complete edge skyline set, leading to the highest transmission latency. PBF, while not using any pruning, often produces a smaller $ESK_{k,1}$ than PRPO, resulting in transmission latency that is lower than PRPO but still higher than EPUS.

Fig.~\ref{fig:number_of_ECNs:c} shows the average computation latency, which is the time required to compute the edge skyline sets on each ECN and the global skyline set on the server node. PRPO has the lowest average computation latency because it only needs to compute the edge skyline set $ESK_{k,1}$ on each ECN using the R-tree pruning method. EPUS has a slightly higher average computation latency than PRPO because it needs to maintain both $ESK_{k,1}$ and $ESK_{k,2}$ on each ECN. PBF has the highest average computation latency because it does not use any pruning techniques and needs to compute the entire edge skyline set $ESK_{k,1}$ on each ECN.

\begin{figure*}[!ht]
	\centering
	\subfigure[Average System Latency]{
		\label{fig:data_dimensionality:a} 
		\includegraphics[width=0.33\textwidth]{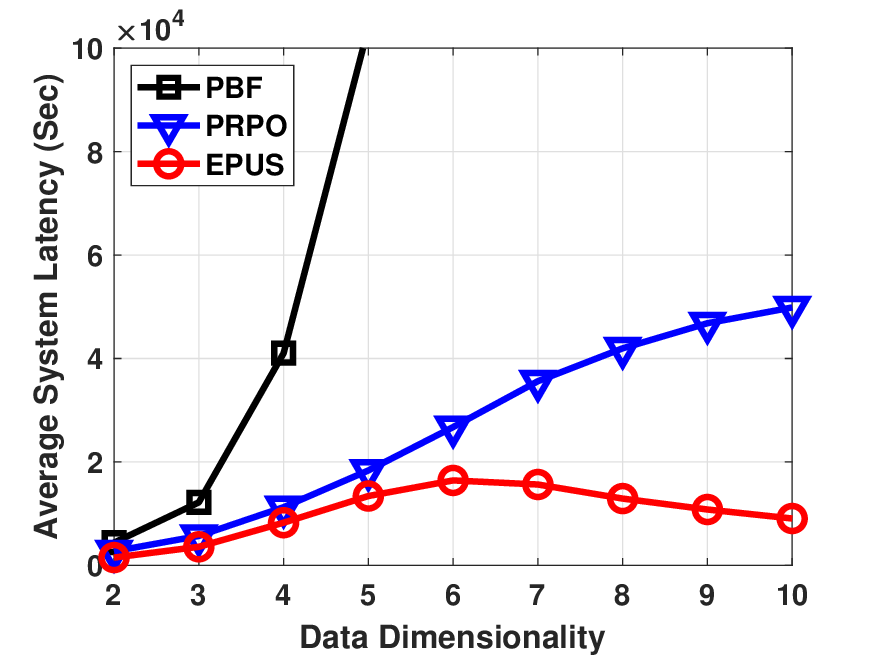}}%
	\subfigure[Average Transmission Latency]{
		\label{fig:data_dimensionality:b} 
		\includegraphics[width=0.33 \textwidth]{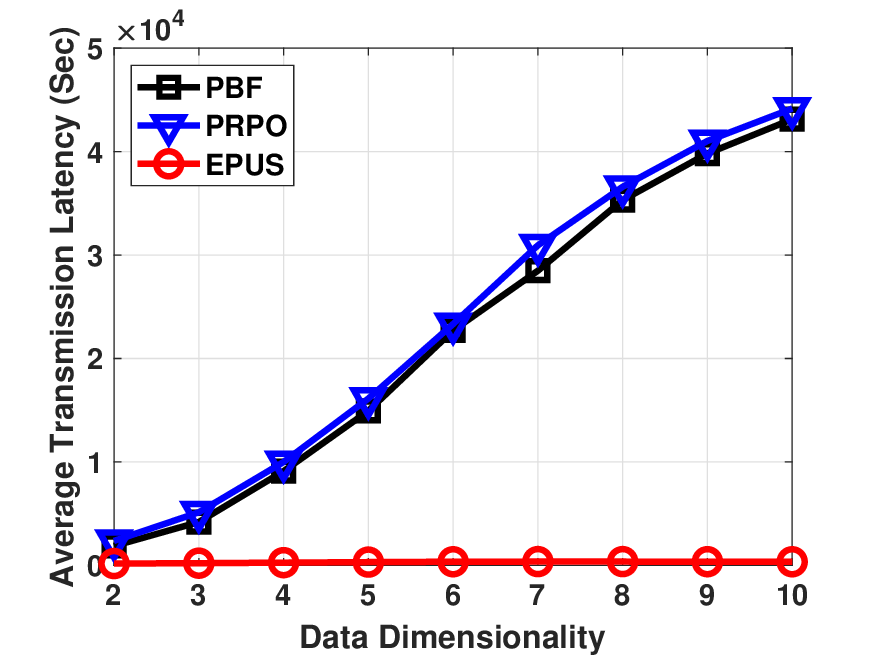}}%
	\subfigure[Average Computation Latency]{
		\label{fig:data_dimensionality:c} 
		\includegraphics[width=0.33 \textwidth]{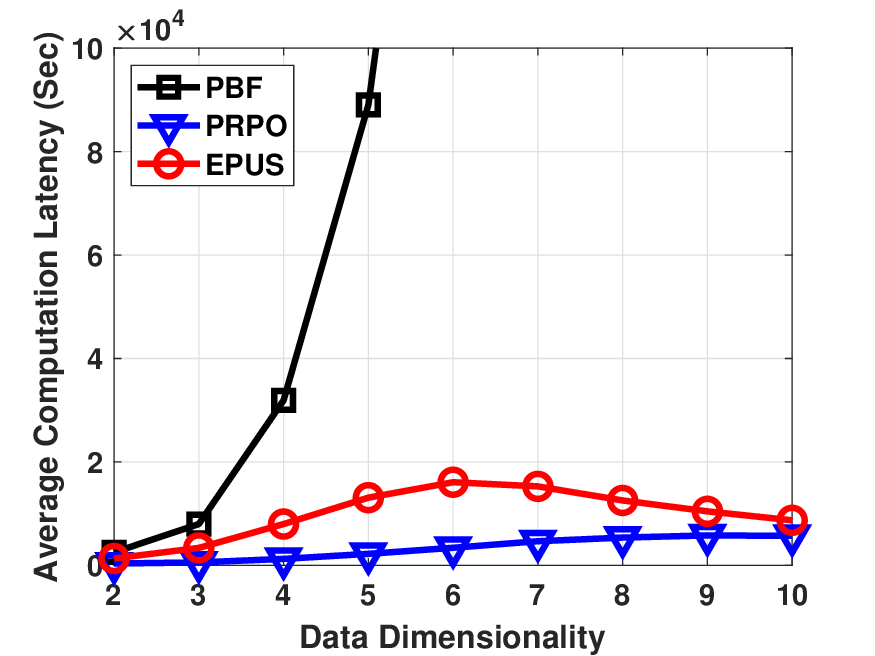}}%
	\caption{The effect of data dimensionality on~\subref{fig:data_dimensionality:a} the average latency and~\subref{fig:data_dimensionality:b} the average transmission latency, and~\subref{fig:data_dimensionality:c} the average computation latency. 
	}
	\label{fig:data_dimensionality} 
\end{figure*}

\begin{figure*}[!t]
	\centering
	\subfigure[Average System Latency]{
		\label{fig:number_instance:a} 
		\includegraphics[width=0.33 \textwidth]{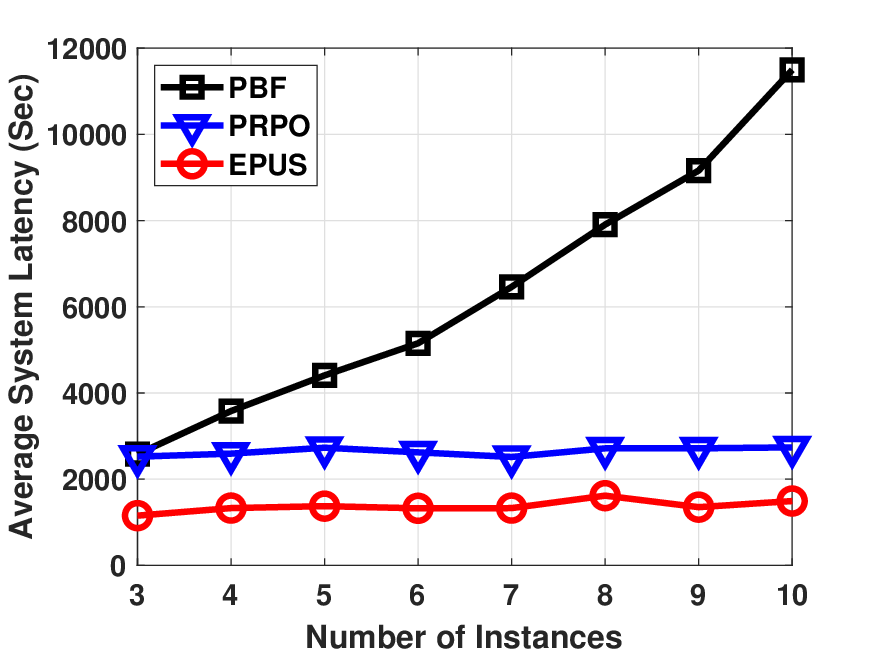}}%
	\subfigure[Average Transmission Latency]{
		\label{fig:number_instance:b} 
		\includegraphics[width=0.33 \textwidth]{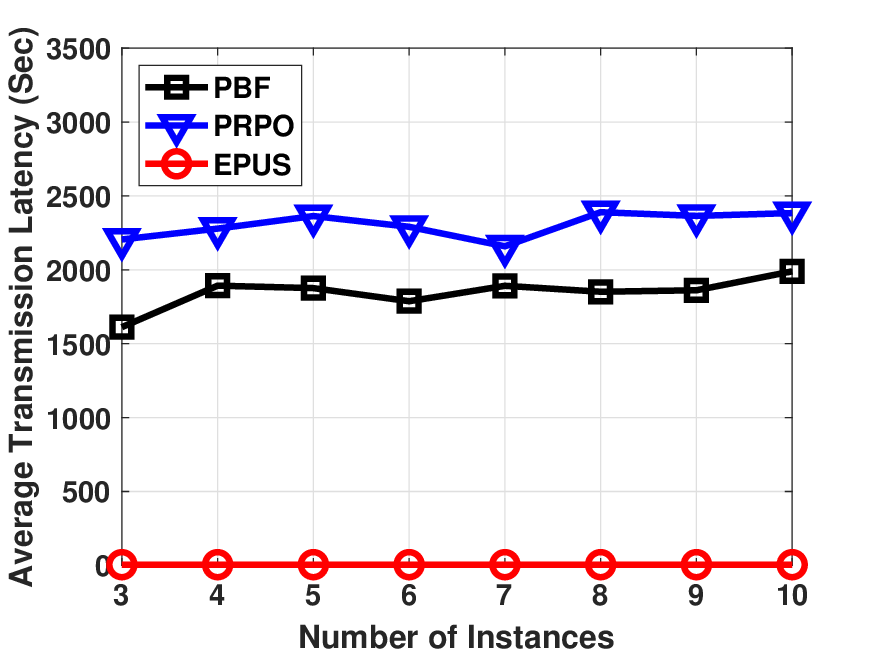}}%
	\subfigure[Average Computation Latency]{
		\label{fig:number_instance:c} 
		\includegraphics[width=0.33 \textwidth]{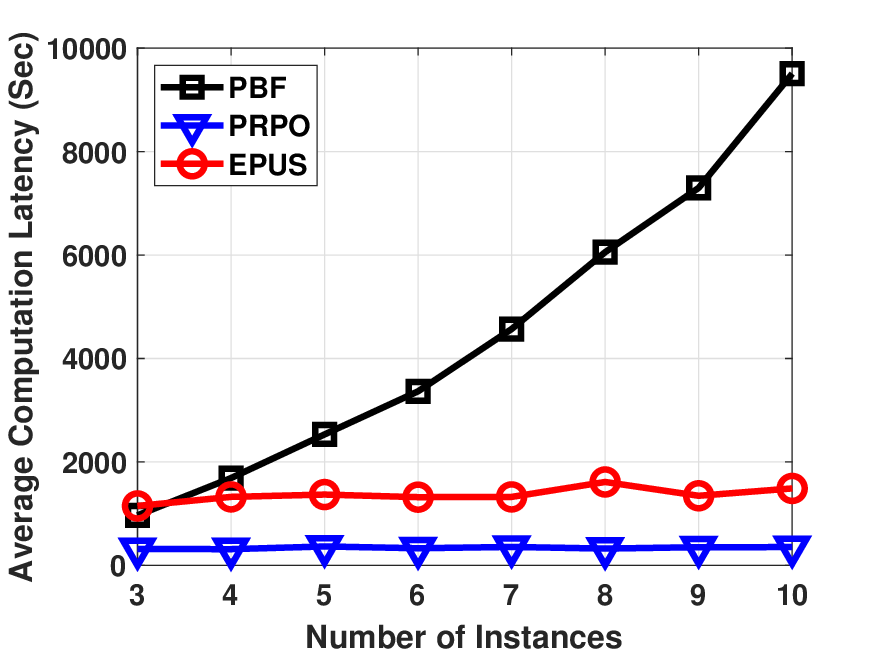}}%
	\caption{The effect of the number of instances on~\subref{fig:number_instance:a} the average system latency and~\subref{fig:number_instance:b} the average transmission latency and~\subref{fig:number_instance:c} the average computation latency.
	}
	\label{fig:number_instance} 
\end{figure*}

\subsubsection{Size of the Sliding Window on Each ECN}
We also investigate the effect of the size of sliding window on each ECN, $|SW_k|$, on the average system latency, average transmission latency, and average computation latency. The results are shown in Fig.~\ref{fig:size_of_sw}. As $|SW_k|$ increases, the average system latency in Fig.~\ref{fig:size_of_sw:a} increases for all three methods. However, EPUS shows a much slower increase compared to PBF and PRPO since it balances the computation and transmission latency more effectively.

Fig.~\ref{fig:size_of_sw:b} presents the average transmission latency as $|SW_k|$ increases. EPUS demonstrates a slight decrease in transmission latency, while both PBF and PRPO show a more pronounced increase. This improvement is attributed to EPUS's ability to efficiently prune irrelevant data and transmit only the updated portions of $ESK_{k,1}$ and $ESK_{k,2}$. In contrast, PBF and PRPO lack effective pruning mechanisms and must transmit the entire, and increasingly larger, edge skyline set $ESK_{k,1}$, resulting in higher transmission latency.

Fig.~\ref{fig:size_of_sw:c} illustrates the average computation latency, which also increases with $|SW_k|$ for all methods. 
EPUS maintains a balance between computation and transmission latency, achieving lower average computation latency than PBF while being slightly higher than PRPO. This is because EPUS requires additional computation to maintain both $ESK_{k,1}$ and $ESK_{k,2}$, but it still benefits from the pruning techniques used in PRPO.

\subsection{Results From A Data Engineering Perspective}
After investigating the performance from a system architecture perspective, we now focus on the impact of data characteristics on the performance of EPUS. We will analyze how the data dimensionality, number of instances, and radius size of data objects affect the average system latency, average transmission latency, and average computation latency.

\subsubsection{Data Dimensionality}
In Fig.~\ref{fig:data_dimensionality}, we analyze the impact of data dimensionality on the performance of EPUS. As the data dimensionality increases, the average system latency in Fig.~\ref{fig:data_dimensionality:a} shows a significant increase for all methods. This is because higher-dimensional data requires more complex computations and larger edge skyline sets, leading to increased processing time.

In Fig.~\ref{fig:data_dimensionality:b}, the average transmission latency also increases with data dimensionality. EPUS continues to outperform PBF and PRPO in terms of transmission latency, as it only transmits the updated portions of $ESK_{k,1}$ and $ESK_{k,2}$, while PBF and PRPO transmit the entire edge skyline set $ESK_{k,1}$, which grows larger with higher dimensionality.

Fig.~\ref{fig:data_dimensionality:c} shows the average computation latency, which also increases with data dimensionality. EPUS maintains a balance between computation and transmission latency, achieving lower average computation latency than PBF while being slightly higher than PRPO. This is due to EPUS's need to maintain both $ESK_{k,1}$ and $ESK_{k,2}$, which requires additional computation compared to PRPO's single edge skyline set.

\begin{figure*}[!t]
	\centering
	\subfigure[Average System Latency]{
		\label{fig:radius_size:a} 
		\includegraphics[width=0.33 \textwidth]{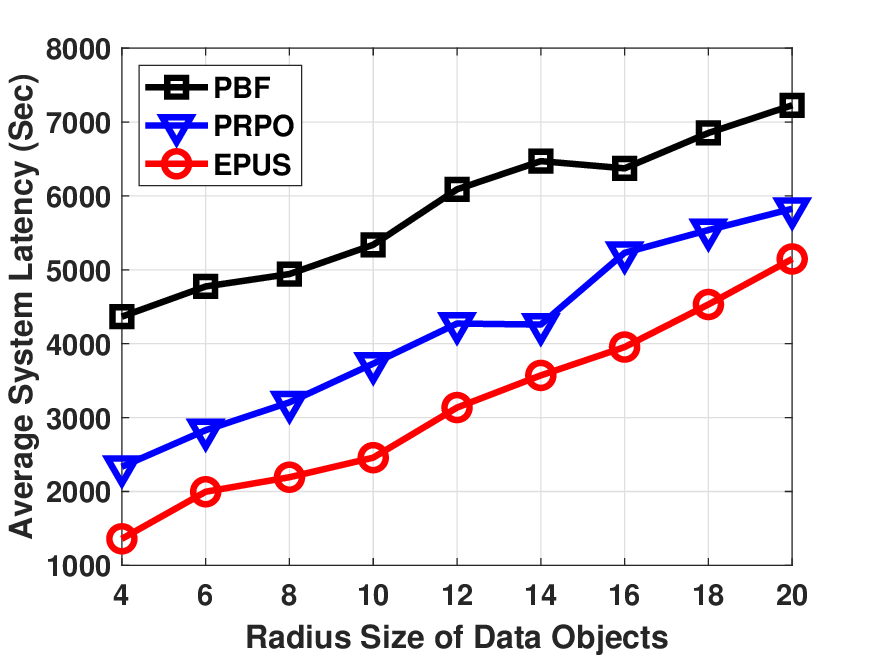}}%
	\subfigure[Average Transmission Latency]{
		\label{fig:radius_size:b} 
		\includegraphics[width=0.33 \textwidth]{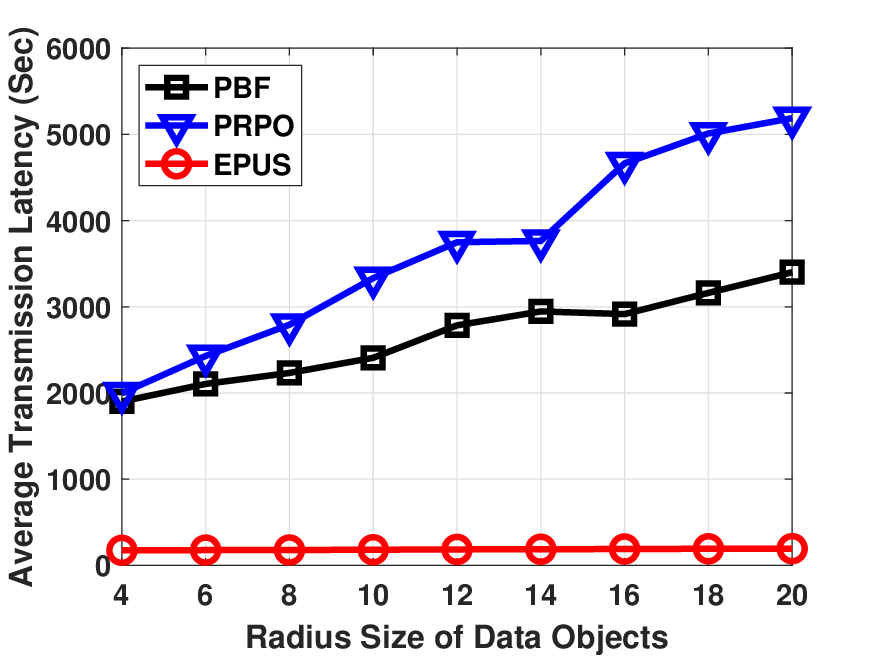}}%
	\subfigure[Average Computation Latency]{
		\label{fig:radius_size:c} 
		\includegraphics[width=0.33 \textwidth]{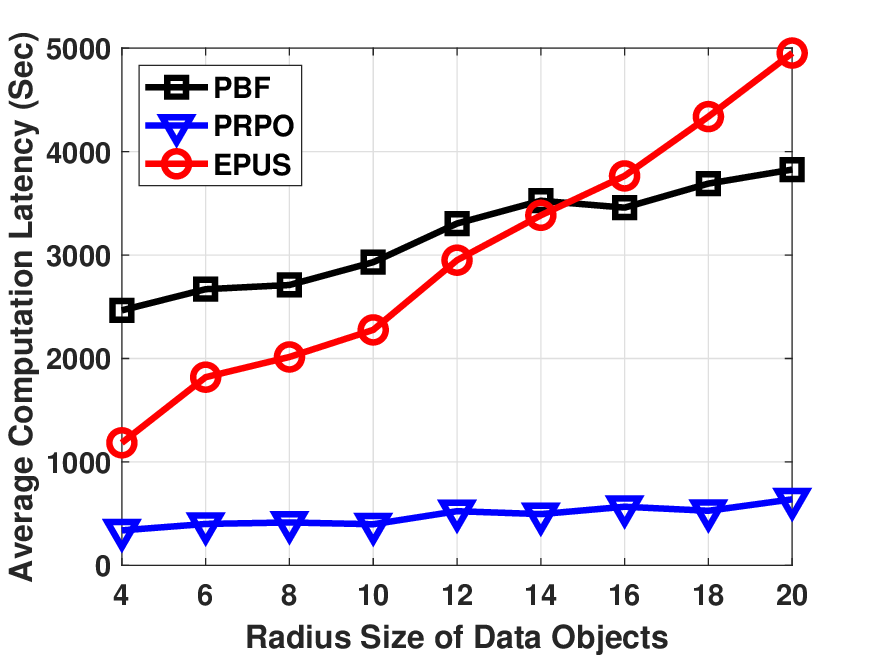}}%
	\caption{The effect of the radius size of data objects on~\subref{fig:radius_size:a} the average system latency and~\subref{fig:radius_size:b} the average transmission latency and~\subref{fig:radius_size:c} the average computation latency.
	}
	\label{fig:radius_size} 
\end{figure*}

\subsubsection{Number of Instances}
In Fig.~\ref{fig:number_instance}, we investigate the impact of the number of instances in each data object on the performance of EPUS. As the number of instances increases, the average system latency in Fig.~\ref{fig:number_instance:a} shows a significant increase for PBF but remains relatively stable for both EPUS and PRPO methods. This is because PBF needs to compute the entire edge skyline set $ESK_{k,1}$ for each ECN and global skyline for the server, which becomes more complex with more instances. Conversely, EPUS and PRPO can leverage R-tree based pruning techniques to avoid checking irrelevant data instances, thus reducing the computation complexity.

Fig.~\ref{fig:number_instance:b} shows that all methods are irrelevant to the number of instances in terms of average transmission latency. This is because the transmission latency is primarily determined by the size of the edge skyline set $ESK_{k,1}$, which does not change significantly with the number of instances. PRPO uses an R-tree to approximately prune irrelevant objects and instances, so it maintains a larger size of $ESK_{k,1}$ compared to PBF, thus resulting in a higher transmission latency.
EPUS continues to outperform PBF and PRPO in terms of transmission latency, as it only transmits the updated portions of $ESK_{k,1}$ and $ESK_{k,2}$.

Fig.~\ref{fig:number_instance:c} illustrates the average computation latency, which increases with the number of instances for PBF but remains relatively stable for EPUS and PRPO. This is because EPUS and PRPO can efficiently prune irrelevant data instances using R-tree based pruning techniques, while PBF needs to compute the entire edge skyline set $ESK_{k,1}$ for each ECN and global skyline for the server in a brute-force manner, leading to higher computation latency.

\subsubsection{Radius Size of Data Objects}
In Fig.~\ref{fig:radius_size}, we analyze the impact of the radius size of data objects on the performance of EPUS. As the radius size increases, the average system latency in Fig.~\ref{fig:radius_size:a} shows a significant increase for all methods. This is because larger radius sizes lead to larger edge skyline sets, which require more complex computations and longer processing time.

In Fig.~\ref{fig:radius_size:b}, the average transmission latency also increases with the radius size. EPUS continues to outperform PBF and PRPO in terms of transmission latency, as it only transmits the updated portions of $ESK_{k,1}$ and $ESK_{k,2}$, while PBF and PRPO transmit the entire edge skyline set $ESK_{k,1}$, which grows larger with larger radius sizes.

Fig.~\ref{fig:radius_size:c} shows the average computation latency, which also increases with the radius size. The increasing trend of radius of data objects on average computation latency for EPUS is higher than PBF and PRPO. EPUS maintains a lower average computation latency than PBF when $r$ is smaller than 14. However, as the radius size increases, the average computation latency of EPUS becomes higher than PBF. This is because EPUS needs to maintain both $ESK_{k,1}$ and $ESK_{k,2}$, which requires twice skyline computation compared to PBF. PRPO has the lowest average computation latency because it only needs to compute the edge skyline set $ESK_{k,1}$ once on each ECN using the R-tree pruning method.

\color{black}

\section{Conclusion}
\label{sec:conclusion}
In this study, we proposed a heuristic algorithm, Edge-Assisted Parallel Uncertain Skyline (EPUS), to efficiently process probabilistic skyline queries over uncertain data streams in edge computing environments. By leveraging the Candidate Skyline Set (CSS) concept, EPUS effectively prunes irrelevant data, reducing average computation latency and transmission cost between edge computing nodes and the main server. Simulation results demonstrate that EPUS outperforms brute-force approaches, particularly in terms of system latency and scalability. However, efficient skyline query processing for high-dimensional uncertain data remains an open challenge for future research.


In the future, we will apply the proposed framework and some customized schemes with domain knowledge to some emerging low-latency multiple criteria decision making applications~\cite{9177297}~\cite{9946428}.


%



\ifCLASSOPTIONcompsoc
\else
\fi


\ifCLASSOPTIONcaptionsoff
  \newpage
\fi



\bibliographystyle{IEEEtran}
\bibliography{reference}

\begin{IEEEbiography}[{\includegraphics[width=1in,height=1.25in,clip,keepaspectratio]{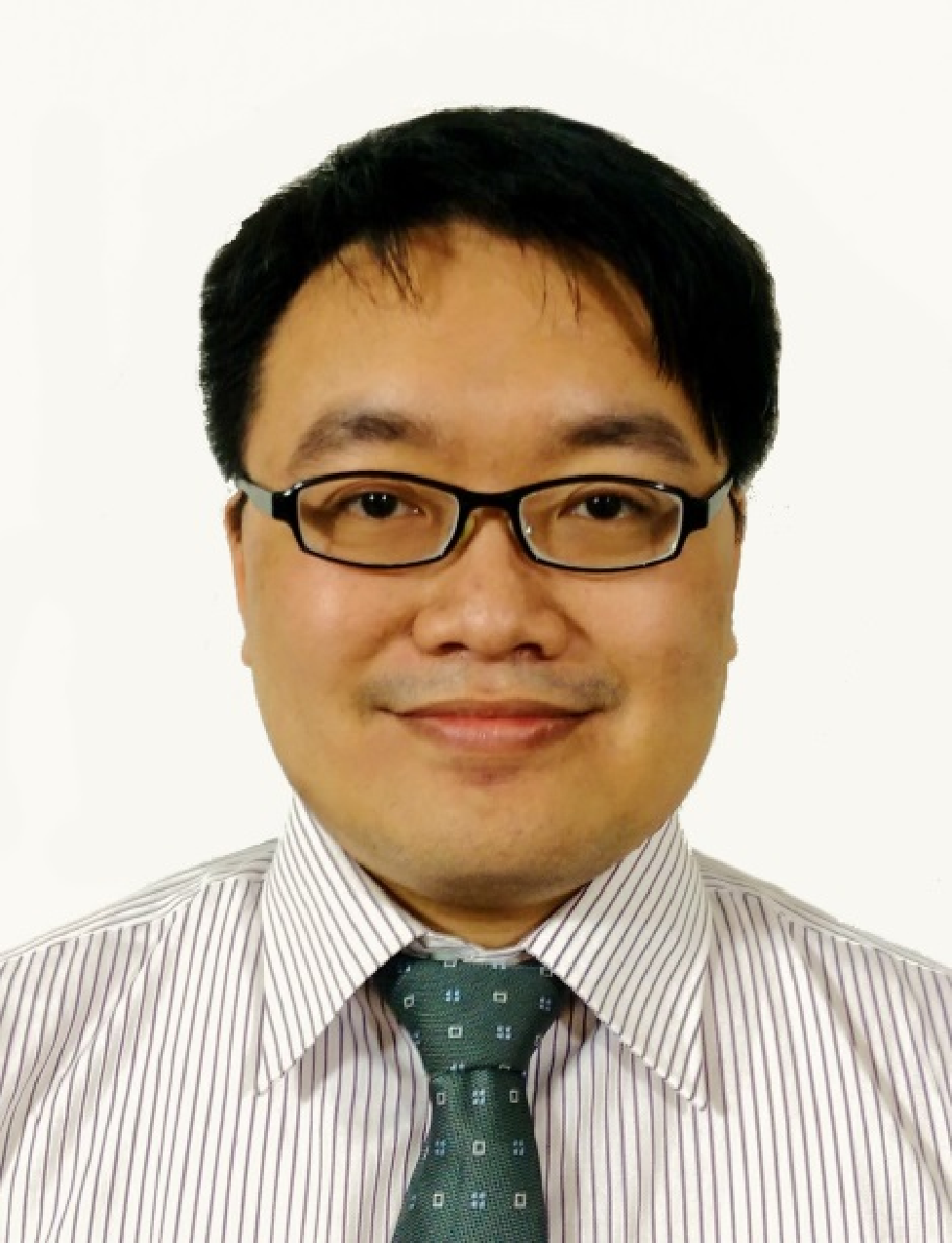}}]{Chuan-Chi Lai}
	(S'13--M'18) received the Ph.D. degree in Computer Science and Information Engineering from National Taipei University of Technology, Taipei, Taiwan, in 2017. He was a postdoctoral research fellow (2017--2020) and contract assistant research fellow (2020) with the Department of Electrical and Computer Engineering, National Chiao Tung University, Hsinchu, Taiwan. From 2017 to 2021, he served as an assistant professor at the Department of Information Engineering and Computer Science, Feng Chia University, Taichung, Taiwan. He is currently an assistant professor at the Department of Communications Engineering, National Chung Cheng University, Chiayi, Taiwan. He is a member of the IEEE Vehicular Technology Society and the IEEE Communications Society.
    His research interests include resource allocation, data management, information dissemination, and distributed query processing for moving objects in emerging applications such as the Internet of Things, edge computing, and next-genaration wireless networks. Dr. Lai received the Postdoctoral Researcher Academic Research Award from the Ministry of Science and Technology, Taiwan, in 2019, Best Paper Awards at WOCC 2021 and WOCC 2018, and the Excellent Paper Award at ICUFN 2015.
\end{IEEEbiography}

\begin{IEEEbiography}[{\includegraphics[width=1in,height=1.25in,clip,keepaspectratio]{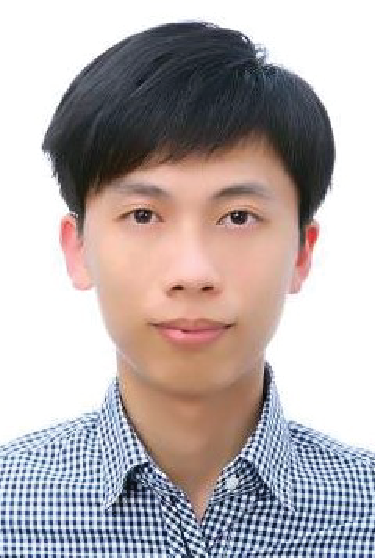}}]{Yan-Lin~Chen}
	is an engineer at Trend Micro Inc., Taipei, Taiwan. He received his MS degree in the Department of Computer Science and Information Engineering, National Taipei University of Technology (Taipei Tech), Taiwan in 2019. He also received a BBA degree in the  Department of Finance and Cooperative Management, National Taipei University, Taiwan. Yan-Lin joined Applied Computing Laboratory in 2017 and  shows his interest in topics related to parallel query processing algorithms for data analytic applications.
\end{IEEEbiography}

\begin{IEEEbiography}[{\includegraphics[width=1in,height=1.25in,clip,keepaspectratio]{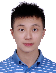}}]{Bo-Xin Liu}
	is currently an engineer at Taiwan Semiconductor Manufacturing Company, Ltd (TSMC), located in Miaoli, Taiwan. He obtained his M.S. degree in Computer Science and Information Engineering from National Taipei University of Technology (Taipei Tech) in 2024. He earned his bachelor's degree in the same field from the National Defense University, Taiwan. In 2022, he joined the Applied Computing Laboratory, where his research interests include parallel query processing and data analytics, particularly in the context of high-performance computing applications.
\end{IEEEbiography}

\begin{IEEEbiography}[{\includegraphics[width=1in,height=1.25in,clip,keepaspectratio]{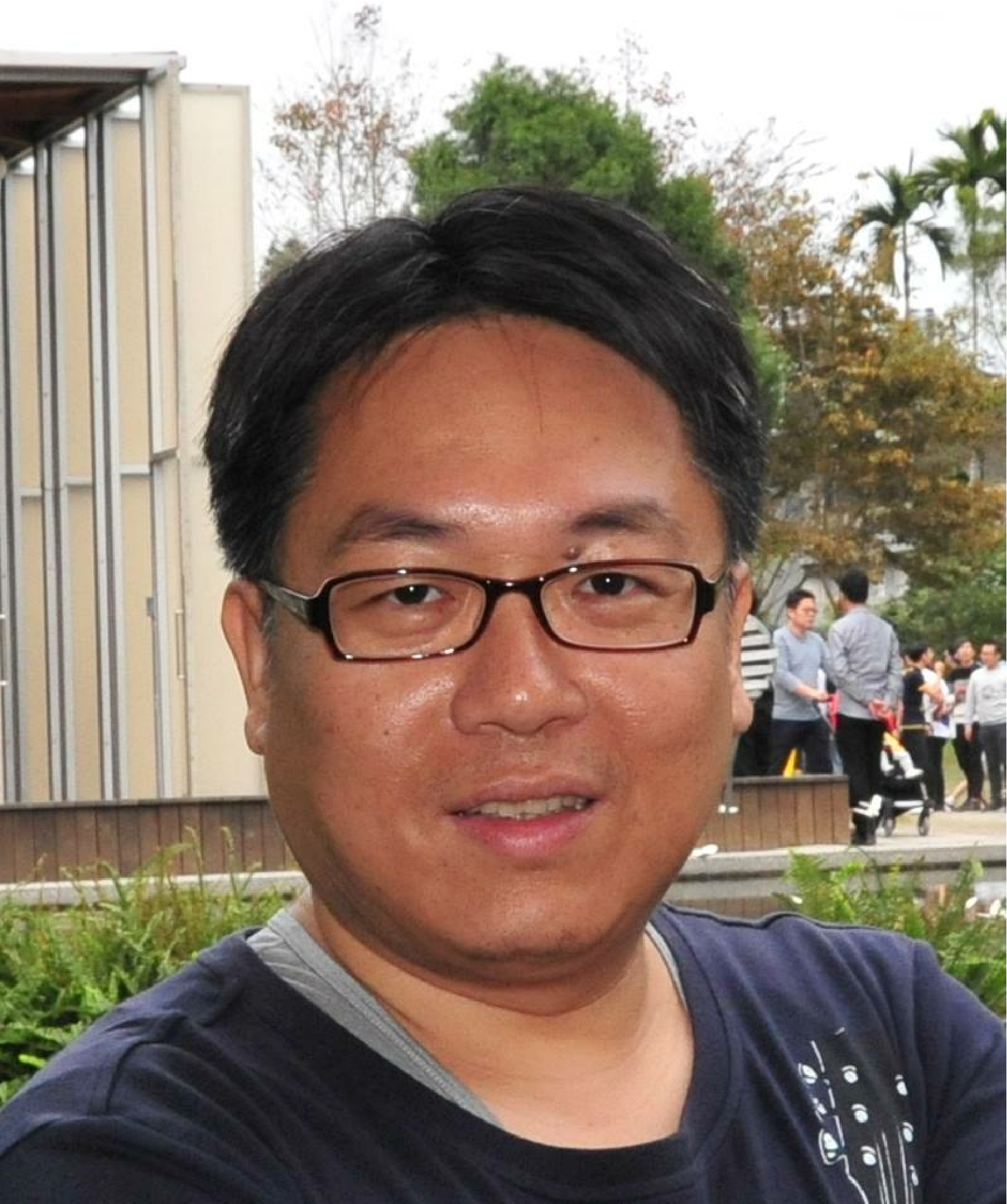}}]{Chuan-Ming Liu}
	(M'03) received the Ph.D. degree in computer science from Purdue University, West Lafayette, IN, USA, in 2002. 
	Dr. Liu is a professor in the Department of Computer Science and Information Engineering, National Taipei University of Technology (Taipei Tech), Taiwan. In 2010 and 2011, he has held visiting appointments with Auburn University, Auburn, AL, USA, and the Beijing Institute of Technology, Beijing, China. He has services in many journals, conferences and societies as well as published more than 100 papers in many prestigious journals and international conferences. His current research interests include big data management and processing, uncertain data management, data science, spatial data processing, data streams, ad-hoc and sensor networks, and location based services.
\end{IEEEbiography}

\end{document}